\documentclass[12pt,preprint]{article}

\usepackage{hyperref}
\usepackage{graphicx}
\usepackage{fullpage,amsmath,amssymb,latexsym}
\usepackage{multirow}
\usepackage{rotating} 
\usepackage{natbib}
\usepackage{url}
\usepackage{graphicx}
\usepackage{esint}
\usepackage[english]{babel}
\usepackage{subfig}
\usepackage{pdfpages}

\title{Jupiter's formation and its primordial internal structure}
\author{Michael Lozovsky$^{1,2}$, Ravit Helled$^{1,2,*}$, Eric D.~Rosenberg$^1$ \& Peter Bodenheimer$^{3}$}
\providecommand{\keywords}[1]{{\textit{Subject headings:}} #1}

\begin{document}
\pdfoutput=1
\maketitle
\begin{center}
\textit{$^1$Department of Geosciences, 
Tel-Aviv University, Tel-Aviv, Israel\\
$^2$Center for Theoretical Astrophysics \& Cosmology, 
Institute for Computational Science, 
University of Zurich, Zurich, Switzerland\\
$^3$UCO/Lick Observatory, University of California Santa Cruz, USA\\
$^*$Corresponding author: rhelled@physik.uzh.ch}
\end{center}
   
\begin{abstract}
The composition of Jupiter and the primordial distribution of the heavy elements are determined by its formation history. 
As a result, in order to constrain the primordial internal structure of Jupiter the growth of the core and the deposition and settling of accreted planetesimals must be followed in detail. 
In this paper we determine the distribution of the heavy elements in proto-Jupiter and determine the mass and composition of the core. 
We find that while the outer envelope of proto-Jupiter is typically convective and has an homogeneous composition, the innermost regions have compositional gradients. 
In addition, the existence of heavy elements in the envelope leads to much 
higher internal temperatures (several times 10$^4$ K) than in the case 
of a hydrogen-helium envelope. 
The derived core mass depends on the actual definition of the core: 
if the core is defined as the region in
which the heavy-element mass fraction is above some limit (say 0.5), 
then it can be much more massive ($\sim$ 15 M$_{\oplus}$) and more extended
(10\% of the planet's radius) than in the case where the core 
is just the region with 100\% heavy elements. In the former case 
Jupiter's core also consists of hydrogen and helium. Our results should be taken into account when constructing internal structure models of Jupiter and when interpreting the upcoming data from the {\it Juno} (NASA) mission. 
\end{abstract}

\par

\section{Introduction}

In the standard model for giant planet formation, core accretion (hereafter, {\it CA}), the formation of a gaseous planet begins with the growth of a heavy-element core \citep[e.g.,][]{Pollack1996,Alibert2005}. The exact mass and composition of the core are unknown, but it is commonly agreed that it should be of the order of at least several Earth masses (M$_{\oplus}$) in order to allow the followup of gas accretion. The core's composition is assumed to be rocky and/or icy. Giant planet formation models also typically assume that the accreted solid material (planetesimals) falls all the way to the center, increasing the core's mass. Indeed, at the beginning of their formation, giant planets are capable of binding only a very tenuous envelope so that in-falling planetesimals essentially reach the core directly, but as the growing planet accretes a gaseous atmosphere, the planetesimals do not necessarily reach the core, but instead, dissolve in the gaseous envelope.
\par

A major science objective in giant planet studies is to relate the observed planetary structure to the origin and evolution of the planet. 
For years now, the existence of a heavy-element (high-Z) core and/or enriched envelope has been taken as support of the {\it CA} model. The connection between planetary origin and internal structure, however, is non trivial, and in fact, different formation mechanisms and birth environments lead to a large range of compositions and internal structures \citep[e.g.,][]{Helled2013}. 
Even within the {\it CA} framework, the primordial internal structures of the planets are not well constrained. 
For giant exoplanets, an estimate of the total mass of heavy elements in the planet is sufficient for planetary characteriszation because (at present) it is difficult to constrain the heavy-element distribution. 
A possible approach to this question employs the tidal Love number and is 
applied to the exoplanet HAT-P 13b by \citet{Kramm2012} and \citet{Buhler2016}. This is different for the case of Jupiter for which we have accurate measurements of the gravitational field, and 
therefore, estimates of its internal density distribution can be determined. 
\par

In this paper, we determine the heavy-element distribution and core mass in proto-Jupiter at different stages during its formation in the {\it CA} model.  
We follow the accreted planetesimals as they enter the planetary envelope and determine their distribution accounting for settling and convective mixing. 
Formation models with different solid-surface densities and planetesimal sizes are considered, as well as different definitions of the planetary core. 
This work aims to provide a more complete theoretical framework for the interpretation of {\it Juno} data. The initial orbit insertion of the  {\it Juno} (NASA) spacecraft was on July 4th. {\it Juno} will measure Jupiter's gravitational moments $J_{n}$ and atmospheric composition below the cloud-level  \citep[e.g.,][]{Helled2014}. These measurements will provide tighter constraints on Jupiter's density distribution, and hence, its internal structure, with the aim of using this information to better understand Jupiter's origin. 
In order to link Jupiter's internal structure and formation history, we first need to understand what is the expected internal structure from formation models, and its sensitivity to the various model assumptions. 
This work aims to explore and put limits on the primordial internal structure of Jupiter, in particular, its core properties. 
\par 

\section{Methods and model assumptions}

\subsection{Planet formation models}
\label{sec:Initial models}

In order to determine the heavy-element
distribution in proto-Jupiter, we use four giant planet formation 
models, whose properties are summarized in Table 1. 
Models are produced using the standard core accretion code described in
\citet{Pollack1996} and updated by 
\citet{Lissauer2008}. 
Model-D is the
model with solid surface density $\sigma=10$ g cm$^{-2}$ described by Movshovitz
et al.~(2010). In this model, the grain opacities were determined
through detailed simulations of grain settling and coagulation. In 
models A through C, the grain opacities were determined approximately
as a fraction $f\ll1$ of interstellar opacities, with $f$ adjusted to 
reproduce the formation times found by Movshovitz et al.~for 
$\sigma=10$ g cm$^{-2}$ and $\sigma=6$ g cm$^{-2}$. 
The formation models provide the distribution of the physical properties of the planetary envelope such as the temperature $T$, pressure $P$, luminosity $L$, opacity $\kappa$, mass $M$, and the total mass of gas (hydrogen and helium in proto-solar ratio) and the total mass of heavy elements as a function of time. 

These formation models are derived under the assumption that all the planetesimals settle to the center and join the core, and the inferred crossover time and crossover mass (mass of solids equals the mass of hydrogen-helium gas $M_Z$ = $M_{H+He}$) correspond to the model under this assumption. If the heavy elements are allowed to stay in the envelope and are accounted for self-consistently in the formation model, the formation history, and therefore also crossover mass and crossover time are expected to change as well.  

As can be seen from the table, the formation timescale and the core mass are sensitive to the assumed solid-surface densities and planetesimal sizes \citep[e.g.,][]{Pollack1996,Movshovitz2010}. Higher solid-surface density $\sigma$ and smaller planetesimal sizes lead to faster growth. 
The crossover time for the different models ranges between 0.9 and 1.5 Myr. These timescales, however, are calculated for envelopes that are metal-free, the time to reach crossover is expected to be shorter if the heavy elements are taken into account in the calculation \citep[e.g.,][]{Venturini2016,Hori2010}, and as a result, these crossover times should be taken as upper bounds. 
\par

\subsection{Accretion and settling of heavy elements}
\label{sec:Settling of heavy material}
Note that
in the models of Pollack et al.~(1996) it was assumed that all the
added planetesimal material eventually sank to the core; here we relax
that assumption. The accreted planetesimals are assumed to be composed of water (H$_2$O), rock (SiO$_2$), and organic material (CHON) 
in relative mass fractions of 0.4:0.3:0.3, respectively. 
The organic material is represented by Hexacosane (C$_{26}$H$_{54}$), which is a paraffin-like substance. The accretion rate of planetesimals is given by the standard fundamental expression 
(e.g., Equation (1) of Movshovitz et al.), with the gravitational enhancement factor $F_g$ given by \citet{Greenzweig1992}. We follow the trajectory of the planetesimal as they pass through the protoplanetary envelope. At each step of the trajectory we compute the motion of the planetesimals in response to gas drag and gravitational forces. The effects of heating, ablation and fragmentation of planetesimals are also included. This calculation provides the distribution of heavy elements in the planetary envelope (see \cite{Pollack1996} for details). 
\par

When the heavy elements (planetesimals) dissolve, they are assumed to be fully vaporized, then, the amount of heavy material that remains at a given layer is limited by its vapor pressure and temperature. Following \citet{Iaroslavitz2007}, we determine the partial pressure of the ablated heavy elements and compare it to the vapor pressure of the layer. If the partial pressure exceeds the vapor pressure, some of the accreted material is assumed to condense and settle to the layer below, leaving the (upper) layer fully saturated. 
The partial pressure of a given ablated material is calculated using the ideal gas law by, 
\begin{equation}
P=\frac{\Re\delta m T}{V\mu} \quad,
\end{equation}
where $\Re$ is the universal gas constant, $\delta m$ is the mass ablated within the layer, $T$ is the temperature, $V$ is the volume of the layer (shell), and $\mu$ is the mean molecular weight of the material. We assume that the vapor pressure of a given substance above the critical temperature becomes infinite. 
The infinite vapor pressures allow any amount of the substance to be accommodated in the vapor without reaching saturation. The vapor pressure for the different materials is taken from \citet{Podolak1988} and \citet{Iaroslavitz2007}: 
\begin{eqnarray}
P^{water}_{vap}&=&e^{-5640.34/T+28.867}, T_{crit} = 647.3 K  \quad,\\
P^{CHON}_{vap}&=&6.46 \cdot 10^{13} e^{-12484.5/T}, T_{crit} = 843.4 K \quad,\\
P^{rock}_{vap}&=&10^{-24605/T+13.176}, T_{crit} = 4,000 K \quad.
\end{eqnarray}
The procedure is then repeated in each successive layer. If the heavy elements reach the innermost layer, they are added to the center of the planet, i.e., the core.

\subsection{The heavy-element distribution}
\label{subsec:Mass profile}
\begin{figure}[!h]\centering
\subfloat{\includegraphics[trim=80 180 100 200,clip,width=0.42\columnwidth]{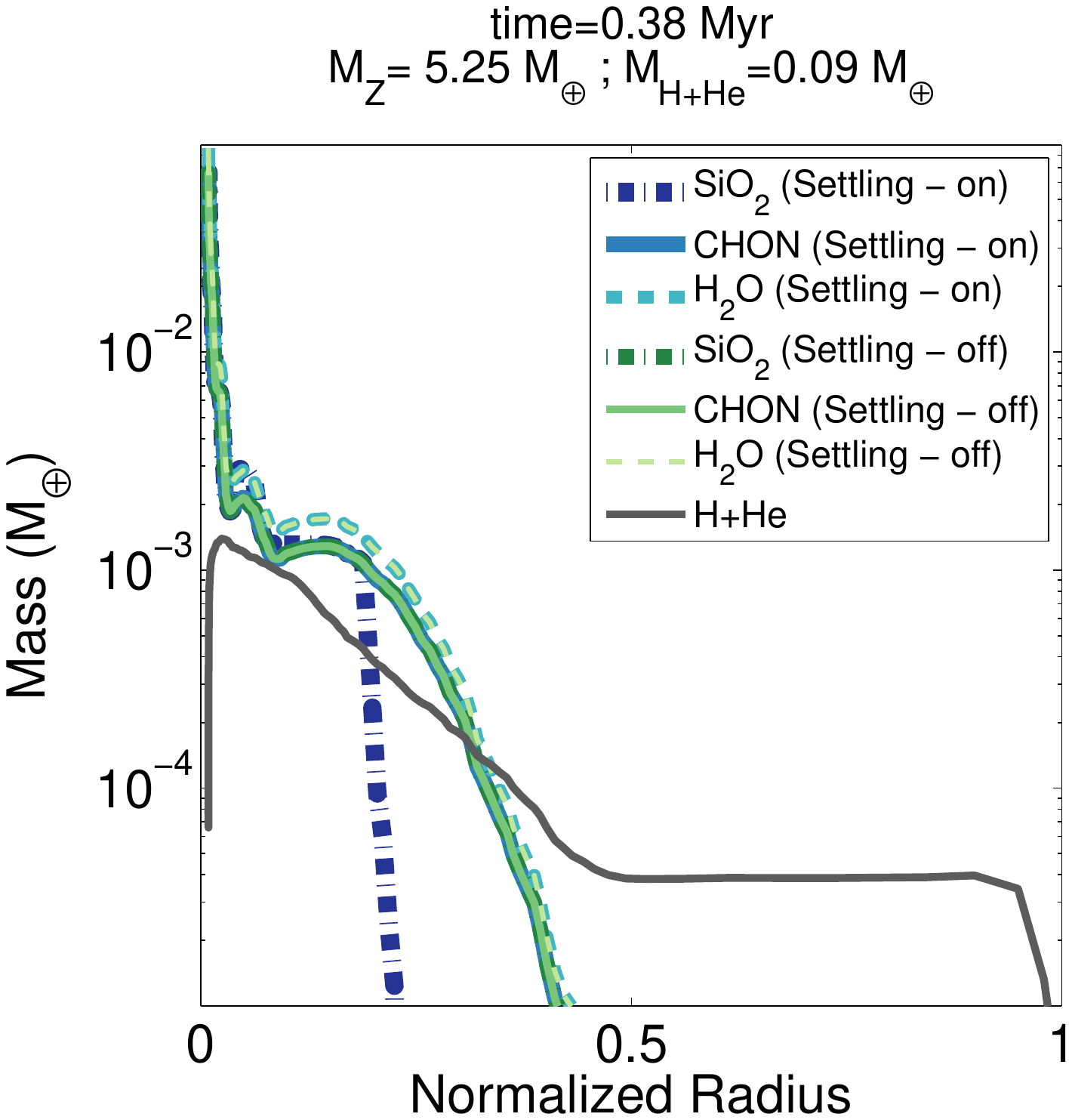}}
\subfloat{\includegraphics[trim=80 180 100 200,clip,width=0.42\columnwidth]{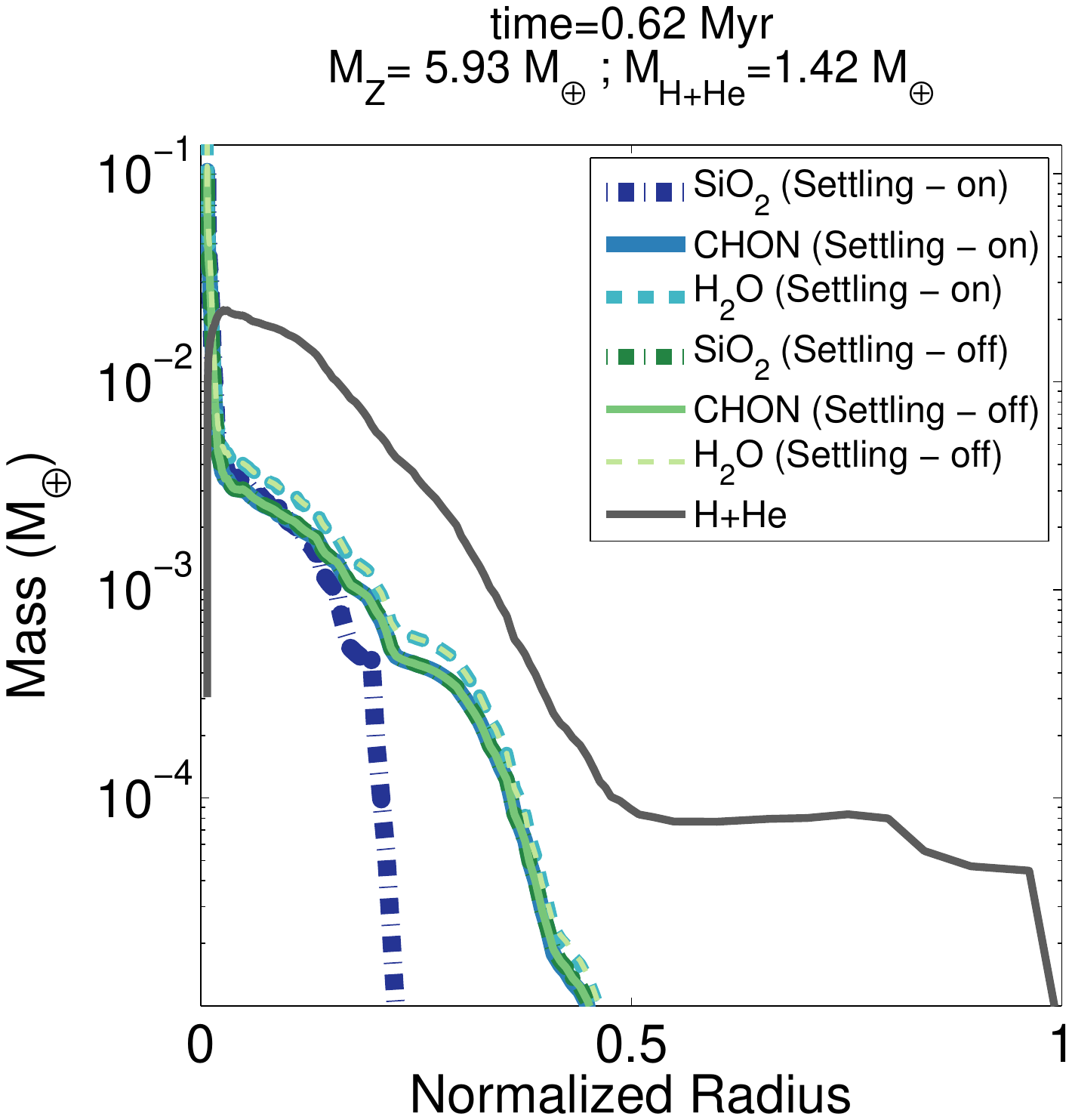}}\\
\subfloat{\includegraphics[trim=80 180 100 200,clip,width=0.42\columnwidth]{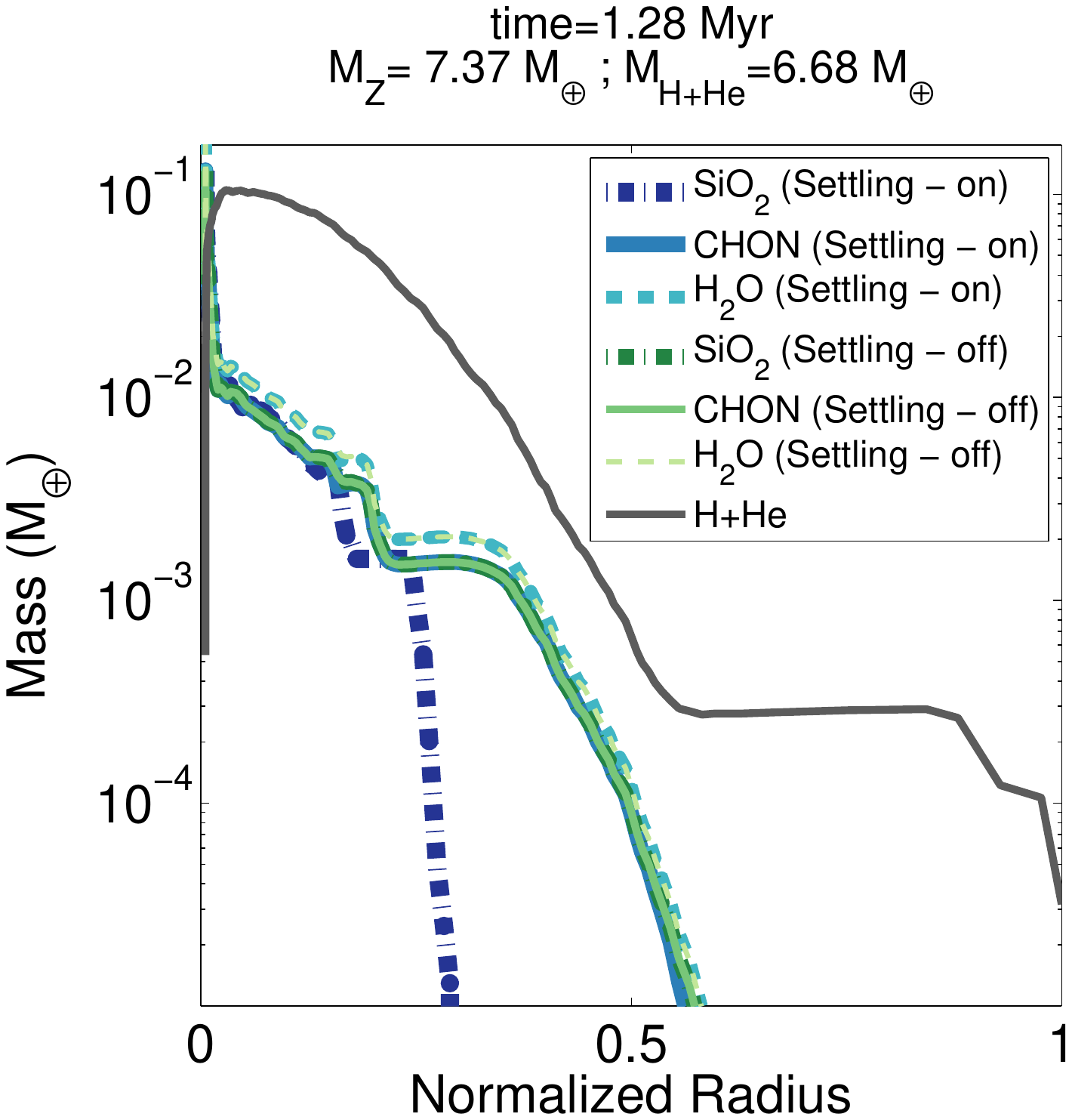}}
\subfloat{\includegraphics[trim=80 180 100 200,clip,width=0.42\columnwidth]{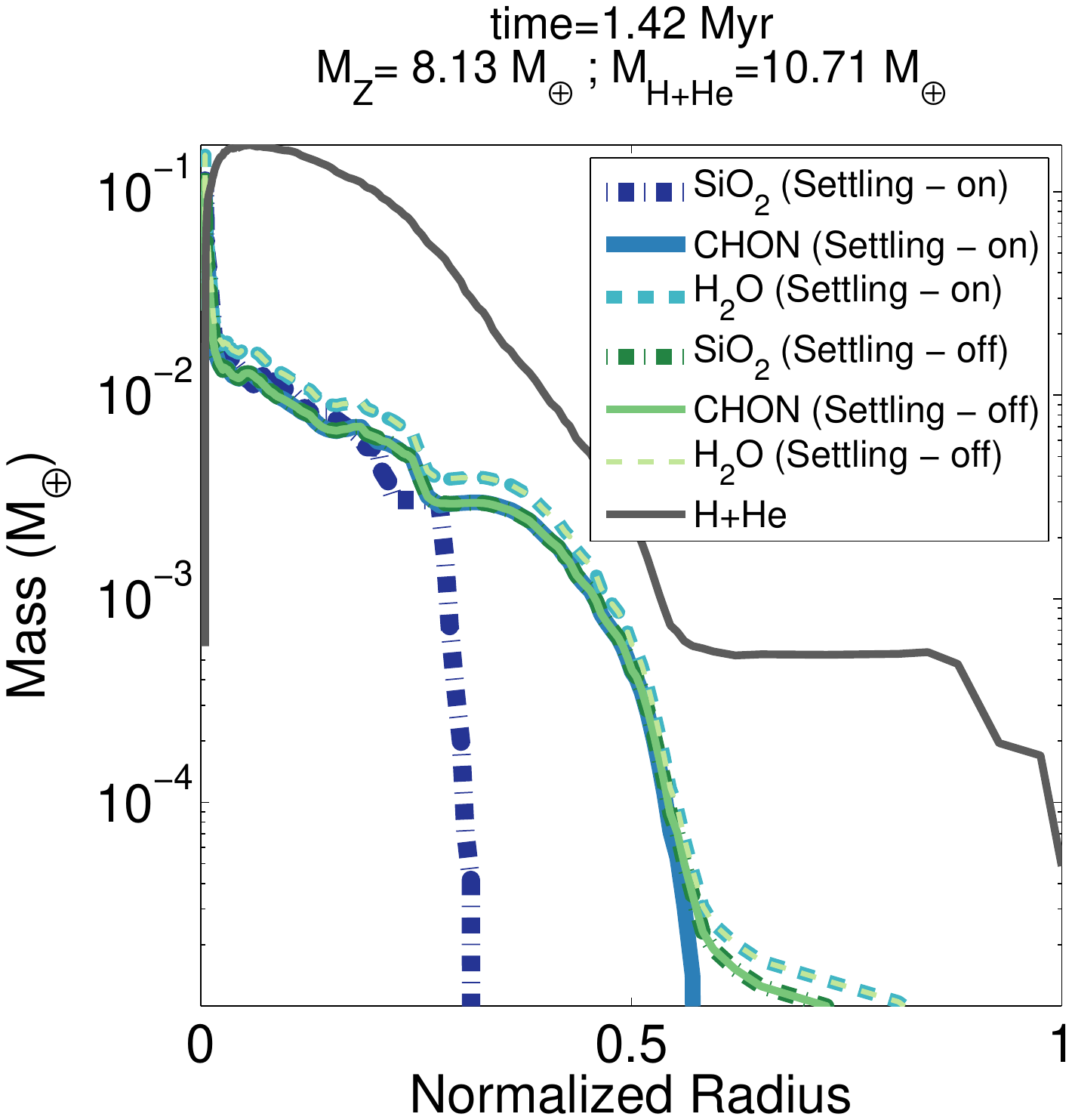}}
\caption{Mass distributions of the different materials versus normalized planetary radius at various times for Model-B. The blue and green curves represent the mass distributions before and after settling is considered, respectively (see text for details).}\label{fig:massProf1}
\end{figure}

The heavy-element mass fraction (Z) at a given layer is calculated by, 
\begin{equation}
Z=\frac{M_{SiO_2}+M_{H_2O}+M_{CHON} }{M_{SiO_2}+M_{H_2O}+M_{CHON}+M_{H+He}} \quad, 
\end{equation}
where $M_{SiO_2}$, $M_{H_2O}$, $M_{CHON}$, and $M_{H+He}$ are the masses of rock, water, organics, and hydrogen+helium, respectively.  
In order to derive the Z-profile (heavy-element distribution) within the proto-planet, we first need to determine the mass distribution of the three different materials: rock (SiO$_2$), water (H$_2$O) and organics (CHON). The different vapor pressures of the different accreted materials lead to different mass distributions after settling is considered. Figure \ref{fig:massProf1} shows the mass distributions of the various components at four different time steps for Model-B. The mass distributions are shown before (green) and after (blue) settling is considered. Rock is represented by the dashed-dotted curve, while organics and water are represented by the solid and dashed curves, respectively. The mass of hydrogen and helium (H+He) is represented by the gray-thin curve. 
\par

Several conclusions can be drawn from the figure. First, in comparison to H+He, even before settling is applied, the heavy elements are more concentrated towards the planetary center. 
This is because planetesimals lose mass and break up only when the envelope's density and temperature are high enough for the planetesimals to experience sufficient gas drag. Second, after settling  is considered, the distribution of the different materials become non-homogenous: while water and organics tend to remain where they are originally deposited, silicates settle toward the center. This may suggest that the (primordial) envelopes of giant planets are more enriched with volatiles while the refractory materials are concentrated in the deep interior. 

\subsection{Convective mixing and EOS}
\label{sec:Mixing due to convection}
Although we find that the distribution of heavy elements due to planetesimal accretion and settling is inhomogeneous, convective mixing can homogenize the envelope's composition. 
The presence of convection in regions with composition gradients is determined by the ratio between the destabilizing temperature gradient and the stabilizing composition gradient (the Ledoux criterion); if the latter is dominant - the region is assumed to be radiative/conductive and no mixing occurs, although layered-convection could develop in such regions \citep{Vazan2015}. 
The stability against convection in each layer accounting for the composition gradients using the Ledoux criterion is given by:
\begin{equation}
\nabla-\nabla_{ad}-\nabla_{X}<0 \quad, 
\end{equation} 
where $\nabla \equiv \frac{dlnT}{dlnP}$, and $\nabla_{ad}$ is the adiabatic gradient, and $\nabla_{X}\equiv\frac{\partial\ln T(p,\rho,X)}{\partial X_{j}}\cdot\frac{dX_{j}}{d\ln P}$. When convection is inefficient  $\nabla \sim \nabla_{rad}$, where $\nabla_{rad}$ is the radiative/conductive gradient. 
\par

In order to compute all the relevant physical properties that are required for the convection criterion calculation (e.g., $\nabla_{X},\nabla_{ad}, \kappa$), we must use an equation of state (EOS). 
The EOS is calculated using the SCVH EOS tables by \citet{Saumon1995} for hydrogen and helium with an extension for low pressure and temperatures. Since there is no EOS available for organic materials, the heavy elements are represented by a mixture of 
H$_2$O and SiO$_2$ using the QEOS method \citep{More1988}.   
More details on the QEOS calculation can be found in \citep{Vazan2013} and references therein. 
\par

We assume that regions with shallow composition gradients (the outer layers) mix if they are unstable to convection according to the Ledoux criterion, resulting in a homogenous composition with the average Z-value. The outermost regions of the protoplanet are not convective but radiative \citep{Guillot1995}, and are too cold to maintain heavy elements in a gaseous phase. As a result, grains are likely to form and settle towards inner regions, where they evaporate and mix with the surrounding gas. In order to account for this effect, we settle the heavy material from atmospheric regions with temperatures. For the silicates we take a critical temperature of 1500 K, and of 650 K for the water and organics (hexacosane). 
This results in two steps in the outer regions of the planet: the outermost region up to a temperature of 650 K is metal-free, there is an intermediate layer of vapor water and organics, and at temperatures higher than 1500K also of vapor rock. 
The outermost part of the proto-planet is then depleted in heavy elements and has Z $\sim$ 0. The heavy elements that were found to be in these outer regions are added to deeper regions and their contribution to the local heavy-element concentration is included. 
\par

The original envelope's temperature (as well as density and pressure) is calculated assuming that all the heavy elements go to the center, and the envelope's composition is a mixture of hydrogen and helium in proto-solar ratio, i.e., the hydrogen-to-helium ratio is set to be 0.705:0.275 \citep[e.g.,][]{Bahcall1995}. 
Since in our calculation the heavy elements can remain in the envelope, the contribution of the heavy materials to the envelope's temperature must be considered. 
In addition, because the temperature profile determines the efficiency of convection, it is important to understand how the heavy elements affect the envelope's temperature. 
Different materials have different effects on the temperature, and as a result, we compare four temperature profiles: the first is the original temperature profile assuming a H+He envelope, and the other three are the temperature profiles that are inferred when accounting for the effect of the heavy elements assuming three different compositions: pure SiO$_2$, pure H$_2$O, and a SiO$_2$-H$_2$O mixture. 
These calculations use the original temperature, pressure and density of the envelope and the distribution of heavy material that is dissolved in the envelope ($X,Y,Z$). We then compute the modified temperature profiles that corresponds to the original pressure of the envelope. 
The temperature as a function of normalized mass for different compositions are presented in Figure \ref{fig:T-profiles, s=10, r=100}. 
\par

It is clear from the figure that the addition of heavy elements to the envelope increases its temperature, with SiO$_2$ having the largest effect.  
The increase in temperature is caused by the change in the opacity when the heavy elements are included. In the innermost regions, the opacity value can increase by about a factor of 100. 
As expected, the temperature profile of the SiO$_2$-H$_2$O mixture falls between the profiles of pure-rock and pure-water. We therefore suggest that the internal temperatures of young giant planets are higher than typically derived by formation models which do not account for this effect. Since the internal temperatures can reach well over 10$^4$ K, it is possible that the small inner core melts and mixes with the envelope as we discuss below. 
Although the existence of heavy elements in the envelope leads to a significant increase in temperature, the {\it temperature gradient} remains about the same, and as a result, the identification of the convective regions in the envelope is not expected to change. The temperature gradient in the convection zones is the adiabatic 
gradient. In order to verify that, we derive the Z-profile for the different calculated temperature profiles and apply the convection criterion for these new conditions. 
The results are shown in Figure \ref{fig:Z-profile for T-profile, s=6, r=0.5 } for formation Model-B. In this calculation the core is defined as the innermost region with heavy element mass fraction larger than 0.5 (see the following section). 
Indeed, the figure suggests that the change of temperature and the assumed composition of the heavy elements do not affect the mixing pattern. 
This ensures that the heavy-element profiles we derive in this work are realistic, and are not expected to differ much if other materials are assumed. In the following sections the heavy elements are represented by a mixture of water and silica (50\%-50\%).

\subsection{The definition of the core}
\label{sec:Core definition}

Defining the core of a giant planet is not trivial. 
Jupiter's core mass is typically inferred from structure models that fit the measured gravitational field of the planet \citep[e.g.,][]{Guillot2005}. Typically, standard 3-layer interior models infer a core mass between zero and $\sim$10 M$_{\oplus}$, while the total heavy element mass is uncertain and is estimated to be between 10 and 40 M$_{\oplus}$ \citep{Guillot1999,Saumon2004,Nettelmann2008,Nettelmann2012}. Alternative models suggest a massive core (about 14 - 18 M$_{\oplus}$) and a smaller enrichment in heavy elements in Jupiter's gaseous envelope \citep{Militzer2008,Hubbard2016}. However, despite the accuracy of interior models, the chemical composition and physical-state of the core cannot be inferred. Moreover, even if a core exists, it is not clear how distinct it is from the layers above it, although for simplicity, it is typically taken to be a separate region, and the core-envelope-boundary (hereafter, CEB) is assumed to have a density (and composition) discontinuity. 
It is also possible that the CEB is "bleary" (and therefore, not well-defined), and the change in density (and composition) is more continuous. In that case, the core could be more extended and could even consist of some hydrogen and helium \citep{Stevenson1982}. 
\par

The original formation models have a small primordial (solid) core in the early stages. The simulations begin with a Mars-size object that accretes planetesimals and gas. The actual core, however, is not modeled. The only thing that is followed is the increase in the core's mass due to the ongoing planetesimal accretion. The innermost region that is physically modeled is the bottom of the envelope, just above the core. Therefore, any possible interaction between the core with the surrounding gaseous envelope is neglected. 
As suggested by the formation models used in this work, once the core mass reaches $\sim$ 1-2M$_{\oplus}$, the planetesimals dissolve in the envelope. 
This is in agreement with other giant planet formation models such as the one presented by Pollack et al.~(1996) and Venturini et al.~(2016). If the core is defined as a pure heavy-element region, the core mass is not expected to increase significantly, suggesting that  
Jupiter's core is smaller than the typical core masses inferred by \textit{CA} models. Thus, as we discuss below, it may be that our perception of the core should be different (and less conservative), and that the cores of giant planets could be defined as the innermost region that is enriched with heavy elements.  In this case, the core is more massive, larger, and diluted with light materials (i.e., H+He). Similar diluted core models
for Jupiter have been considered by \citet{Fortney2010}. 
\par

In this work the core is taken to be the innermost region with a heavy-element mass fraction (Z) that is larger than some critical value. We consider two critical values: Core-i with Z$\geq$0.9, and Core-ii with Z$\geq$0.5. In practice, due to the simplifications assumed for computing the disruption and mixing of the accreted material, it can happen that during the runaway gas accretion phase, the Z-profiles dilute too much. This causes the region of Z $\geq$ Z$_{crit}$ to shrink. 
However, this is a numerical artifact, because the planetesimal disruption model assumes that the envelope is metal-free. In reality, once a steep Z-profile exists in the envelope, the additional accretion of H+He should stay mainly on the top of the planet, leading to an onion-like structure \citep{Stevenson1982}. Following this argument, and in order to not introduce a numerical bias in the mass of the core, we add, to the region of Z $\geq$ Z$_{crit}$, the necessary layers to the core region in order to reach (at least) the same mass of core we had in the previous time step. 
As we show below, under our core definition, the core is not necessarily distinct from the envelope due to the existence of composition gradients within the planet. This is rather different from the "standard" view of Jupiter's core as a pure-Z region and the existence of a density discontinuity between the core and the envelope. 
\par

The rate of solid accretion during the phase of runaway gas accretion is not well-understood. After crossover is reached, the proto-planet starts to accrete gas rapidly, and the planetary composition from this point on depends on the composition of the accreted gas as well as of the planetesimal accretion rate during this formation phase, both are poorly known \citep[e.g.,][]{Alibert2005,Lissauer2008,Helled2014}. As a result, more robust predictions in terms of composition and internal structure can be made up to crossover time. Nevertheless, in section \ref{sec:Full formation model} we also present calculations in which the planetary formation goes all the way to a Jupiter-mass assuming that the accreted gas has a solar composition, and that no planetesimals are accreted during runaway gas accretion. 
Thus, on the contrary, if 
during runaway accretion a large fraction of heavy elements 
were accreted, the final composition of the envelope would be 
super-solar super-solar, and in that case our results should be taken as a lower bound for the envelope (atmospheric) enrichment of Jupiter. 
In addition, it is possible that during rapid gas accretion some material from the outer 
part of the core will mix into the envelope, and thus reduce the mass of the core. This, however, must be modeled in detail before conclusions on core erosion during formation can be made. 
\par

\section{Results}
\label{ch:Results and discussion}

Figures \ref{fig:Z05Z09, s=6, r=100} - \ref{fig:Z05Z09, s=10, r=100} show the distributions of the high-Z material (in mass fraction, $Z$) as a function of normalized mass at different times for the four different formation models.  
The blue curve represents the original distribution of heavy elements after settling is also included, while the dashed-dotted orange and dashed black curves show the Z-profiles after convective mixing is considered for Core-i, Core-ii, respectively.  
The time, total mass of heavy elements (M$_Z$), total mass of hydrogen and helium (M$_{H+He}$), and calculated core mass (M$_{core}$) are given in the header of each panel. 
It can be seen that at early times, the protoplanet is composed of mostly heavy elements, while only the outermost layers are depleted in heavy elements (blue curve). 
At each time step, we first derive the composition gradient, then the innermost layers with a heavy-element mass fraction $Z$ that is larger than the critical value are added to core, and the rest of the envelope is allowed to mix according to the Ledoux criterion. As discussed above, if the mass of the core region is smaller than the core mass calculated in the previous model, we add a few layers above the core region until the correct core mass is reached. Otherwise, we simply collect all the layers with heavy-element mass fraction that is larger than the critical value, and derive the new core mass.  
As time progresses, the heavy-element distribution becomes more gradual and but in the inner region the gradient is steep enough to inhibit convective mixing. 
In all cases, the bottom of the envelope reaches very high temperatures (over $10^4$ K) due to the existence of heavy elements. 
\par

It is interesting to note that for the formation models with $\sigma = 6$ g cm$^{-2}$ the distribution of the heavy elements is more gradual than the ones for the case of $\sigma=10$ g cm$^{-2}$. 
This seems to be linked to the total formation timescale, in particular, the length of the slow gas accretion phase. When $\sigma=10$ g cm$^{-2}$ the planetary structure is more similar to a standard core+envelope structure. Model-D is the closest to having a core+envelope structure since the solid surface density is high, and the planetesimals are large, so even with dissolution of planetesimals in the envelope, most of the accreted heavy elements tend to be in the deep interior. 

For the case of Core-ii, the calculated core mass is larger, especially at later times, but the core's density is lower because in this case the core consists of a larger fraction of hydrogen and helium. 
In addition, the core has a gradual change in the heavy-element mass fraction, which is decreasing towards the core's outer region. 
Both Core-i and Core-ii consist of hydrogen and helium and have composition gradients. This suggests that the cores of giant planets can have very different physical properties (mass, composition, radius, etc.) and can be more complex than the standard cores that are typically assumed. 
\par

The crossover times and the corresponding inferred core masses are summarized in Table \ref{tab:Crossover Z} and are denoted by $t_{cross}$ and $M_{cross}$. 
As the planetesimal's size increases, the time to reach crossover increases as well, while for higher solid-surface densities, the formation process is shorter and therefore the time to crossover decreases. This is because small planetesimals tend to form smaller cores, and high solid-surface densities correspond to higher solid accretion rates, which reduce the formation timescale. 
It should be kept in mind, however, that the crossover times presented here correspond to formation models that assume a metal-free envelope, and the actual crossover time could be significantly shorter when the heavy elements in the envelope are included \citep[e.g.,][]{Venturini2016,Hori2010}. 
Based on the work of \cite{Venturini2016}, $t_{cross}$ is expected to change by a factor of a few.  
The derived core masses are between about 5 and 20 $M_{\oplus}$, 
with the core mass depending on the critical value that is applied as well as on the assumed planetesimal size and solid-surface density. 
These values are significantly higher than the $\sim$2 M$_{\oplus}$ that is expected from a pure heavy-element core, and at the same time also different from a massive core of a similar mass assuming that all the accreted planetesimals go to the center. Thus, these cores are massive, extended, and have compositional gradients and are not pure-Z. 
\par

Figure \ref{fig:Z05Z09, summary} shows the envelope's (H+He) mass and the core masses for the two core cases as a function of time until crossover is reached for the four formation models. 
As can be seen from the figure, the calculated core mass is typically smaller than the total heavy element mass for Core-i, but is larger than the total heavy-element mass for Core-ii. 
For the models with $\sigma = 6$ g cm$^{-2}$, the derived core masses are smaller by more than a factor of two than for the cases with $\sigma = 10$ g cm$^{-2}$. 
Another interesting thing to note is that the H+He curve and the curve of Core-ii meet at about the same time as the original M$_Z$, although this may suggest that the crossover time is not expected to change much, 
it is clear that the core properties in our models are very different since Core-ii can contain a large fraction of H+He near crossover as we show in the following section.  
\par

\subsection{The composition of the core}
\label{sec:Core composition}

Since in our analysis the core is defined by a critical Z-value of the innermost layers, we can derive the core's composition as a function of time. Figure \ref{Pies, s=10, r=1} shows the derived core composition at various times. The top and bottom correspond for formation models B and C, respectively.  
We find that as time progresses, the mass of the core and the percentage of hydrogen and helium increase. While the initial core has a similar composition to that of the planetesimals, as it grows in mass, the overall composition changes because of the different distributions of the materials when settling is considered.
The core is found to consist of a larger fraction of rock compared to water and CHON in Model-C and Model-D (formation models with $\sigma$=10 g/cm$^2$) , and has almost the initial proportions for Model-A and Model-B (formation models with $\sigma$=6 g/cm$^2$). 
The derived core compositions for the four formation models at crossover time are listed in Table \ref{tab:Core Compose}. 
\par

It is interesting to note that although the derived core masses are different, the composition of the core is about the same. Moreover, we find that the core's composition is more sensitive to the assumed solid-surface density than to the planetesimals' size. In addition, the difference between the case of Core-i and Core-ii is not very large, which suggests that there are not many internal layers with heavy-element fractions between Z$\geq$0.9 and Z$\geq$0.5, i.e., the composition gradient is relatively steep. 
This is indeed demonstrated in the figures that show the distribution of heavy elements within the planets at various times. 
\par

\subsection{Modeling the formation up to a Jupiter Mass}
\label{sec:Full formation model}

In the previous sections, the formation of Jupiter was followed only up to crossover time (i.e., when $M_{H+He}=M_{Z}$ in the original formation models). 
Here we follow the planetary formation including runaway gas accretion up to the stage when Jupiter's mass is reached for Model-A and Model-C. The distribution of the heavy elements at various times is shown in Figure \ref{fig:Z05Z09 - All the way, r6} and Figure \ref{fig:Z05Z09 - All the way, r10}, respectively. 
The accreted gas is assumed to have a proto-solar composition.  
We consider the two different core definitions.  After crossover is reached, the gas accretion rate increases rapidly and the proto-planet accretes large amounts of gas (H+He). The total mass of the planet increases rapidly and the core becomes very small relative to the envelope. 
The most massive cores we derive are for Model-C, with a mass of about 13 M$_{\oplus}$ and 20 M$_{\oplus}$, for the cases of Core-i and Core-ii, respectively.  
\par

As can be seen from the figures, when we allow the outer regions to mix by convection the original heavy-element distribution (blue-curve) can break into a few different regions, with the outermost regions becomes nearly metal-free due to grain settling. 
Towards the end of the formation, convective regions are found to be separated by stable (radiative/conductive) layers, resulting in "stairs". These stairs could be a numerical artifact, but they might also indicate that giant planets could have several convective regions that are separated by thin radiative layers \citep[see e.g.,][]{Guillot1994}. 
Regions with high fraction of heavy elements that are found to be stable against convection could in principle develop layered-convection \citep[e.g.,][]{Leconte2012,Vazan2016}. The importance of the primordial internal structure of Jupiter on its planetary evolution has been recently investigated by our group \citep[see][]{Vazan2015,Vazan2016}, and we hope to combine these two aspects self-consistently in the future. 
In the innermost regions with the steep composition gradients, slow mixing could essentially occur by double-diffusive/layered convection at later stages in the planetary evolution. We therefore suggest that future giant planet formation and evolution models should account for this effect, as it can change the predicted distribution of heavy elements in proto-Jupiter and the cooling history of the planet  \citep[e.g.,][]{Vazan2016}. 
\par 

\begin{figure}[!h]\centering
\subfloat{\includegraphics[trim=10 145 00 60,clip,width=0.42\columnwidth]{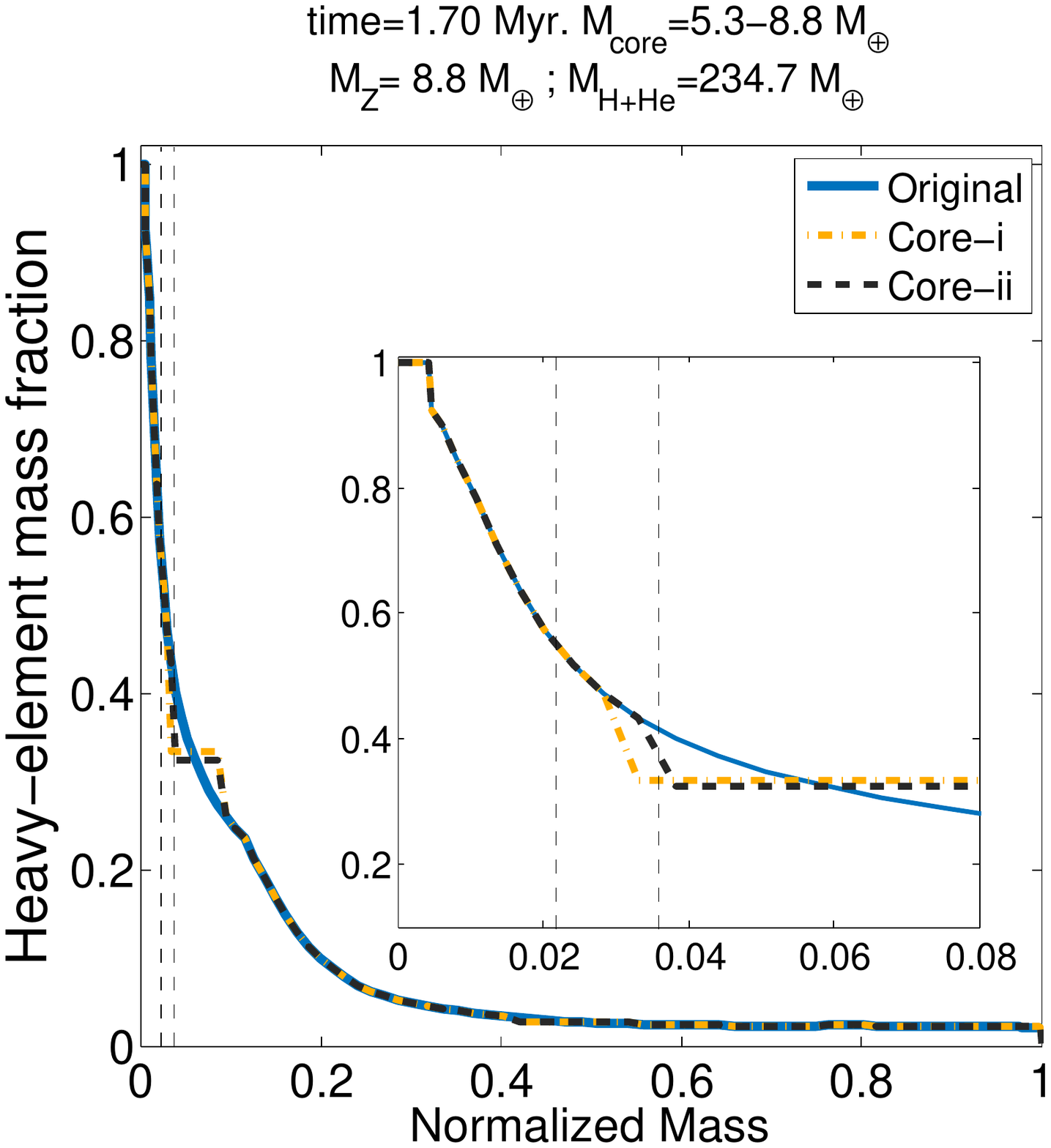}}
\subfloat{\includegraphics[trim=10 145 00 60,clip,width=0.42\columnwidth]{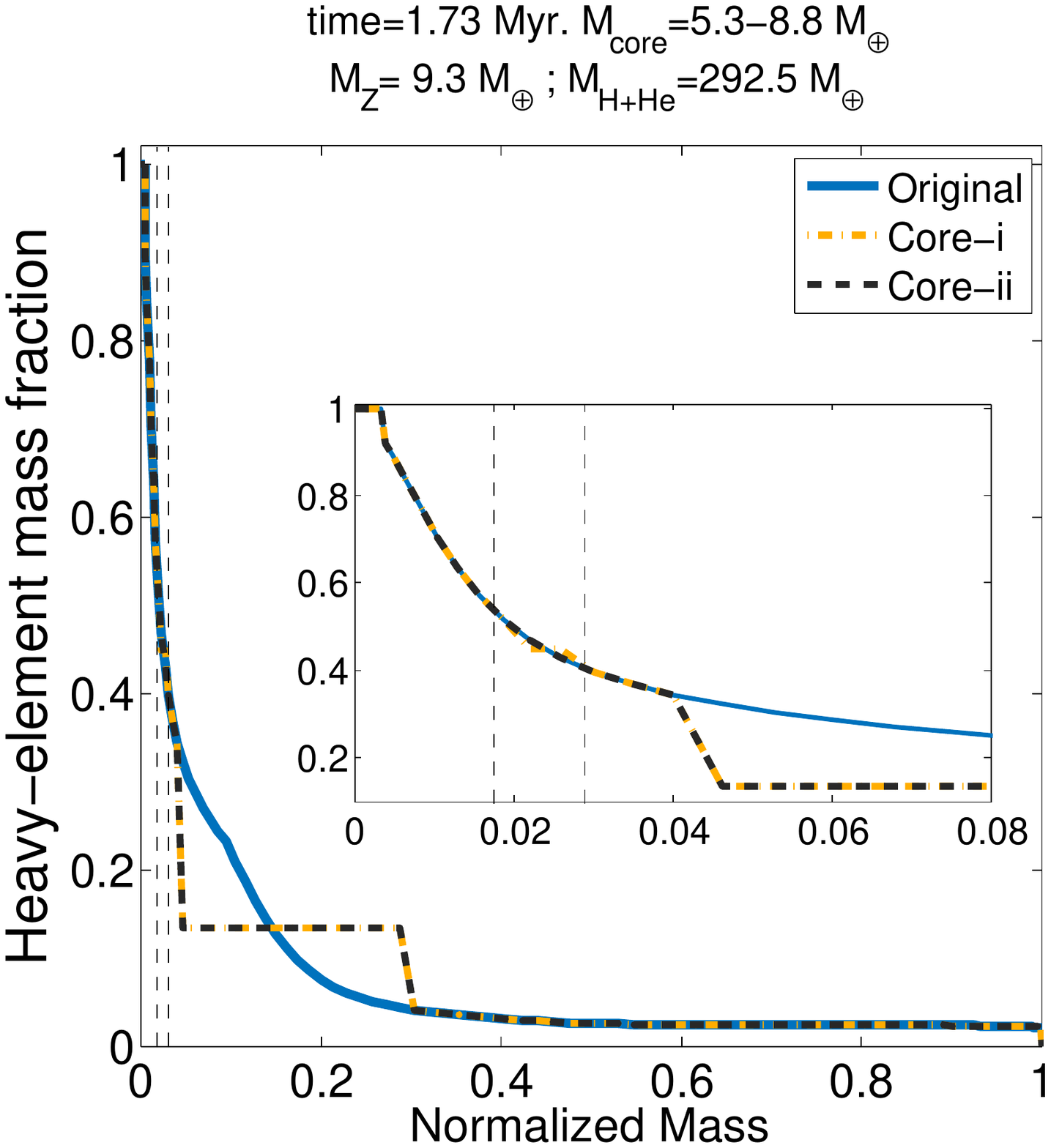}}\\
\vspace{-0.8cm}
\subfloat{\includegraphics[trim=10 145 00 60,clip,width=0.42\columnwidth]{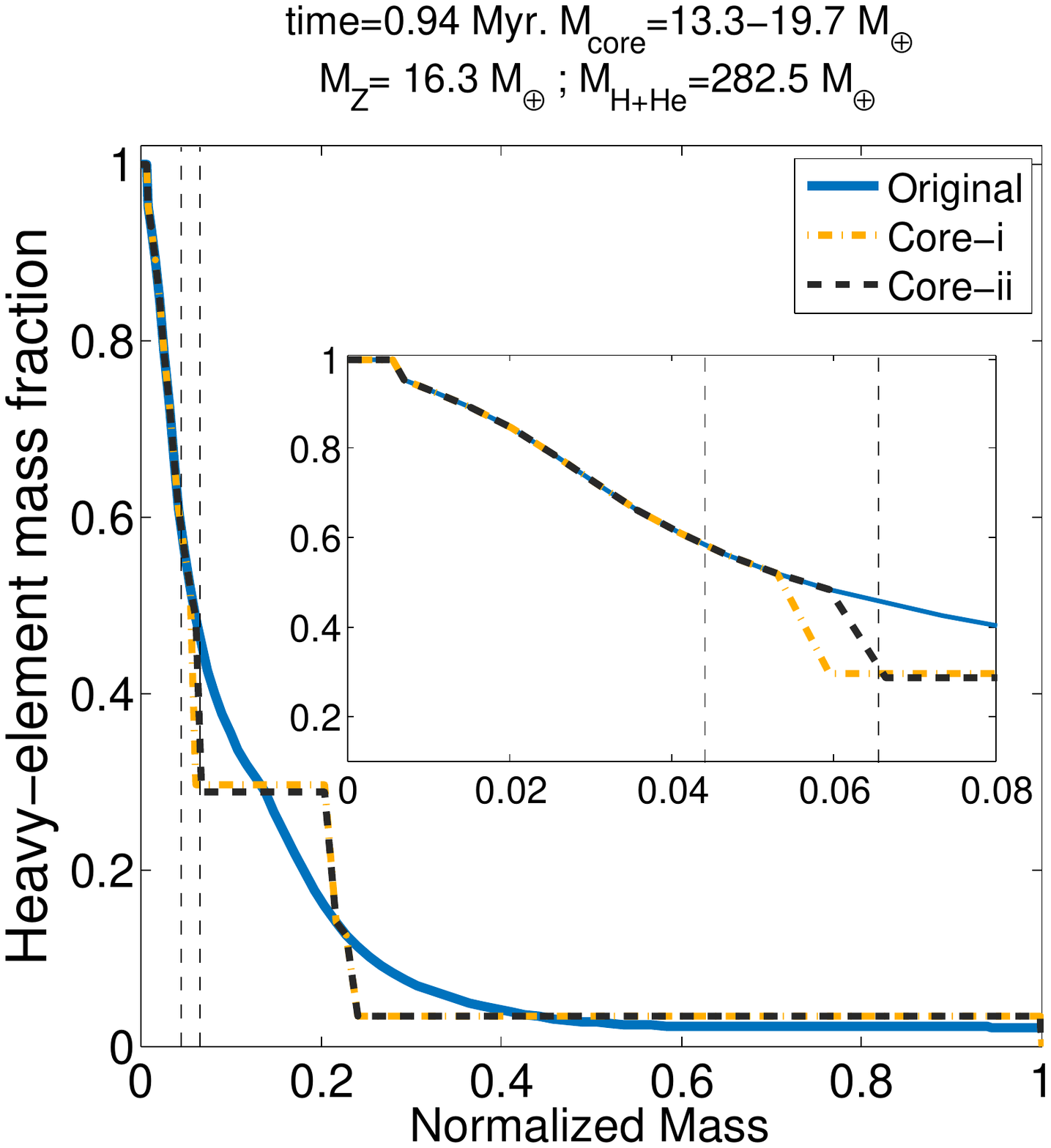}}
\subfloat{\includegraphics[trim=10 145 00 60,clip,width=0.42\columnwidth]{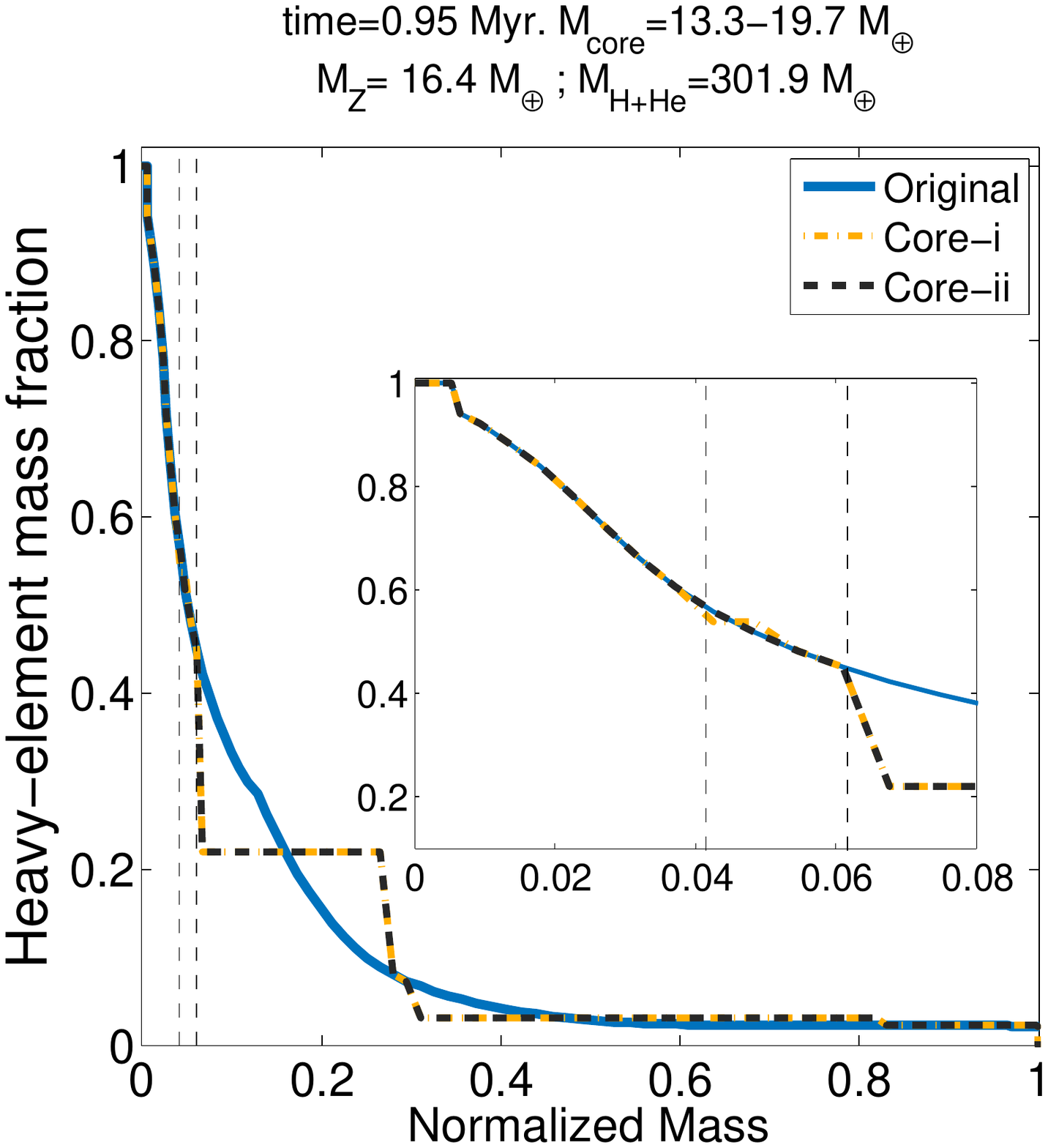}}\\
\caption{The distribution of heavy elements in models of proto-Jupiter as they
approach the final mass of 1 Jupiter-mass at the end of rapid gas accretion.
Top panels: Model-A; bottom panels: Model-C. Dash-dotted orange lines:
Core-i; dashed black lines: Core-ii. }\label{fig:Final Plot-Z0509}
\end{figure}

Figure \ref{fig:Final Plot-Z0509} shows the final models for Core-i and Core-ii for Model-A (top) and Model-C (bottom). 
The "final model" in our case is defined by the mass of the planet, i.e., when the planet reaches a mass of $\sim 1M_J$ and gas accretion is terminated.  
In both cases, there is an inner region with a composition gradient (above the small primordial core). 
While this region is of the order of 5\% of the planet's radius, this configuration can affect the long-term evolution of the planet. The envelope above the core is found to be fully mixed, suggesting an adiabatic outer envelope. 
The inner regions with composition gradients can mimic the existence of a massive core. As a result, this option should be accounted for when inferring Jupiter's core mass from gravitational data. 
\par

\section{Discussion \& Conclusions}

We present a calculation of Jupiter's formation and primordial internal structure in the {\it CA} model, including the enrichment of the planetary envelope as a result of planetesimal accretion and settling. 
We follow the distribution of heavy elements within the proto-planet during its growth, taking into account the re-distribution of heavy elements due to settling and convective mixing in the outer envelope. It is found that a substantial amount of the ablated heavy material remains in the planetary envelope. Since different assumed solid-surface densities and planetesimal sizes lead to different core masses, it is clear that the birth environment of the planet has an impact on its final internal structure. 
\par

If the core is defined by a pure heavy-element region that is not interacting with the enriched envelope, Jupiter's core is predicted to have a mass of $\sim$ 1-2M$_{\oplus}$. After that core mass is reached, the accreted planetesimals dissolve in the envelope. 
Here, we consider an alternative definition of Jupiter's core, accounting for two different critical values for Z and investigate how they affect the predicted internal structure and core mass of proto-Jupiter. 
We find that under this definition Jupiter's core is massive ($\sim$ 15 -20 M$_{\oplus}$) but also consists of a non-negligible fraction of H+He. 
Naturally, when the core is defined by Z$>$0.5, the inferred core mass is larger, and the core is more extended and has larger fractions of H+He. At crossover, the fraction of H+He for the core definition of  Z$>$0.9 is typically 4-6 \%, and in the case of Z$>$0.5, 11-15 \%. 
Our analysis suggests that traditional planet formation models considering a heavy-element core surrounded by a H+He envelope are over-simplified, and future formation (and structure) models should account for the existence of heavy elements in the envelope, and their distribution self-consistently. This will lead to a more accurate prediction of the primordial internal structure of giant planets, and in particular, their core masses. We suggest that a natural next step is to follow the evolution of the planet and investigate whether the composition gradients will persist in Jupiter today \citep[see][]{Vazan2016}. 
\par

While in most \textsl{CA} models the terms "heavy material" and "core" are essentially the same, we show that there is an important difference between the two, because most of the accreted heavy elements remain in the planetary envelope, and the core mass is significantly smaller than the total heavy-element mass. When convective mixing is considered, we find that the heavy elements are homogeneously distributed along the envelope. Thus, the innermost regions which have a steep enough composition gradient can remain stable against convection. These regions can also consist of hydrogen and helium and could be viewed as "extra-extended cores". The  "extended cores" are of the order of 20 M$_{\oplus}$ in mass,  but with lower density than that of a pure-Z core due to the existence of H+He. Such a core should be considered in future internal structure models of Jupiter. 
The results of this work are relevant for the data interpretation from the {\it Juno} (NASA) mission to Jupiter. We suggest that Jupiter's core can be significantly smaller than predicted by standard core accretion models. At the same time, we also find that the innermost region of Jupiter can be enriched in heavy elements, and mimic a massive core, with the difference that the core is larger and has a relatively low mean density. We hope that with {\it Juno's} accurate gravitational measurements it will be possible to discriminate between the no/small core case, and the extended massive (but not very dense) core case for Jupiter.  
Since our model suggests that composition gradients may exist in Jupiter's deep interior, such a (non-adiabatic) configuration should be considered in structure models. A non-adiabatic interior can affect the temperature profile within the planet and the efficiency of convection. 
\par

Our study emphasizes the importance of determining the heavy-element accretion rate during runaway gas accretion. The predicted composition of the Jupiter (and giant planets in general) depends on the {\it assumed} gas composition and the solid accretion rate that is expected during runaway gas accretion. 
If during runaway a significant amount of heavy elements are accreted onto the planet, the composition of the outer envelope could be enriched as well. It is therefore extremely important to determine the heavy-element accretion rate during the last stages of giant planet formation, since it is directly linked to the prediction of the planetary composition \citep[see][and references therein for details]{Helled2014}.
\par

The work presented here should be taken only as a first step towards more advanced and detailed investigations of Jupiter's formation and primordial internal structure. It is clear that the existence of heavy elements in the envelope can affect the growth history of giant planets, the physical properties of the envelope, and the predicted primordial internal structure \citep[e.g.,][]{Hori2010,Venturini2016} and these should be simulated and included in future giant planet formation models.  
Future studies could also include the impact of the heavy elements on the (gas+dust) opacity in the envelope as well as re-condensation of the heavy elements in the envelope and the formation of clouds for various species, and the possibility of layered-convection in regions where composition gradients exist.  Moreover, this work has not included miscibility of the materials in hydrogen \citep[e.g.,][]{Wilson2012,Soubiran2015}. This could also affect the evolution of the planet and the predicted internal structure.    
Finally, our study demonstrates the importance of simulating the core in more detail. It would be desirable to model the core using physical equations of state, and to investigate the core-envelope interaction. A more detailed analysis of the core's physical properties, its cooling rate, and the properties of the core-envelope boundary are crucial for constraining the internal structure of Jupiter. Such studies can put tighter limits on the predicted core mass (and physical properties) of Jupiter's primordial core.

\subsection*{Acknowledgments}
We thank D.~Stevenson, A.~Vazan, A.~Kovez, J.~Venturini, and M.~Podolak for valuable discussions and suggestions.  
R.~H.~acknowledges support from the Israel Space Agency under grant 3-11485, and from the United States - Israel Binational Science Foundation (BSF) grant 2014112. P.~B.~received support from a grant from
the NASA Origins program.
\newpage

\bibliographystyle{apj} 

\newpage
\begin{table}[ht]\centering
    \begin{tabular}{ | c | c | c | c | c |}
    \hline
    Formation Model & $\sigma$ & planetesimal size & M$_{\text{core}}$ at crossover time & crossover time \\
    &  ($g/cm^{2}$) &  ($km$) & ($M_\oplus$) &  ($Myr$) \\ \hline
    Model-A & 6 & 100  & 7.5 & 1.54 \\ \hline
    Model-B & 6 & 0.5& 7.6 & 1.33 \\ \hline
		Model-C & 10 & 1 & 15.7 & 0.90 \\ \hline
    Model-D & 10 & 100 & 16.0 & 0.94 \\ \hline
		
    \end{tabular}\label{tab:init_cond}\caption{Properties of the four baseline formation models.}
		\end{table}

\begin{center}
\begin{table}[h]
\begin{tabular}{| l | l | l | l | l |}
\hline
\multicolumn{1}{|c|}{Formation Model} & \multicolumn{1}{c|}{$M_{core}$, Core-i} & \multicolumn{1}{c|}{$t_{cross}$, Core-i} & \multicolumn{1}{c|}{$M_{core}$, Core-ii} & \multicolumn{1}{c|}{$t_{cross}$, Core-ii} \\
\multicolumn{1}{|c|}{} & \multicolumn{1}{c|}{($M_{\oplus}$)} & \multicolumn{1}{c|}{(Myr)} & \multicolumn{1}{c|}{($M_{\oplus}$)} & \multicolumn{1}{c|}{(Myr)} \\ \hline
Model-A & 5.30 & 1.39 & 8.09 & 1.55 \\ \hline
Model-B & 5.74 & 1.20 & 8.08 & 1.33 \\ \hline
Model-C & 13.28 & 0.86 & 19.72 & 0.90 \\ \hline
Model-D & 13.17 & 0.92 & 17.98 & 0.96 \\ \hline
\end{tabular}%
\caption{The inferred core mass and modified crossover time ($M_{Core}=M_{H+He}$ instead of the original $M_Z=M_{H+He}$) for the different formation models and various core definitions.}
\label{tab:Crossover Z}
\end{table}
\end{center}

\begin{center}
\begin{table}[h]
    \begin{tabular}{ || c | c | c | c | c ||}
    \hline
        {\bf Core-i}& &&&\\ \hline
  Formation Model  & $H+He $ (\%) & SiO$_2$ (\%)& H$_2$O (\%)& CHON (\%)\\ \hline       \hline
    Model-A  &  4.7 & 28.6& 38.2& 28.6\\ 
    \hline
    Model-B  & 5.2 & 28.5& 37.9&  28.4\\ 
    \hline
		Model-C  &3.7& 37.3 & 30.1& 28.9 \\ 
		\hline
    Model-D & 5.2 & 36.2& 30.1& 28.4 \\ 
    \hline
    \hline
      {\bf Core-ii}& &&&\\ \hline
 Formation Model & $H+He $ (\%) & SiO$_2$ (\%)& H$_2$O (\%)& CHON (\%)\\ \hline
    Model-A & 21.0 & 23.7 & 31.6 & 23.7 \\ \hline
    Model-B &  19.0 & 24.4 & 32.3 & 24.2\\ \hline
		Model-C &  19.2 & 31.6 & 25.0 & 24.2 \\ \hline
    Model-D & 14.0 & 33.2& 27.0 & 25.8 \\ \hline	
				
    \end{tabular}
		\caption{The composition of Jupiter's core at crossover for Core-i (top) and Core-ii (bottom) for the four formation models (see text for details).}\label{tab:Core Compose}
\end{table}
\end{center}

\newpage

\newpage
\begin{figure}[!h]\centering
\subfloat{\includegraphics[trim=80 180 100 200,clip,width=0.42\columnwidth]{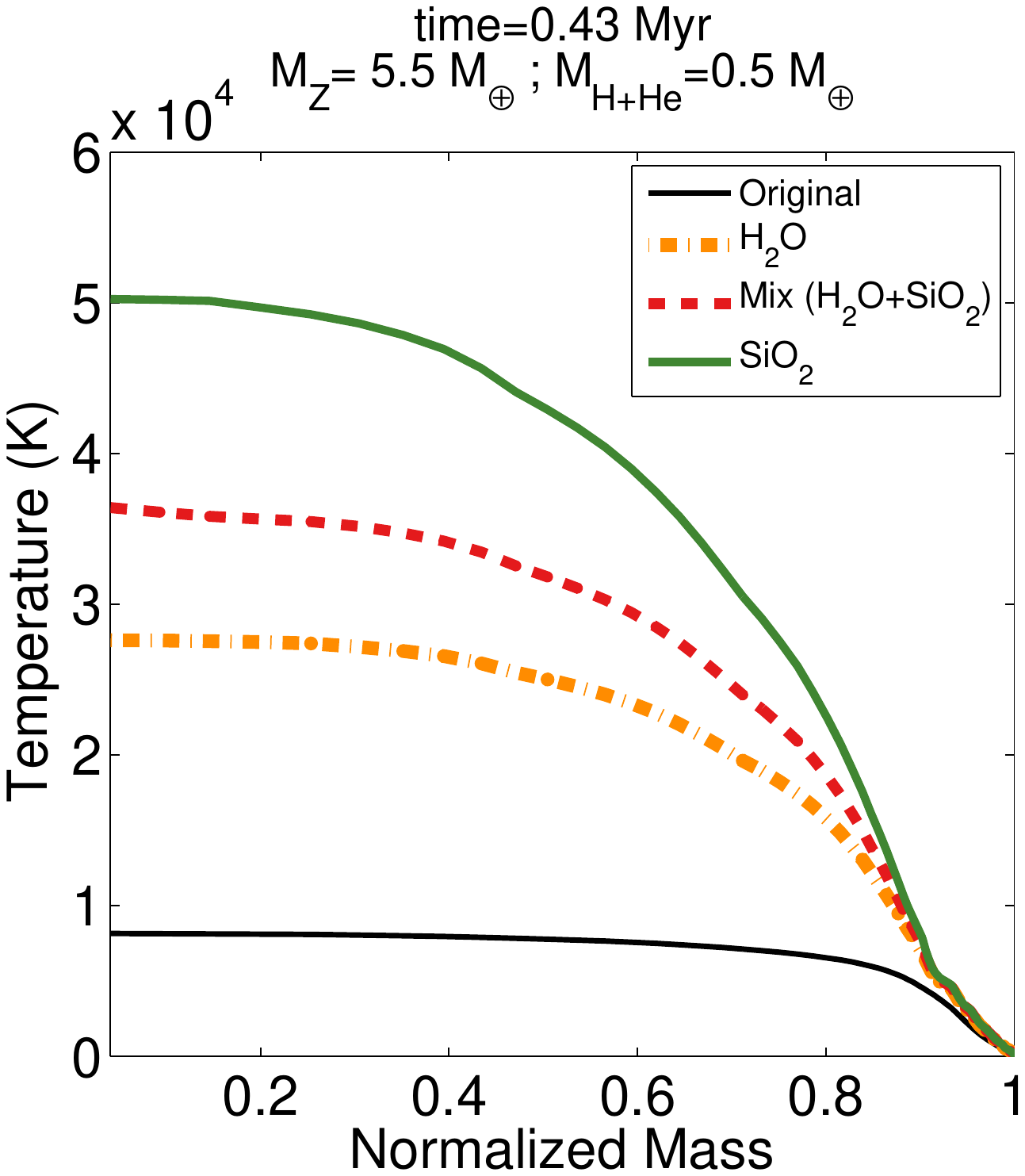}}
\subfloat{\includegraphics[trim=80 180 100 200,clip,width=0.42\columnwidth]{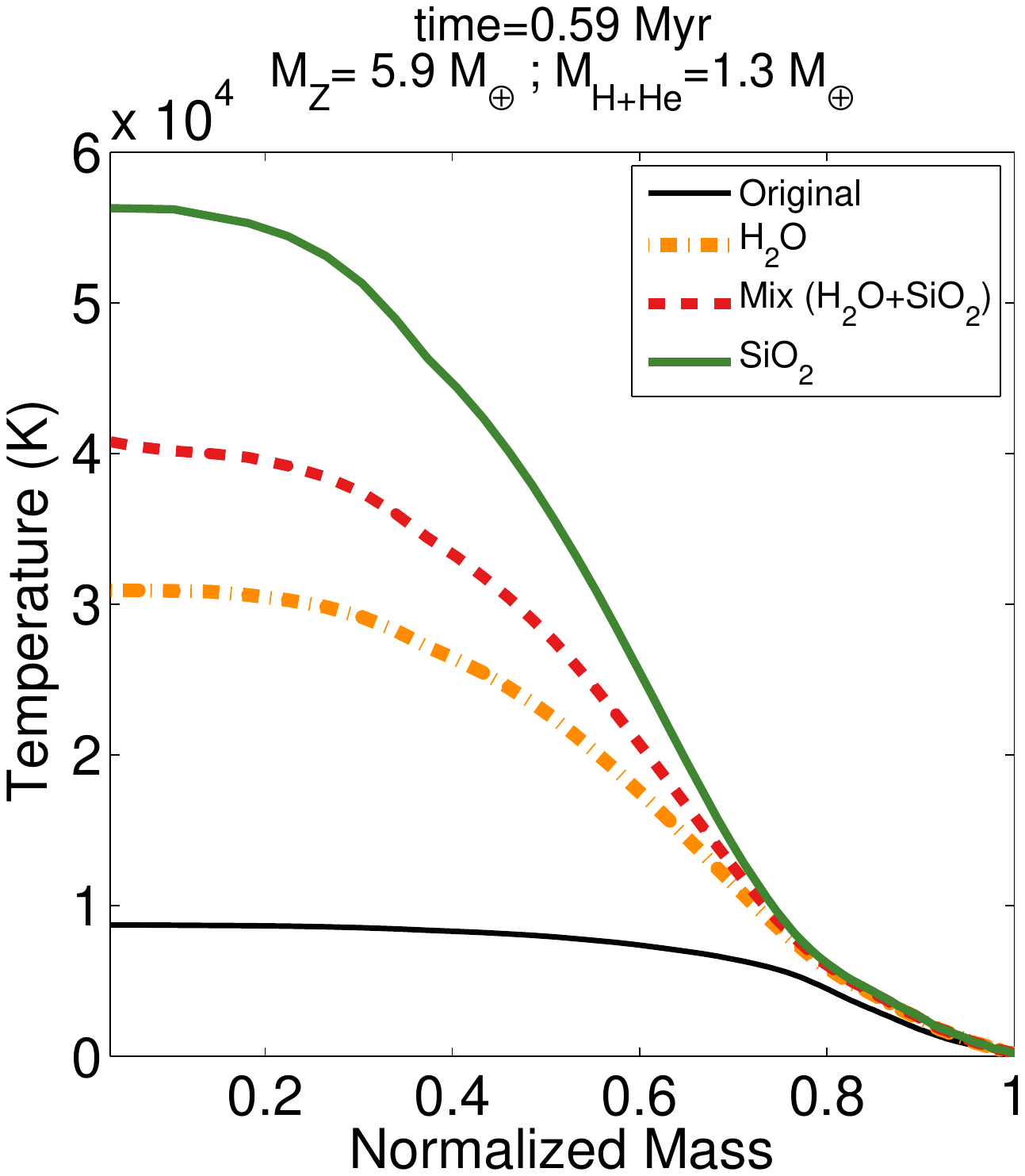}}\\
\subfloat{\includegraphics[trim=80 180 100 200,clip,width=0.42\columnwidth]{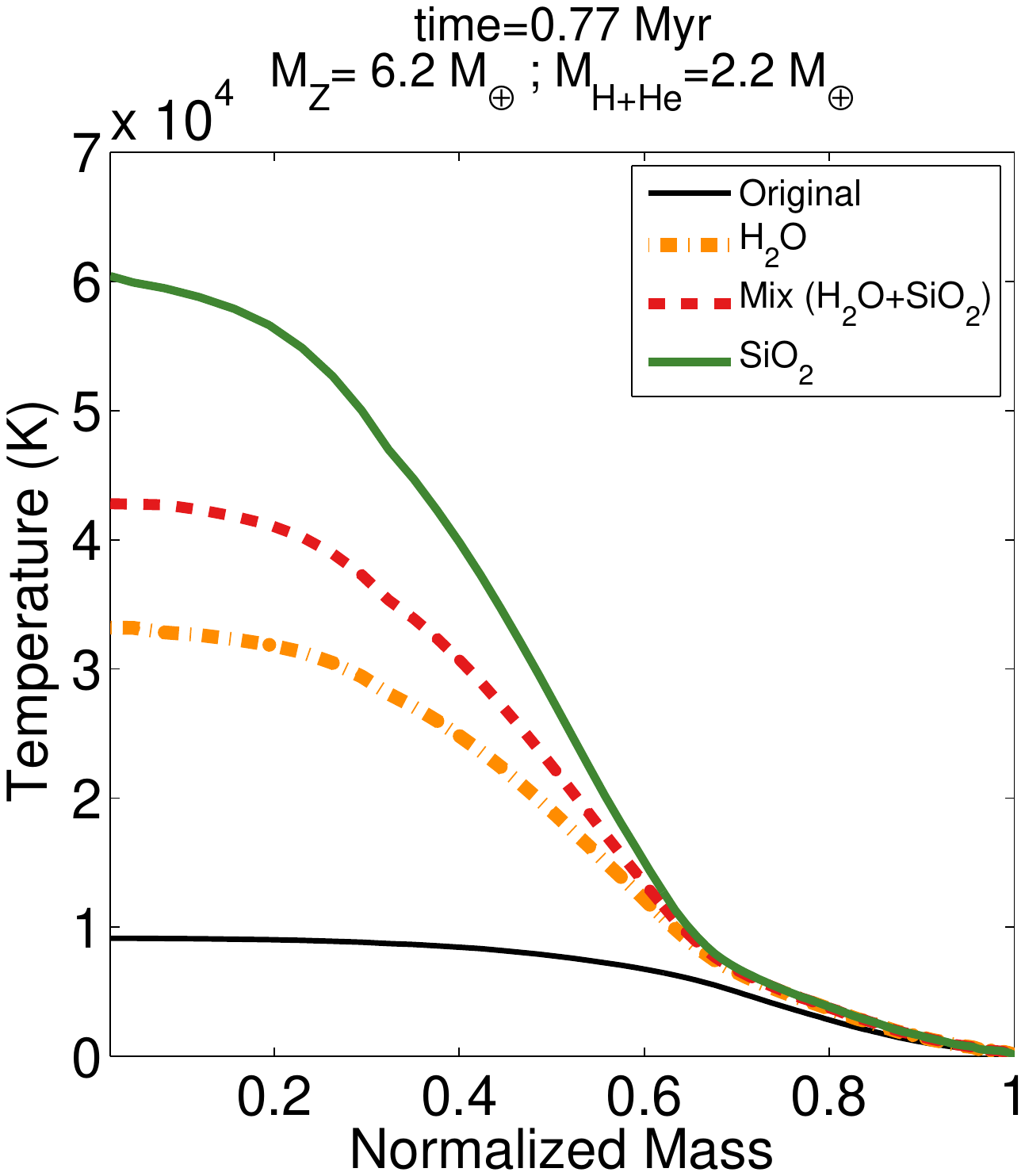}}
\subfloat{\includegraphics[trim=80 180 100 200,clip,width=0.42\columnwidth]{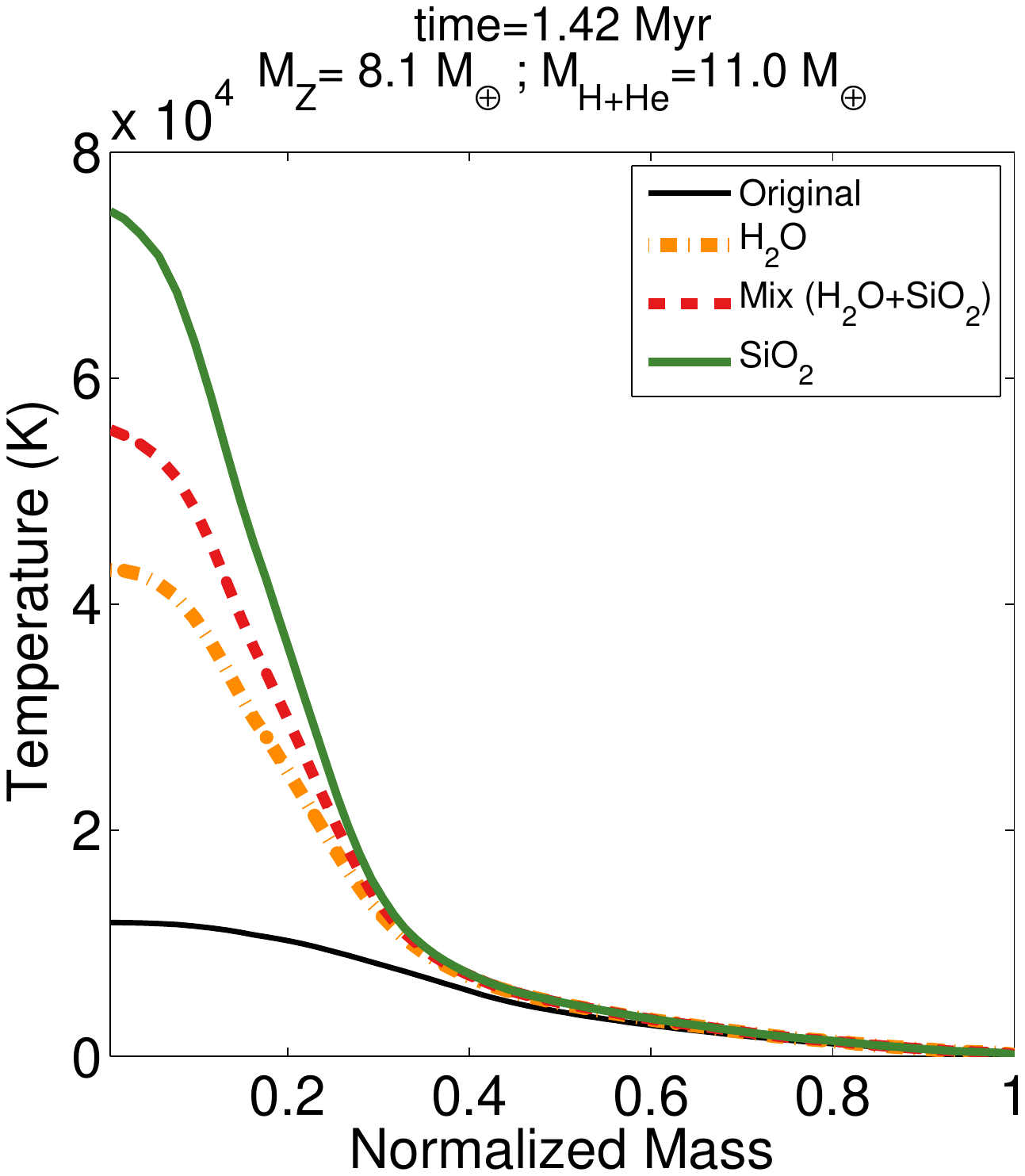}}\\
\caption{The temperature profile as function of normalized mass for four different times for Model-B. Shown are four different temperature profiles: the original temperature profile (without heavy elements, solid-black), pure water (dashed-orange), mixture of water and silica (dashed-red), and pure silica (solid-green).}\label{fig:T-profiles, s=10, r=100}
\end{figure}

\newpage
\begin{figure}[!h]\centering
\subfloat{\includegraphics[trim=80 180 100 200,clip,width=0.42\columnwidth]{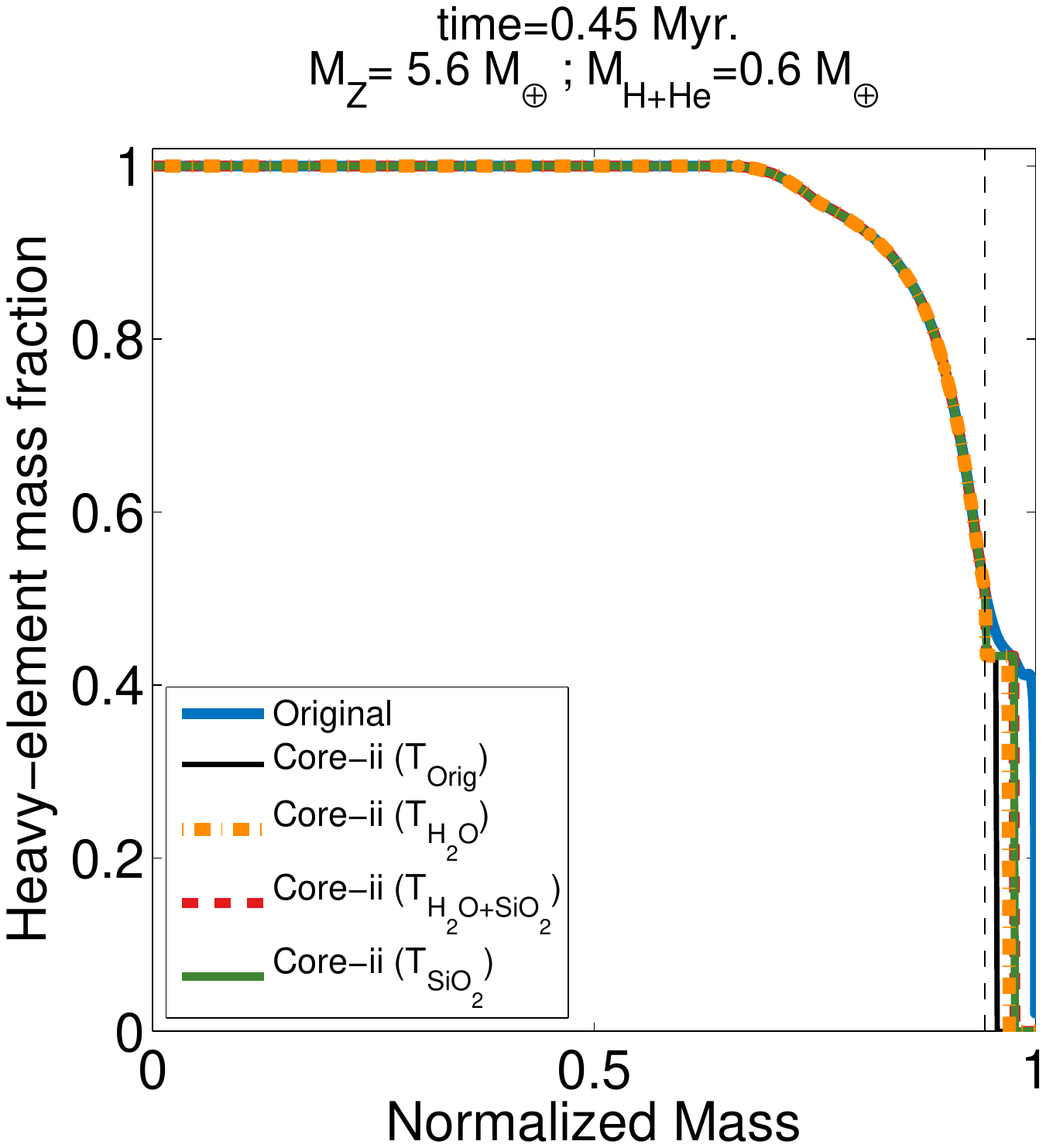}}
\subfloat{\includegraphics[trim=80 180 100 200,clip,width=0.42\columnwidth]{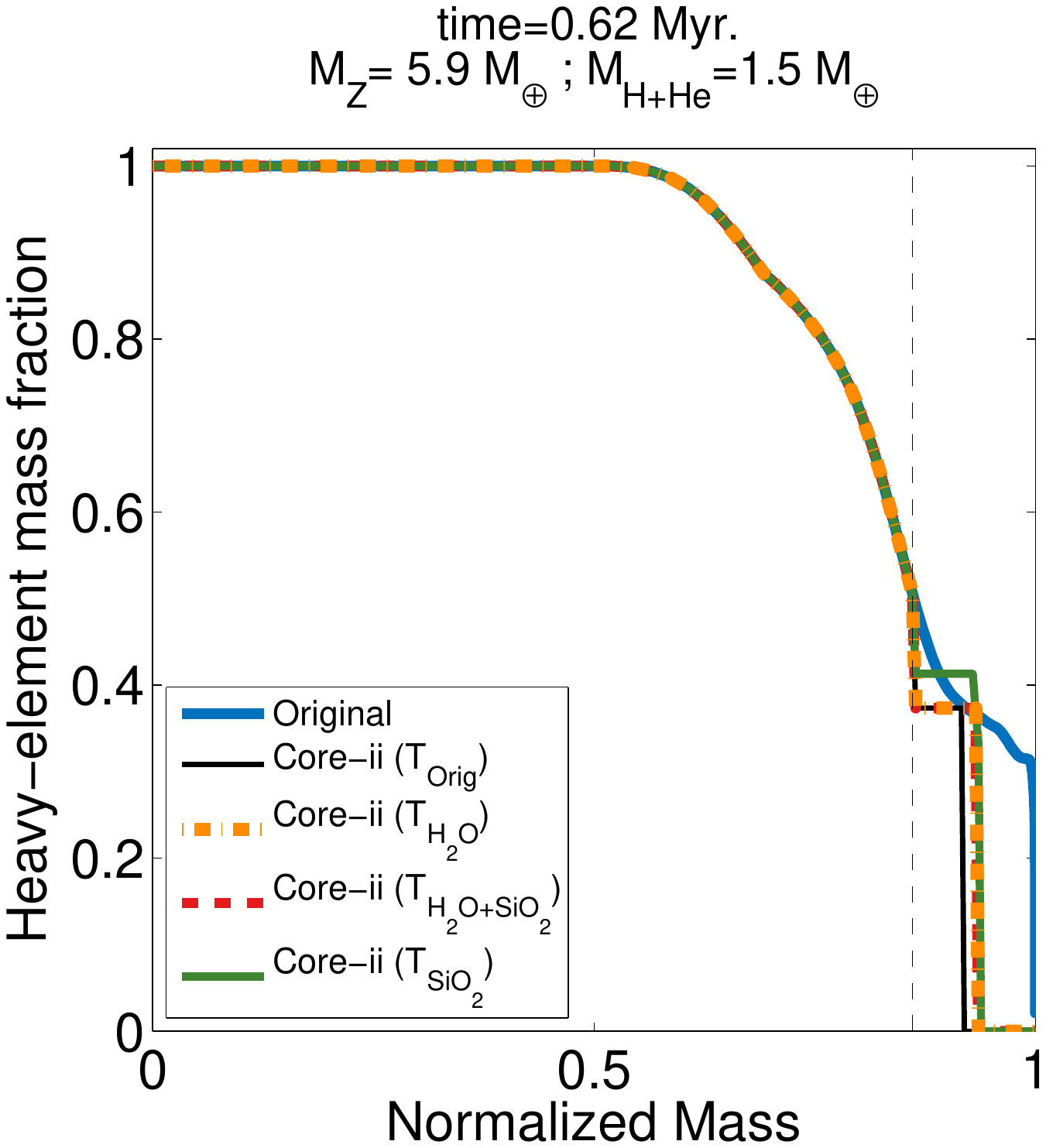}}\\
\subfloat{\includegraphics[trim=80 180 100 200,clip,width=0.42\columnwidth]{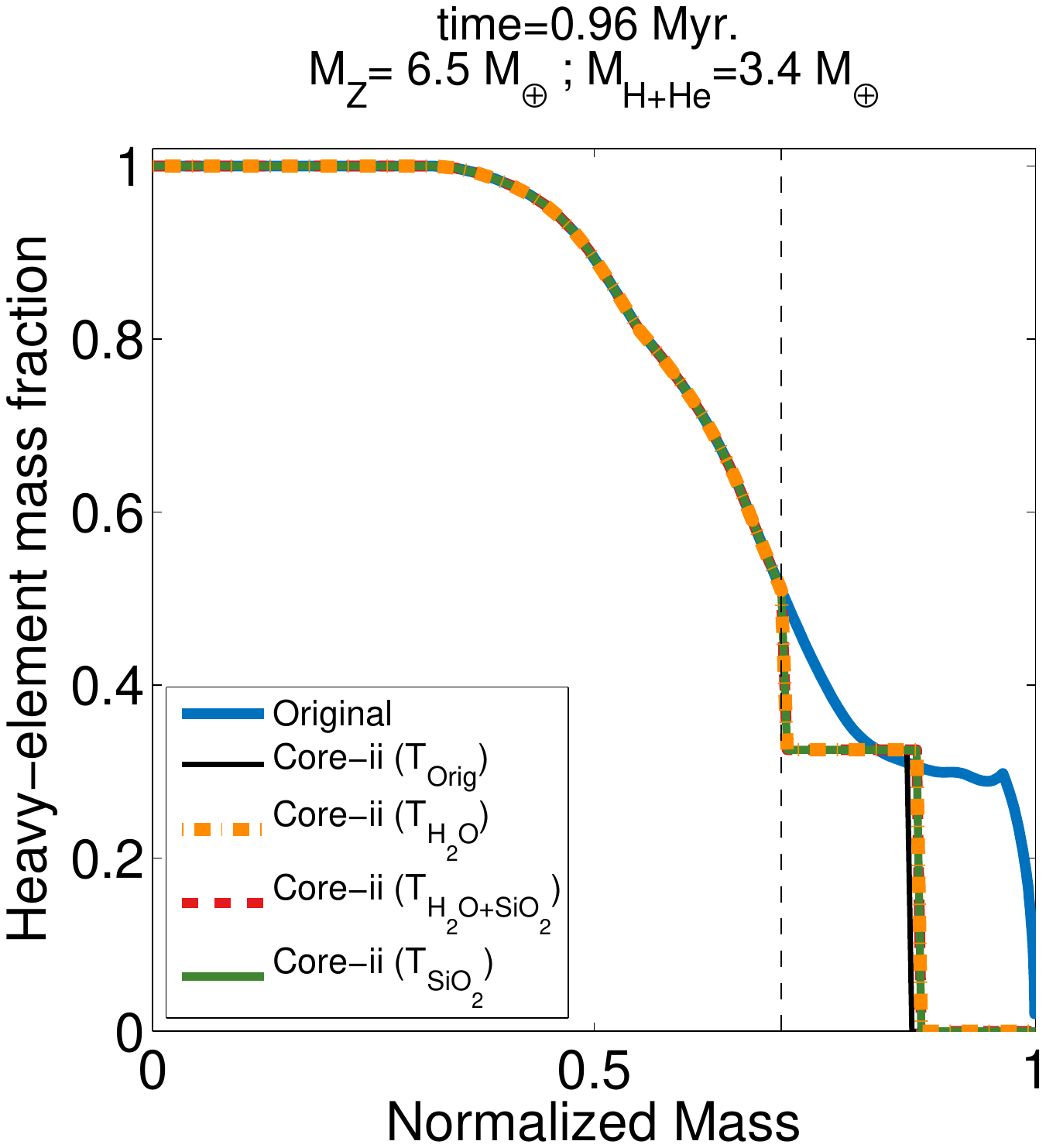}}
\subfloat{\includegraphics[trim=80 180 100 200,clip,width=0.42\columnwidth]{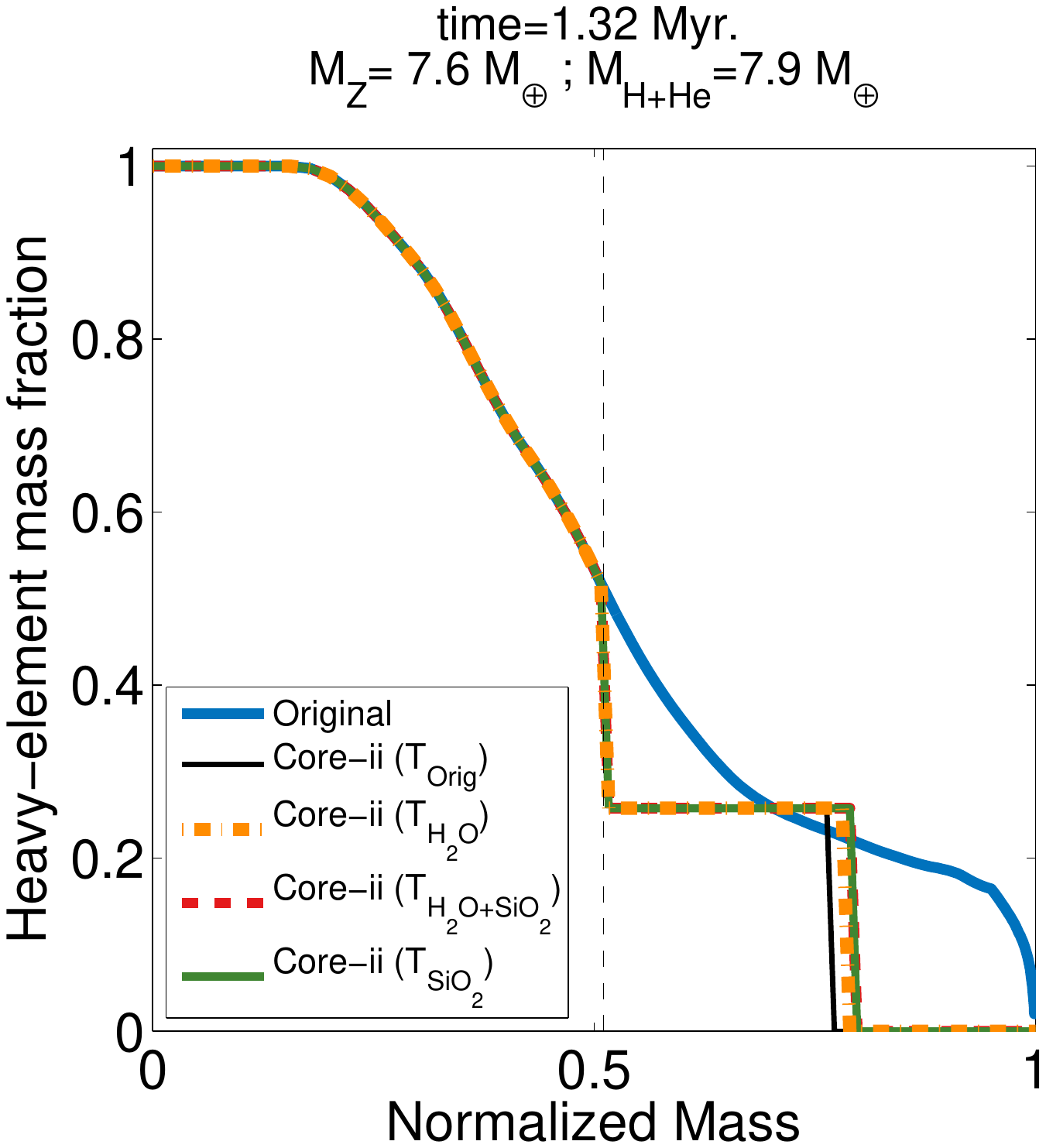}}\\
\caption{ The heavy-element distribution as a function of normalized mass at different times for Model-B when using the different temperature profiles for different compositions.} \label{fig:Z-profile for T-profile, s=6, r=0.5 }
\end{figure}

\newpage
\begin{figure}[!h]\centering
\subfloat{\includegraphics[trim=80 180 100 200,clip,width=0.42\columnwidth]{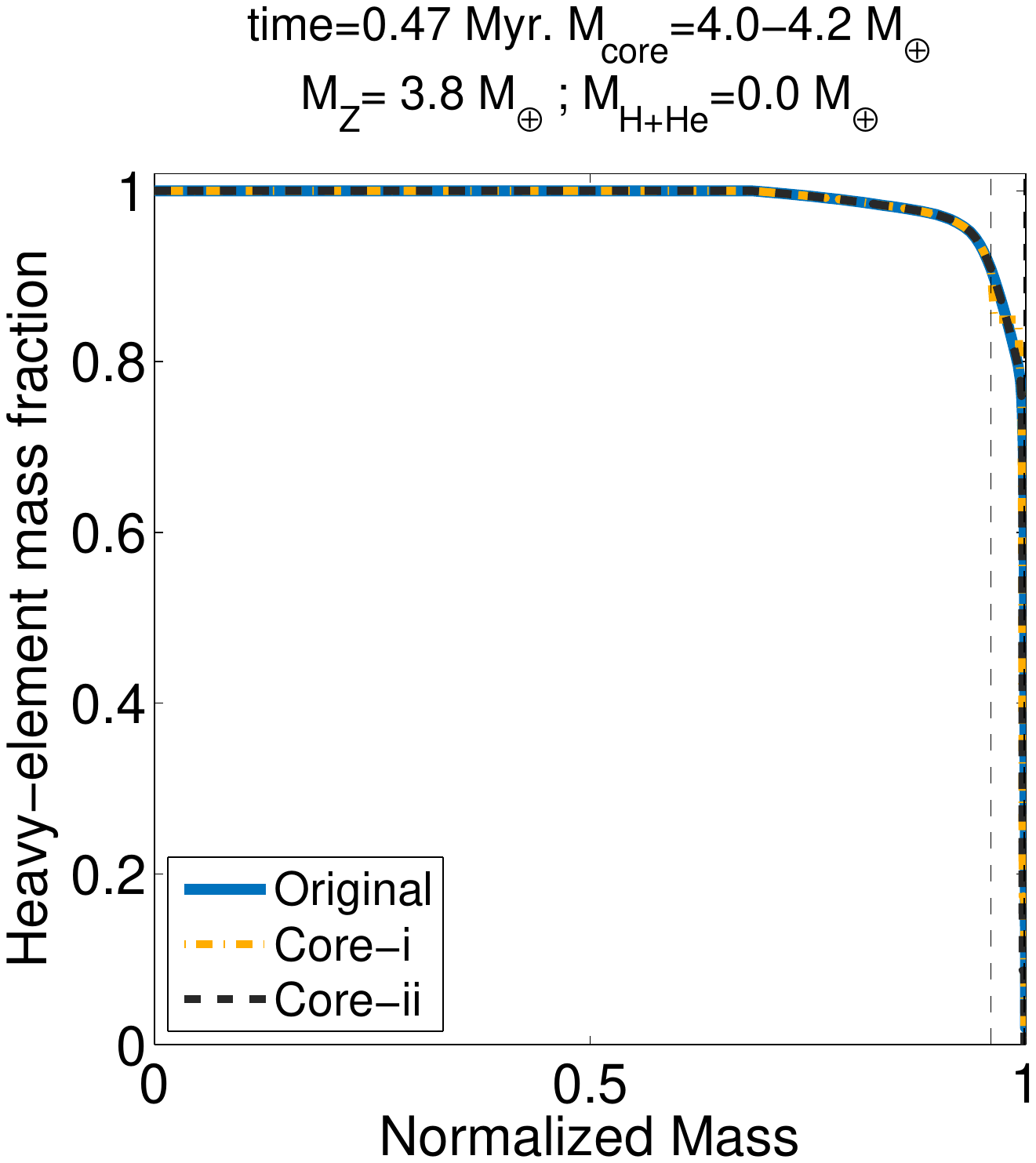}}
\subfloat{\includegraphics[trim=80 180 100 200,clip,width=0.42\columnwidth]{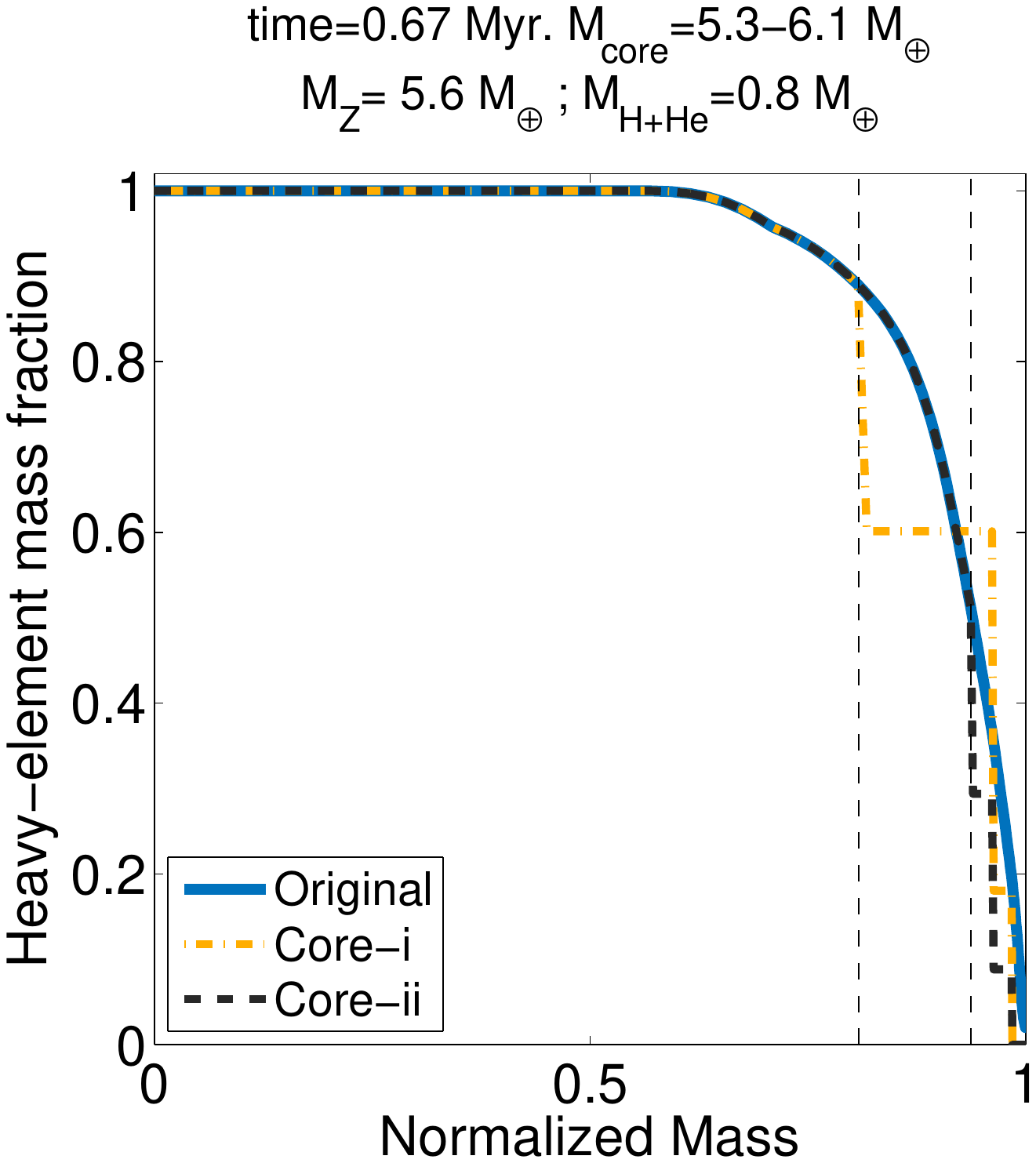}}\\
\subfloat{\includegraphics[trim=80 180 100 200,clip,width=0.42\columnwidth]{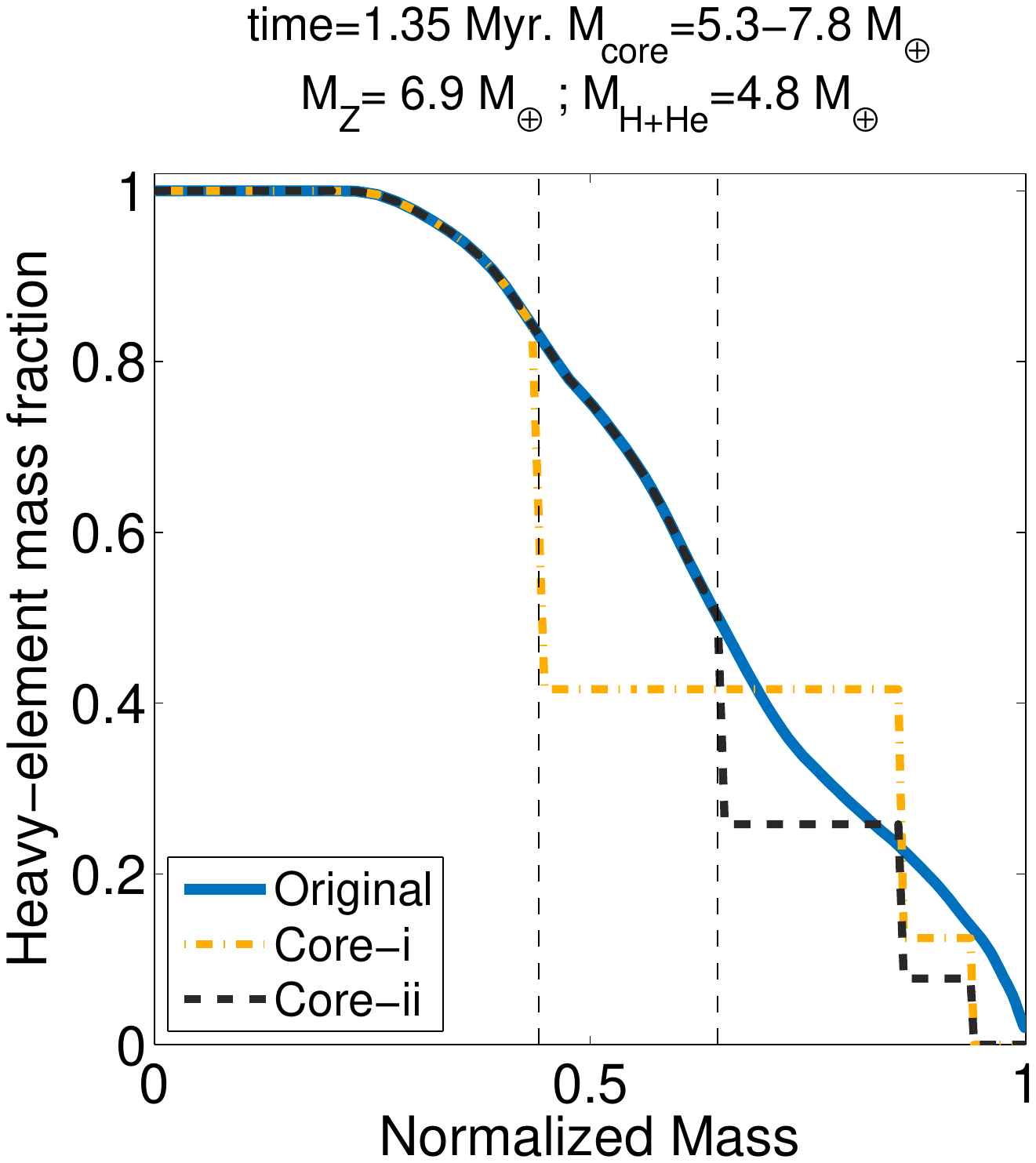}}
\subfloat{\includegraphics[trim=80 180 100 200,clip,width=0.42\columnwidth]{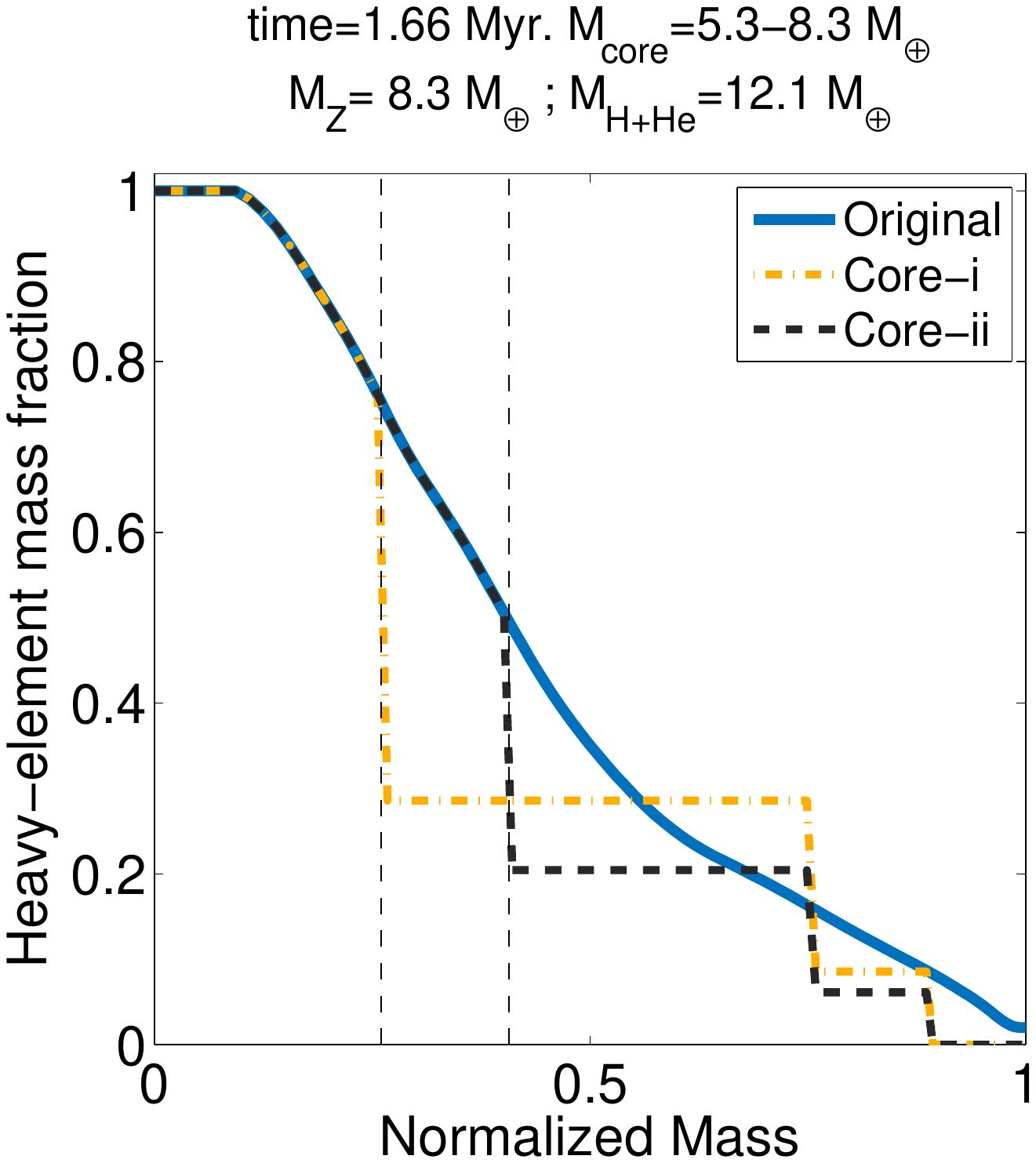}}\\
\caption{The distribution of the high-Z material (as a function of normalized mass) at different times for Model-A: $\sigma$=6 g/cm$^{2}$, planetesimal size: 100 km for the two core definitions. The dashed-dotted orange and dashed-black curves show the Z-profiles for Core-i, Core-ii, respectively. 
The core region is represented by dashed-black line.}\label{fig:Z05Z09, s=6, r=100}
\end{figure}

\newpage
\begin{figure}[!h]\centering
\subfloat{\includegraphics[trim=80 180 100 200,clip,width=0.42\columnwidth]{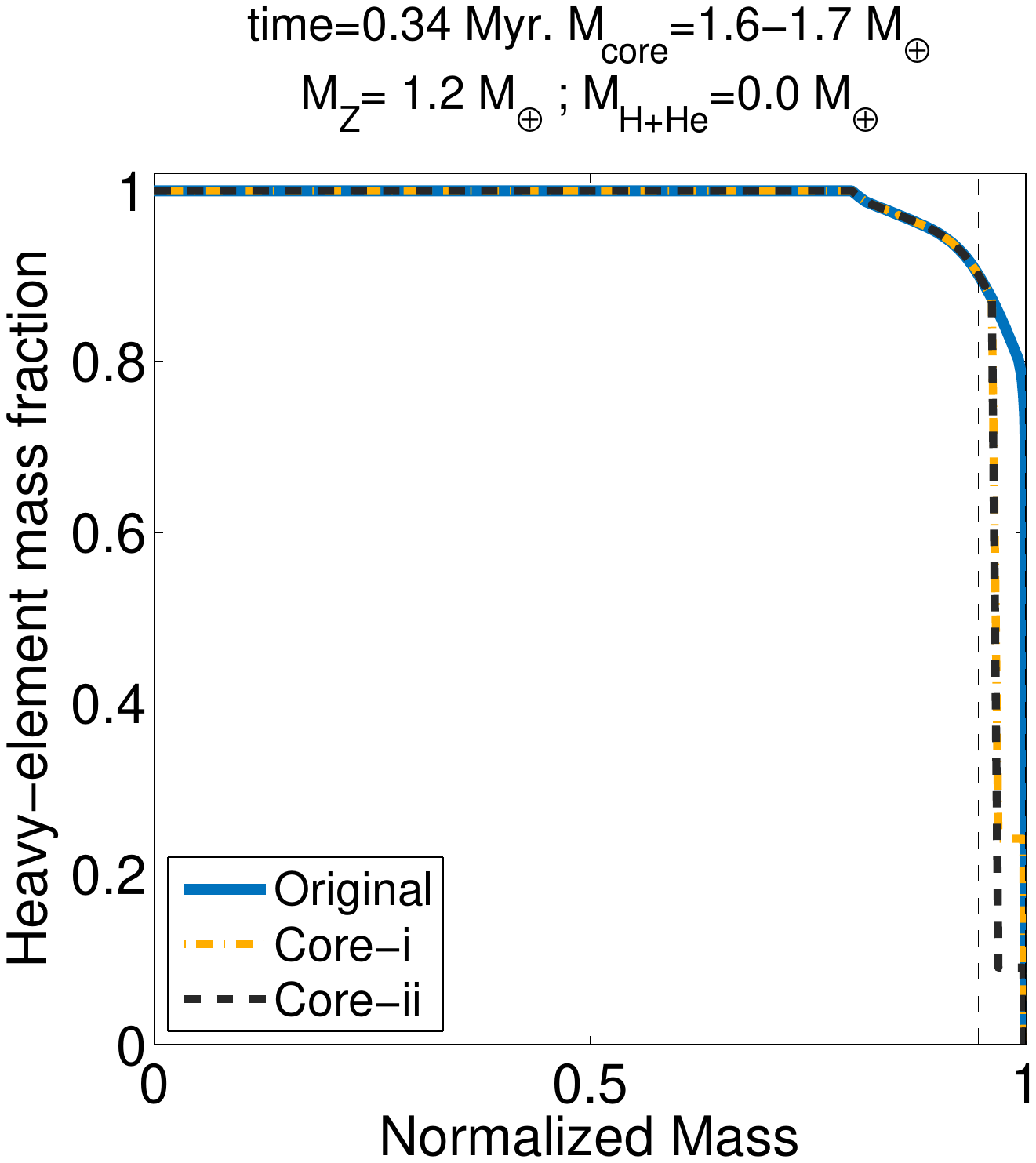}}
\subfloat{\includegraphics[trim=80 180 100 200,clip,width=0.42\columnwidth]{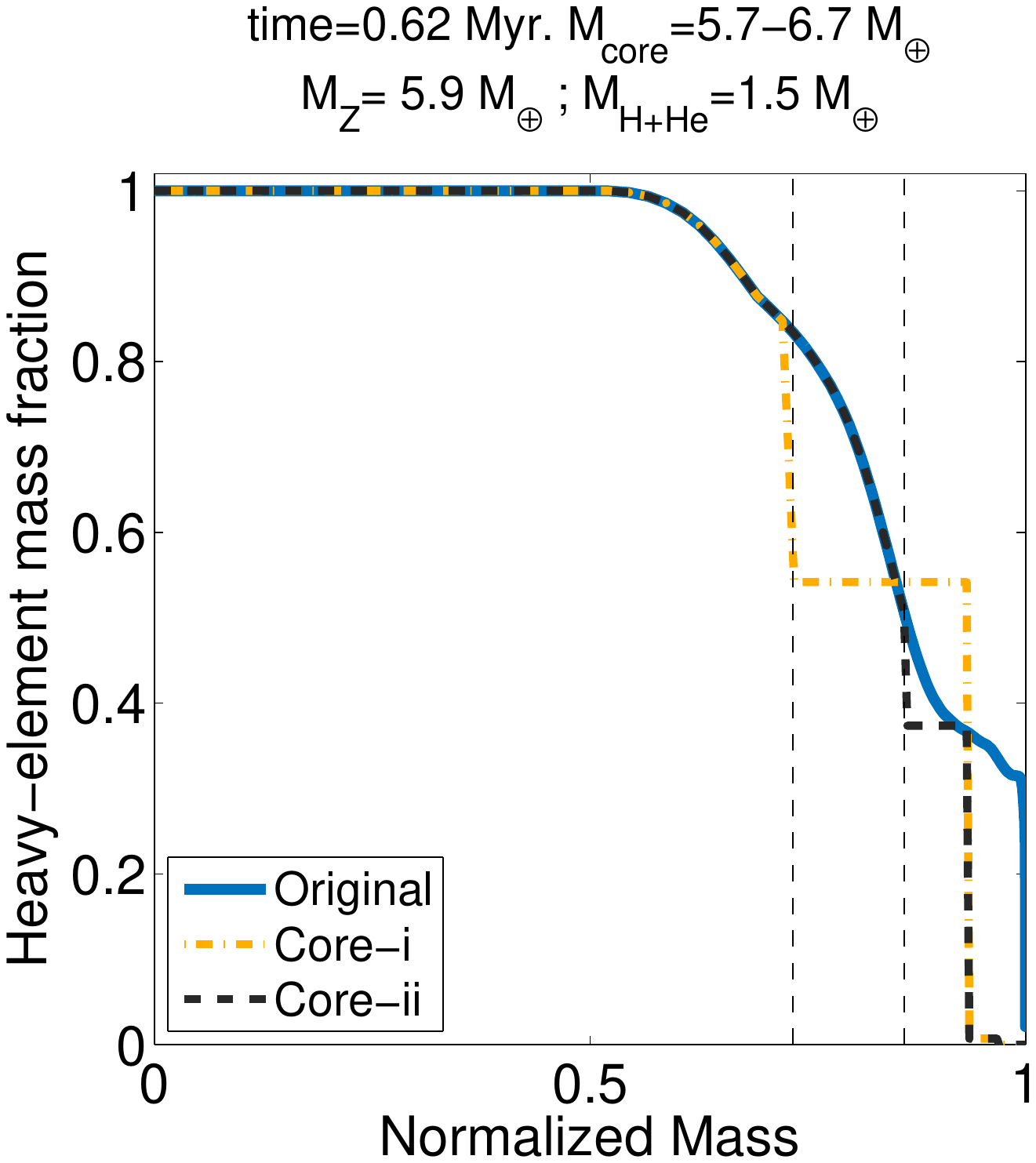}}\\
\subfloat{\includegraphics[trim=80 180 100 200,clip,width=0.42\columnwidth]{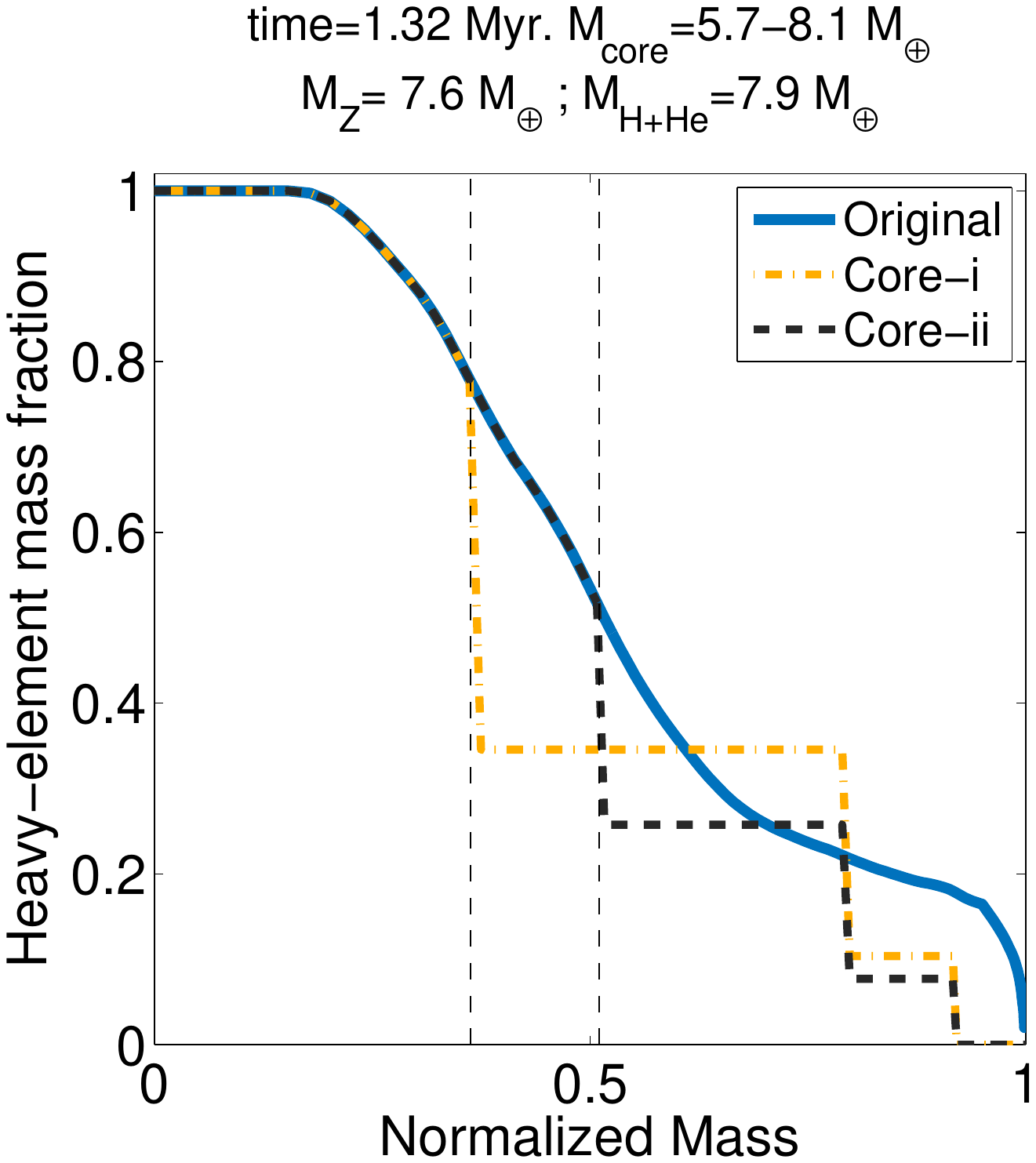}}
\subfloat{\includegraphics[trim=80 180 100 200,clip,width=0.42\columnwidth]{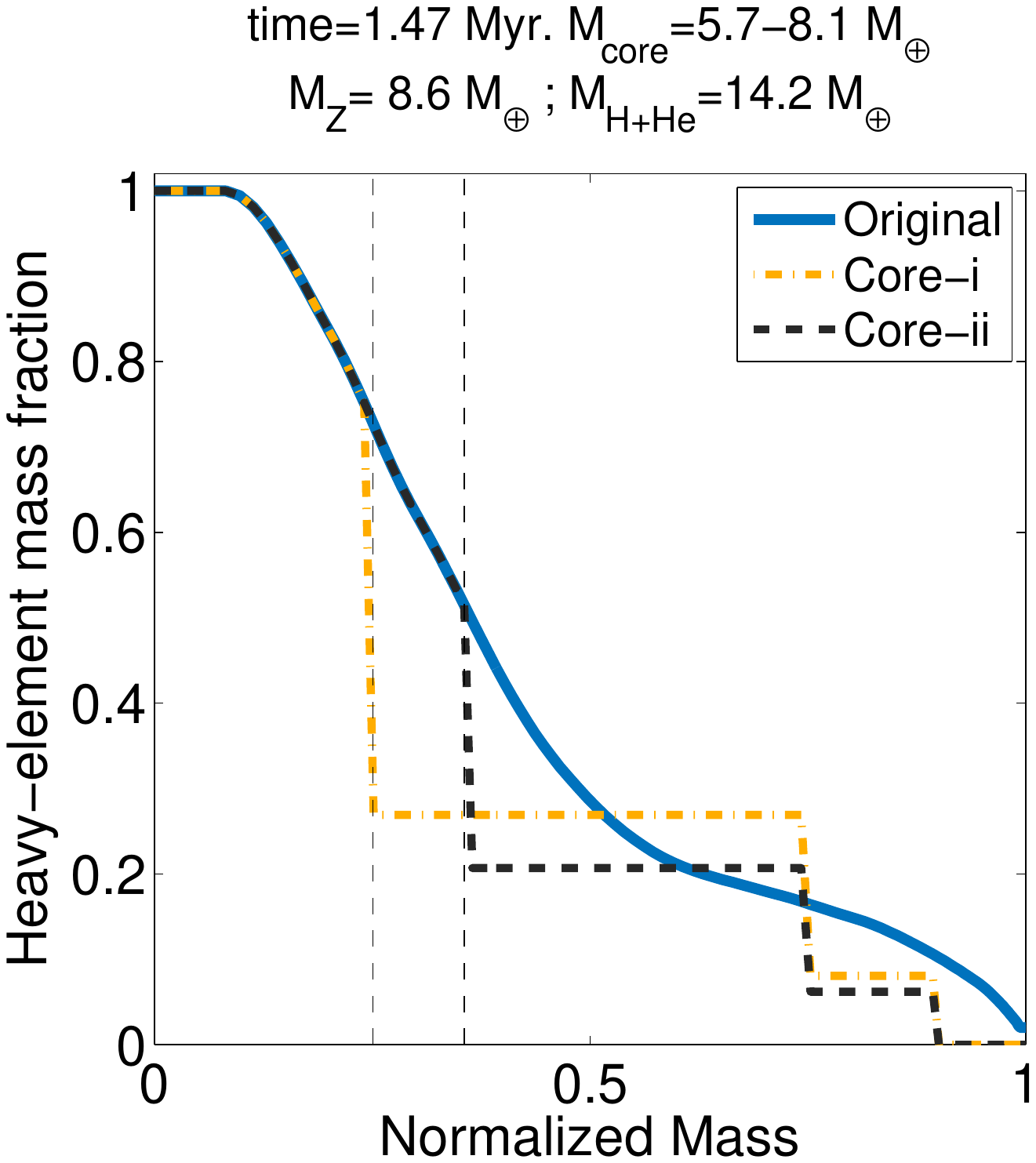}}\\
\caption{Same as Figure \ref{fig:Z05Z09, s=6, r=100} for Model-B: $\sigma$=6 g/cm$^{2}$, planetesimal size: 0.5 km.}\label{fig:CoreNoCore s=6, r=0.5}
\end{figure}

\newpage
\begin{figure}[!h]\centering
\subfloat{\includegraphics[trim=80 180 100 200,clip,width=0.42\columnwidth]{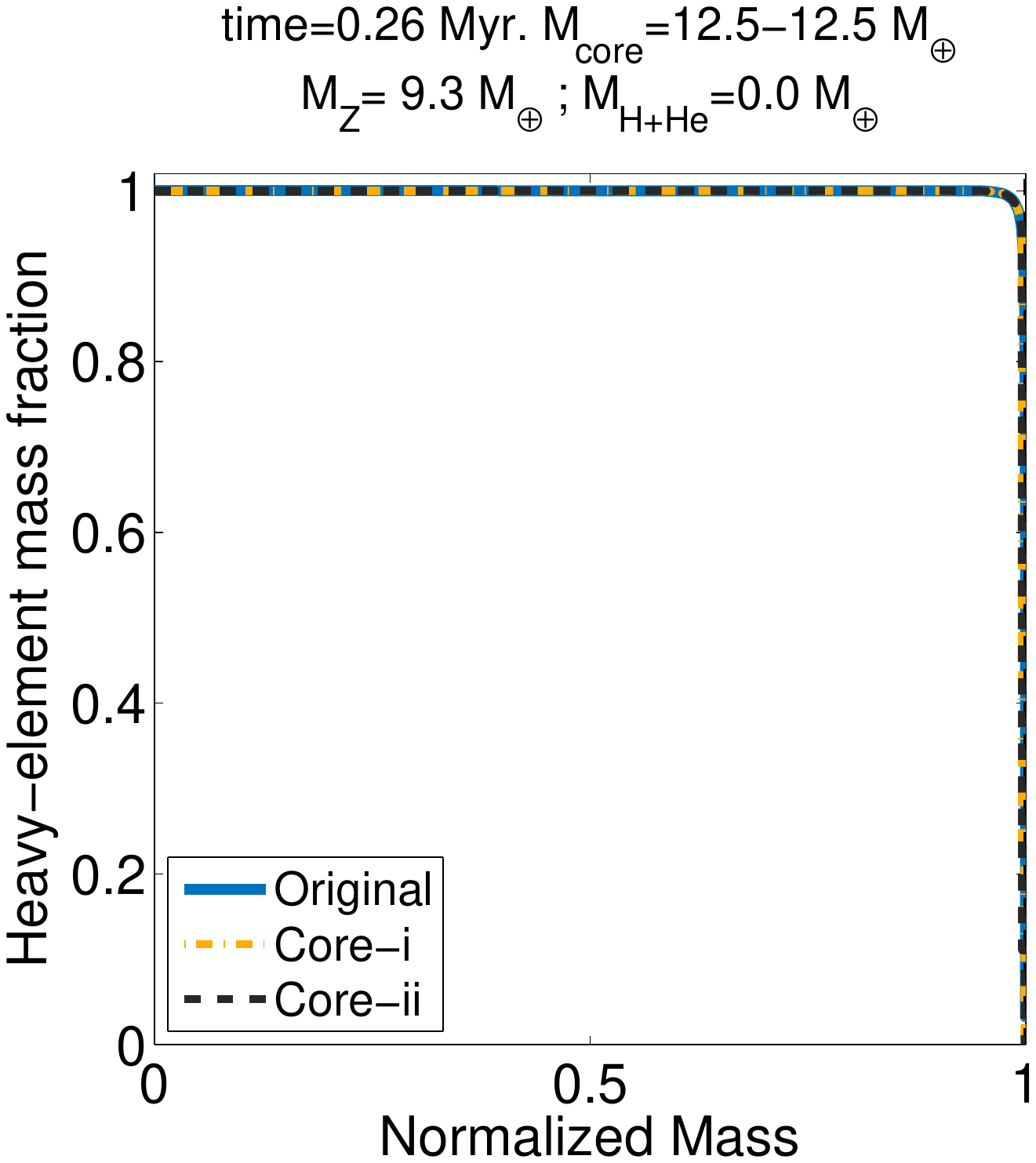}}
\subfloat{\includegraphics[trim=80 180 100 200,clip,width=0.42\columnwidth]{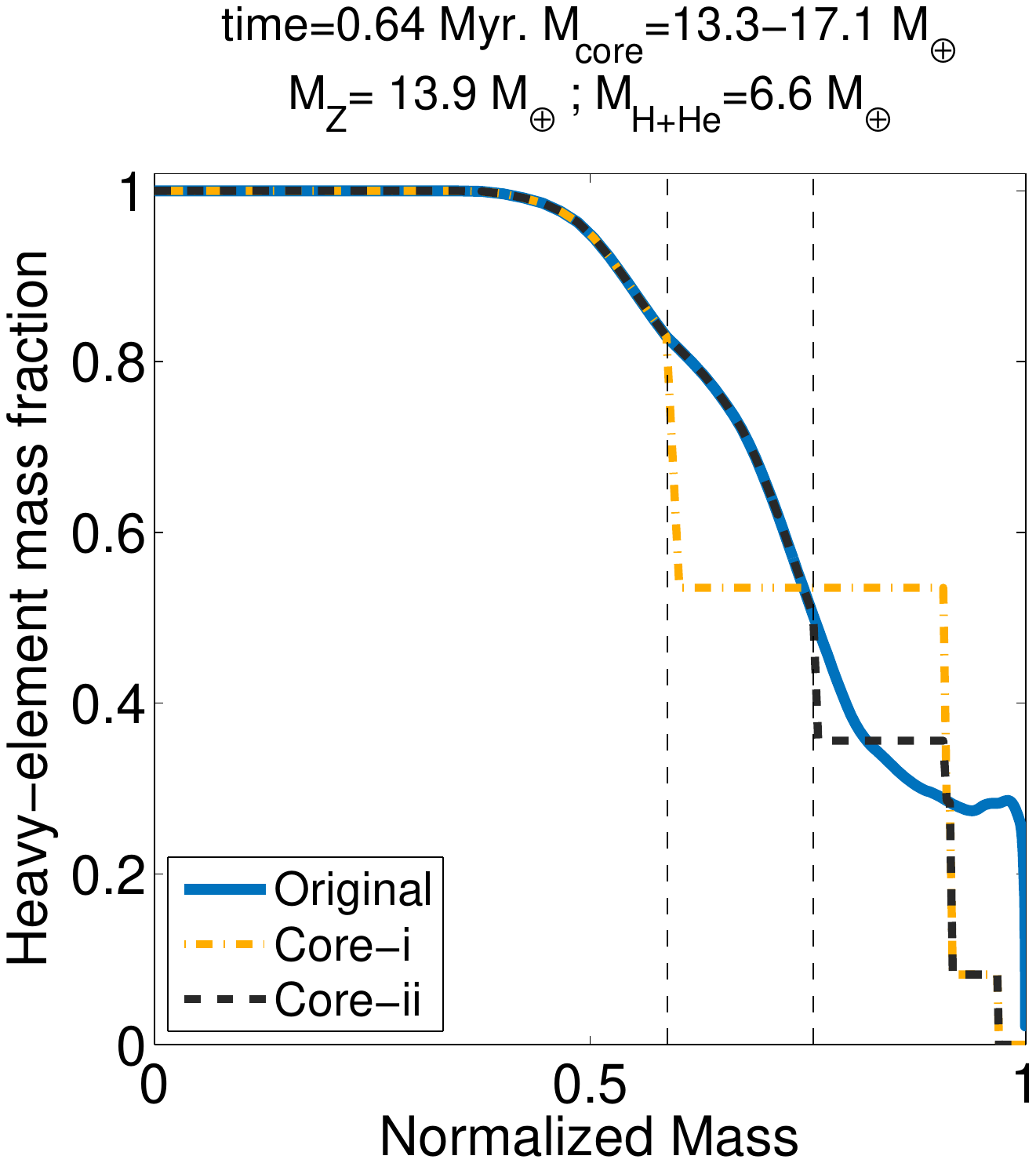}}\\
\subfloat{\includegraphics[trim=80 180 100 200,clip,width=0.42\columnwidth]{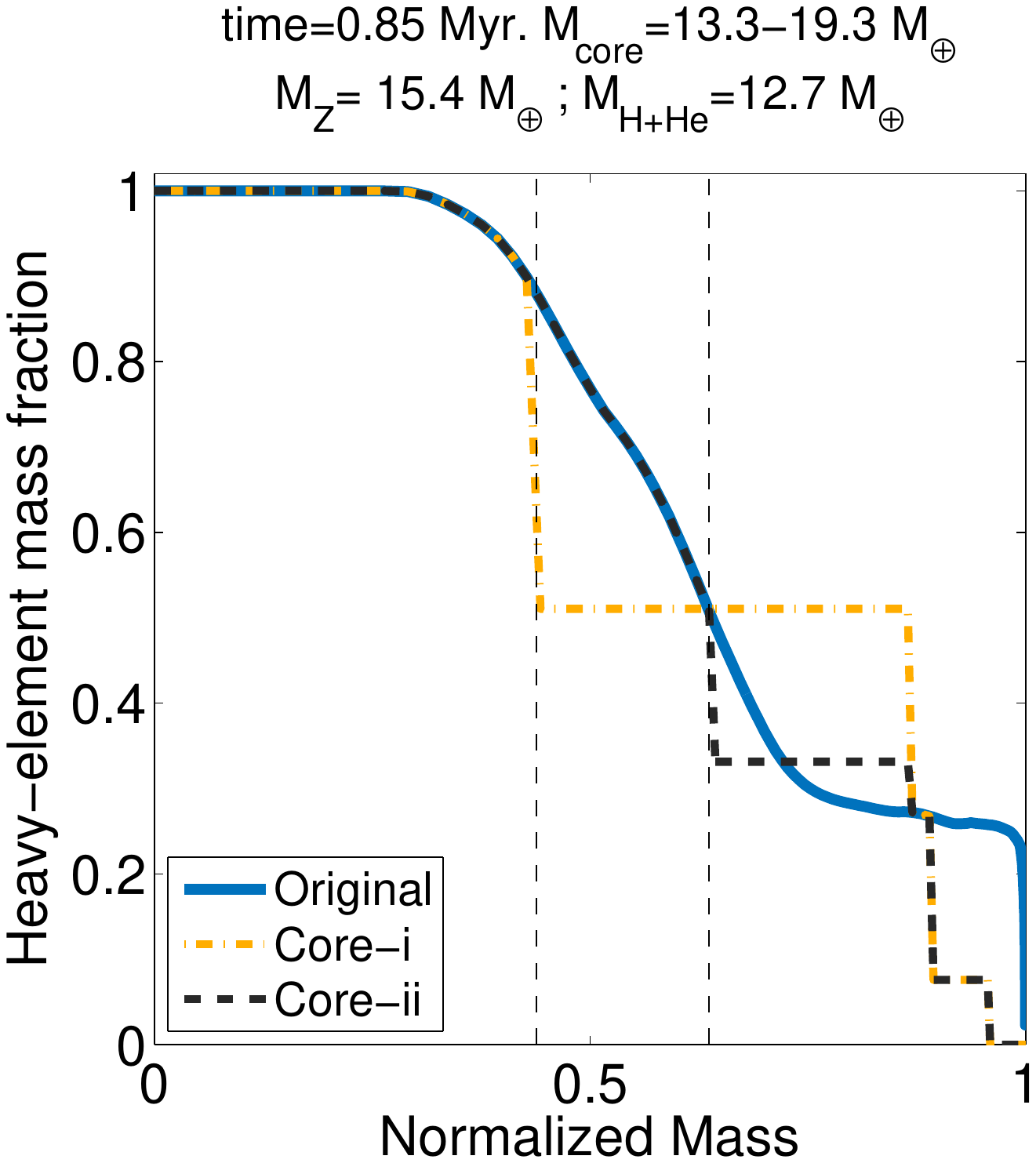}}
\subfloat{\includegraphics[trim=80 180 100 200,clip,width=0.42\columnwidth]{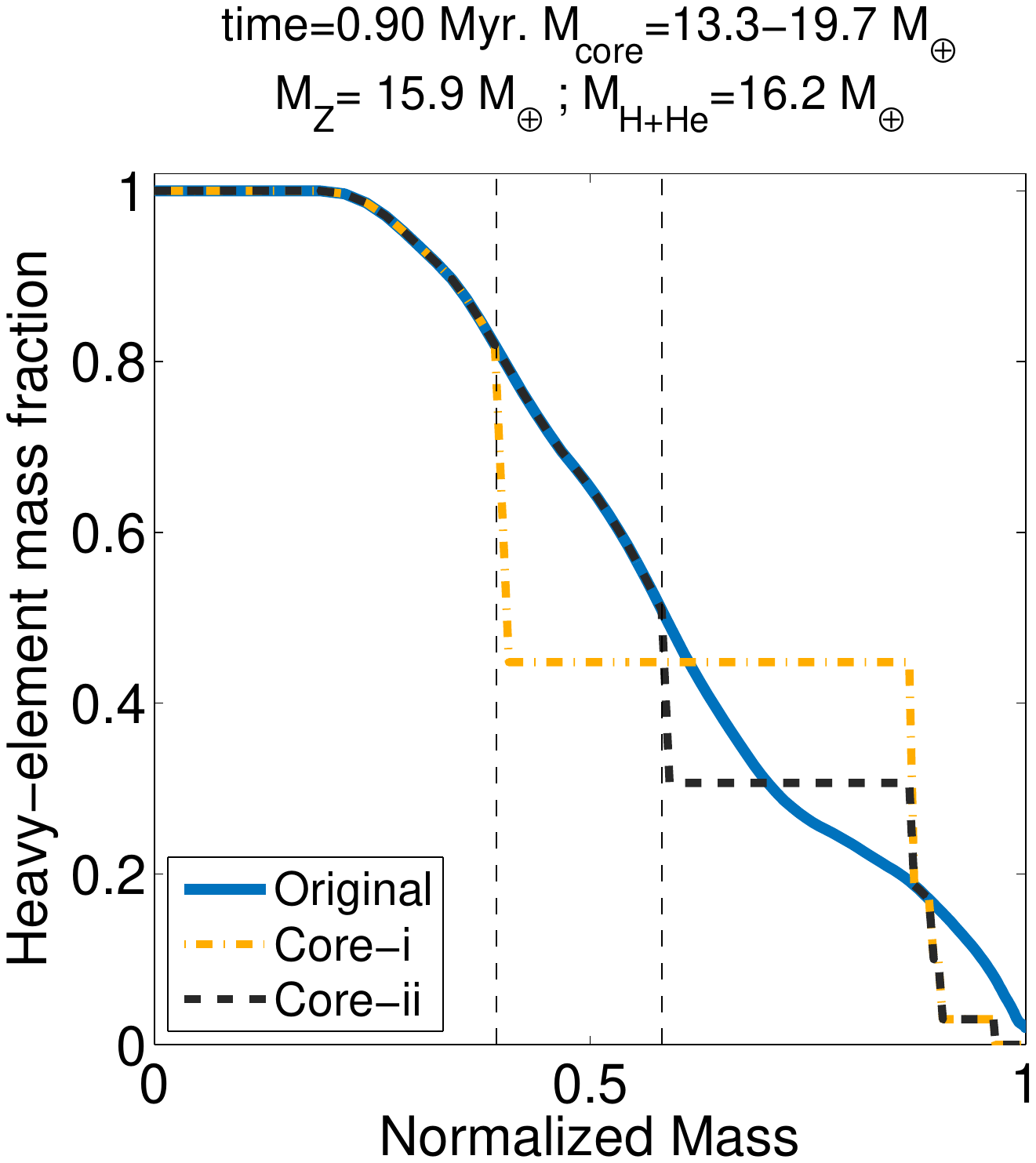}}
\caption{Same as Figure \ref{fig:Z05Z09, s=6, r=100} for Model-C:  $\sigma$=10 g/cm$^{2}$, planetesimal size: 1 km.}\label{fig:Z05Z09, s=10, r=1}
\end{figure}

\newpage
\begin{figure}[!h]\centering
\subfloat{\includegraphics[trim=80 180 100 200,clip,width=0.42\columnwidth]{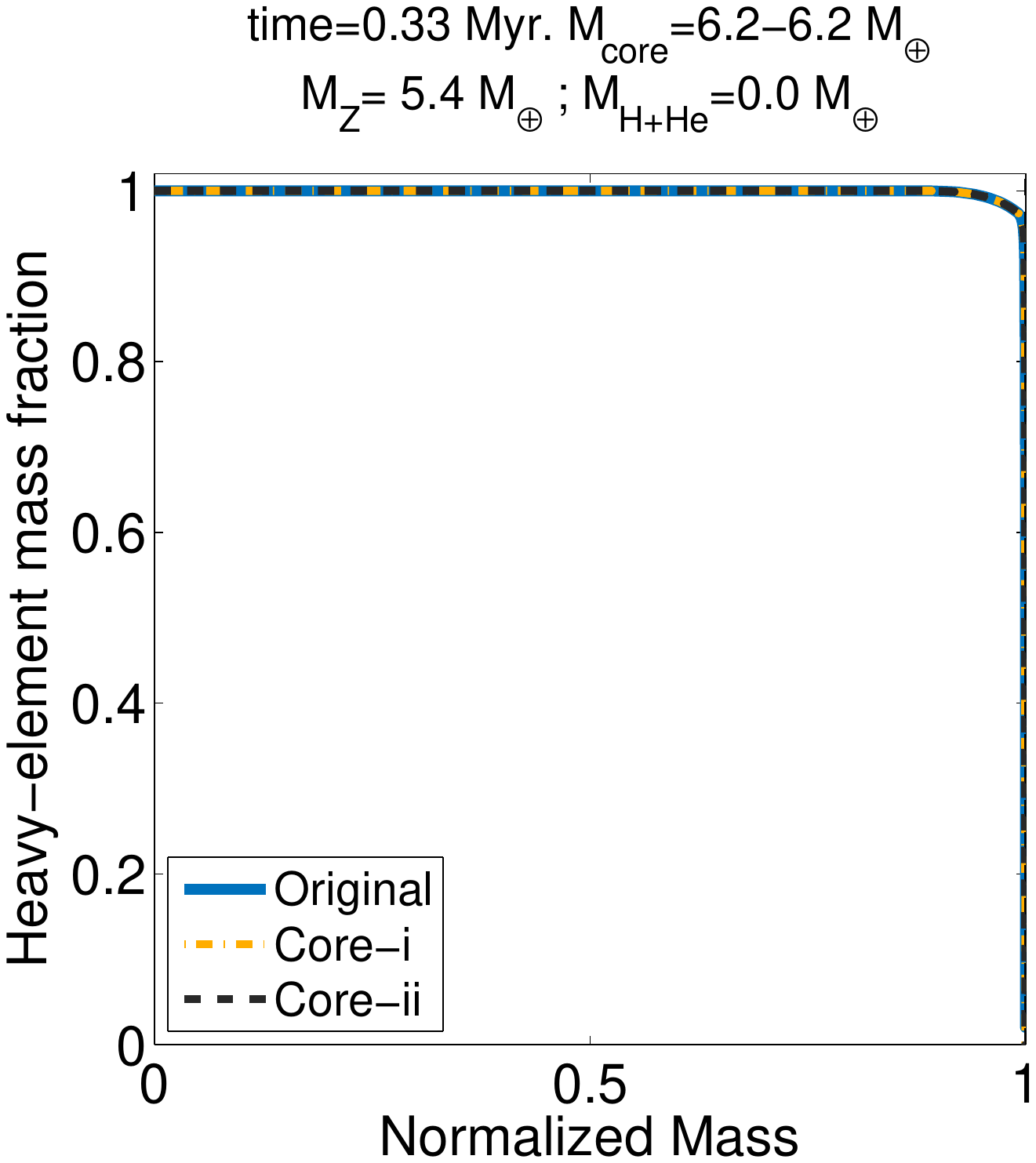}}
\subfloat{\includegraphics[trim=80 180 100 200,clip,width=0.42\columnwidth]{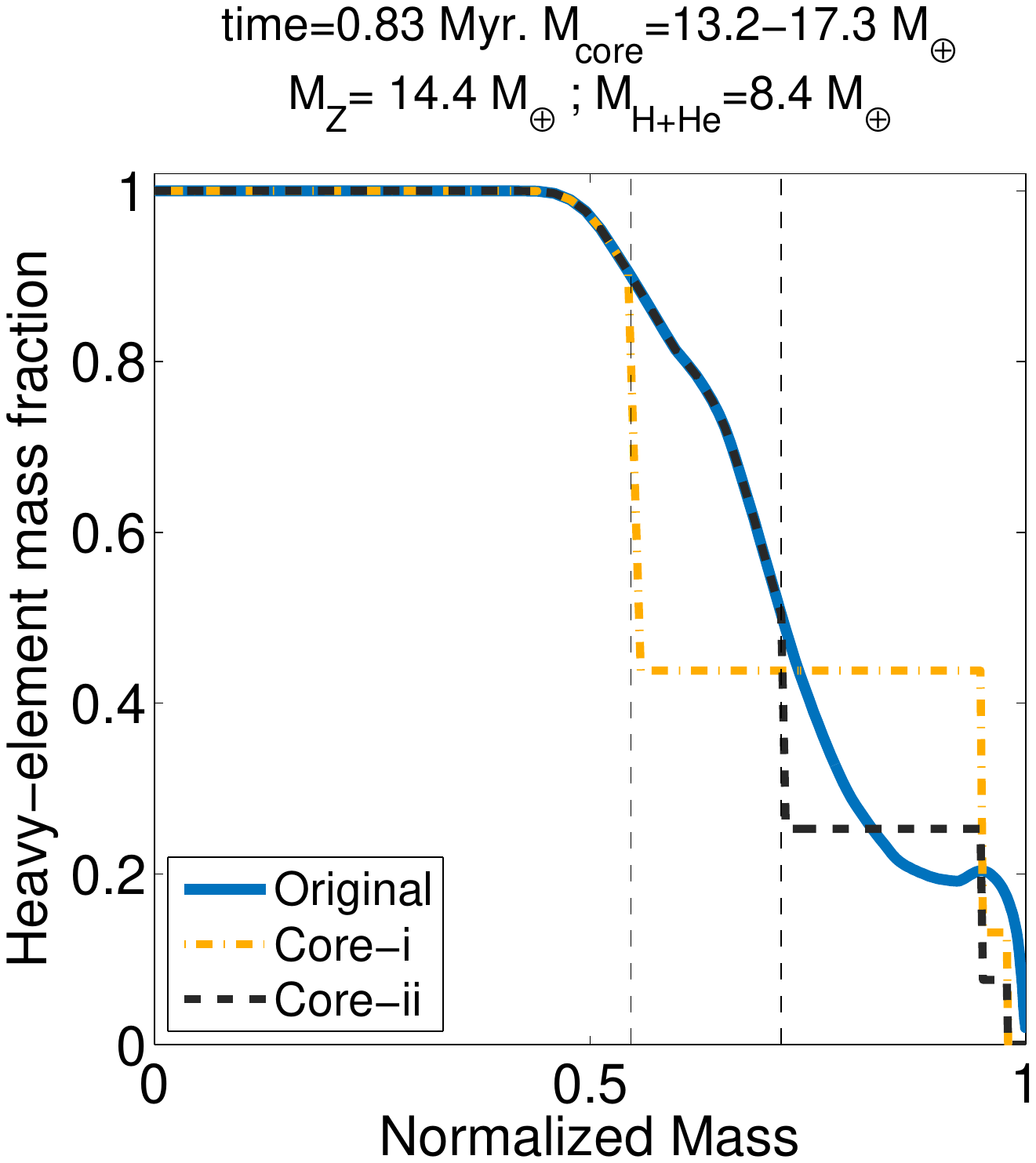}}\\
\subfloat{\includegraphics[trim=80 180 100 200,clip,width=0.42\columnwidth]{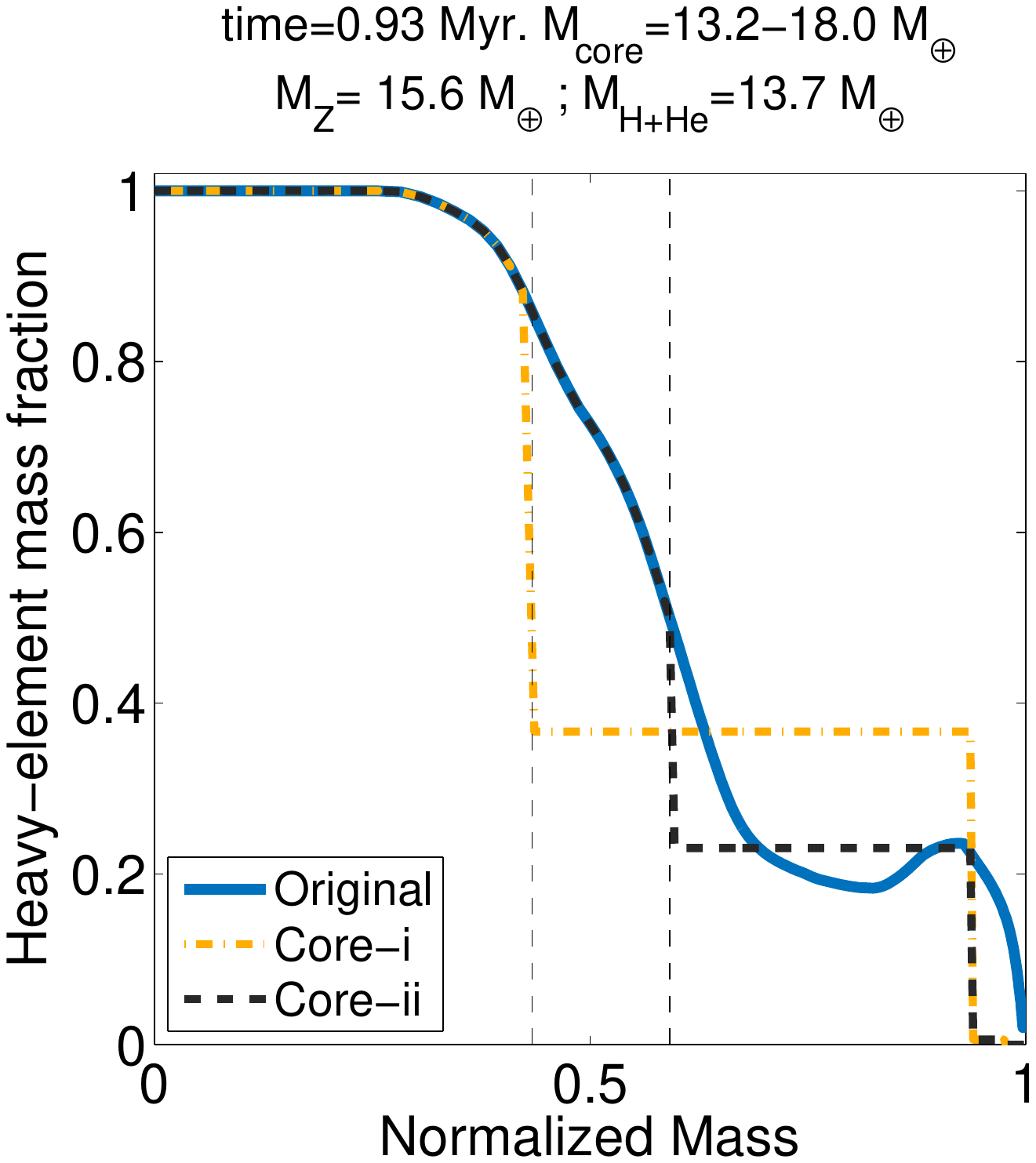}}
\subfloat{\includegraphics[trim=80 180 100 200,clip,width=0.42\columnwidth]{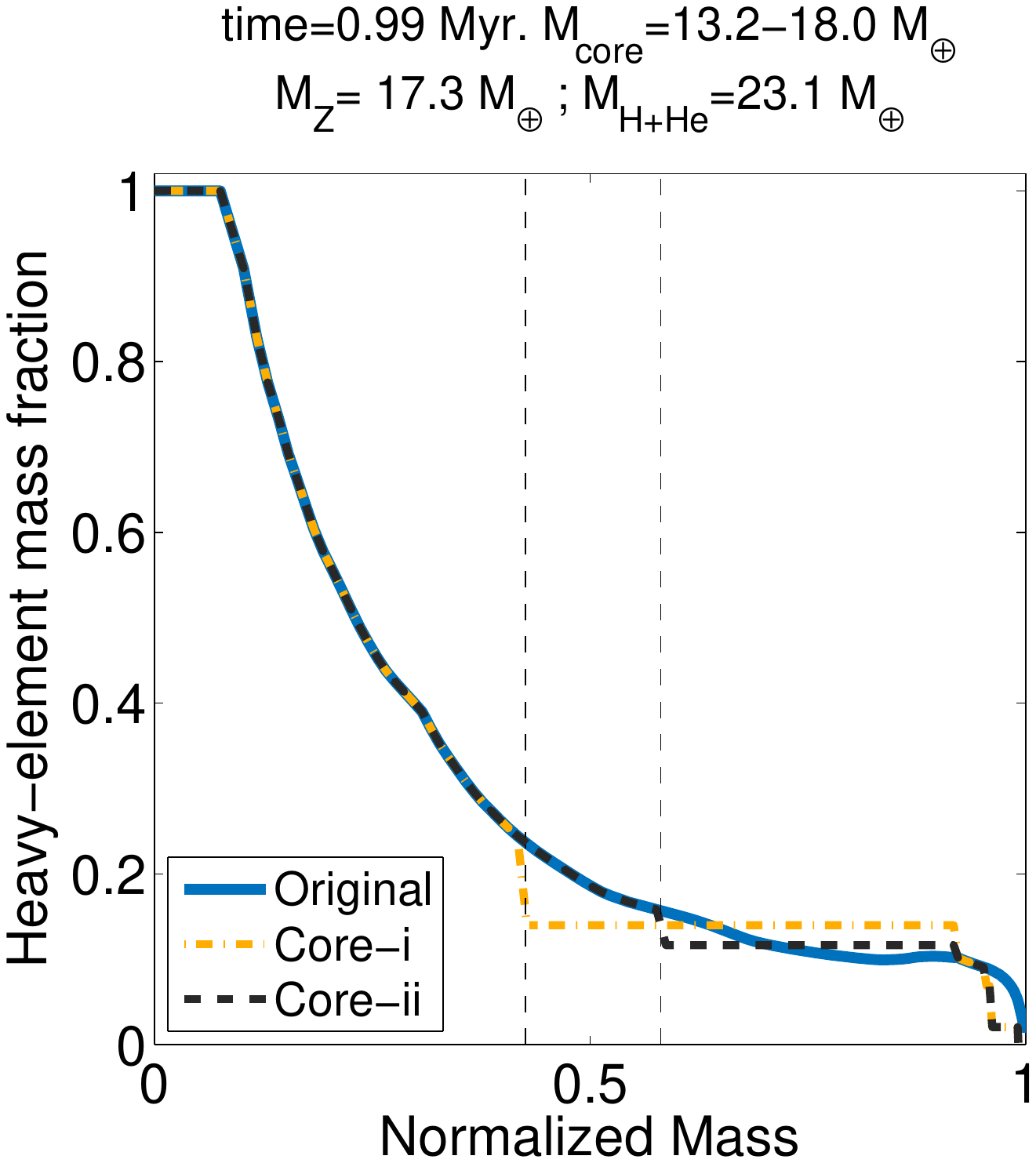}}\\
\caption{Same as Figure \ref{fig:Z05Z09, s=6, r=100} for Model-D: $\sigma$=10 g/cm$^{2}$, planetesimal size: 100 km.}\label{fig:Z05Z09, s=10, r=100}
\end{figure}

\newpage
\begin{figure}[!h]\centering
\subfloat{\includegraphics[trim=0 120 30 140,clip,width=0.55\columnwidth]{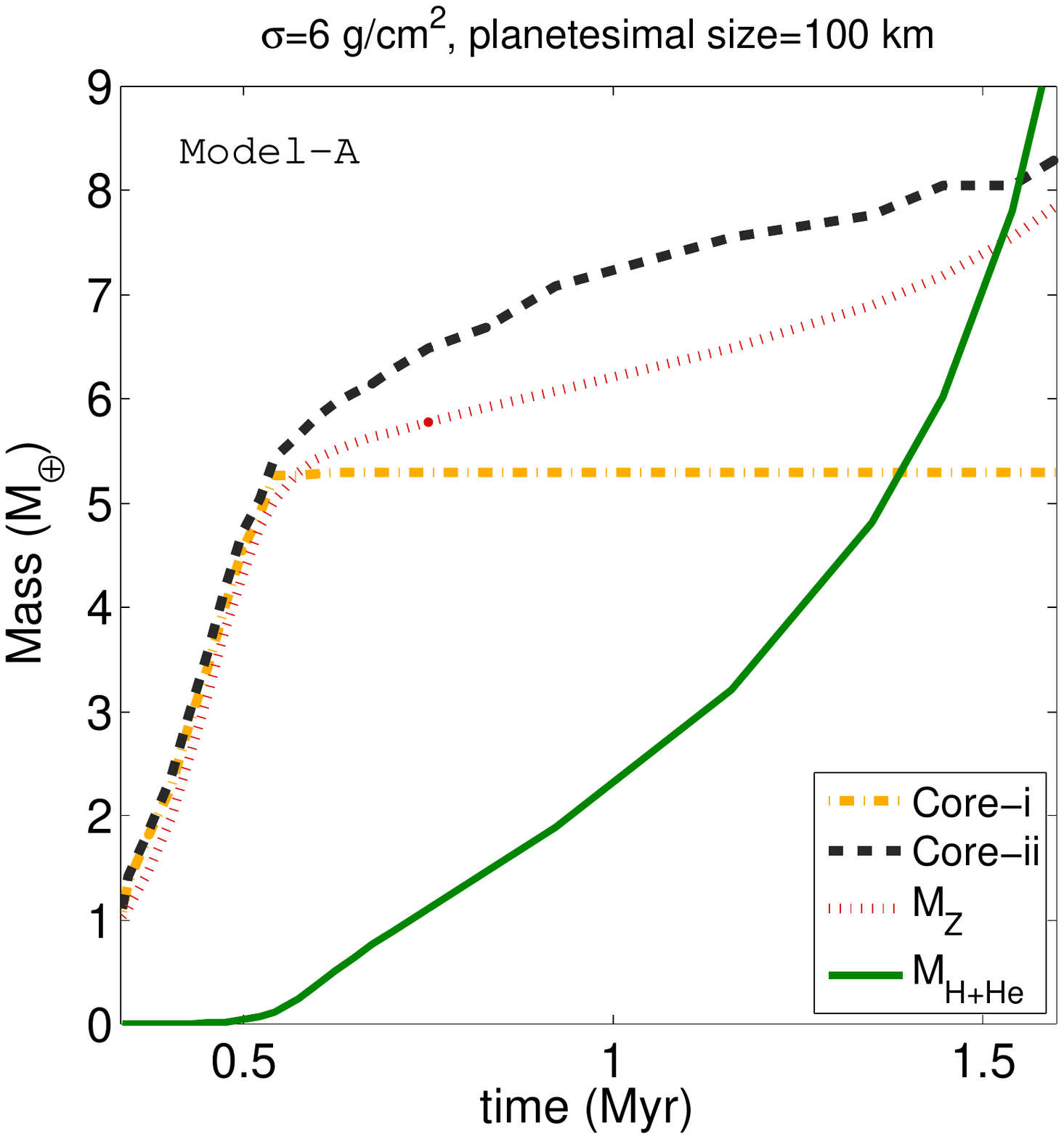}}
\subfloat{\includegraphics[trim=0 120 30 140,clip,width=0.55\columnwidth]{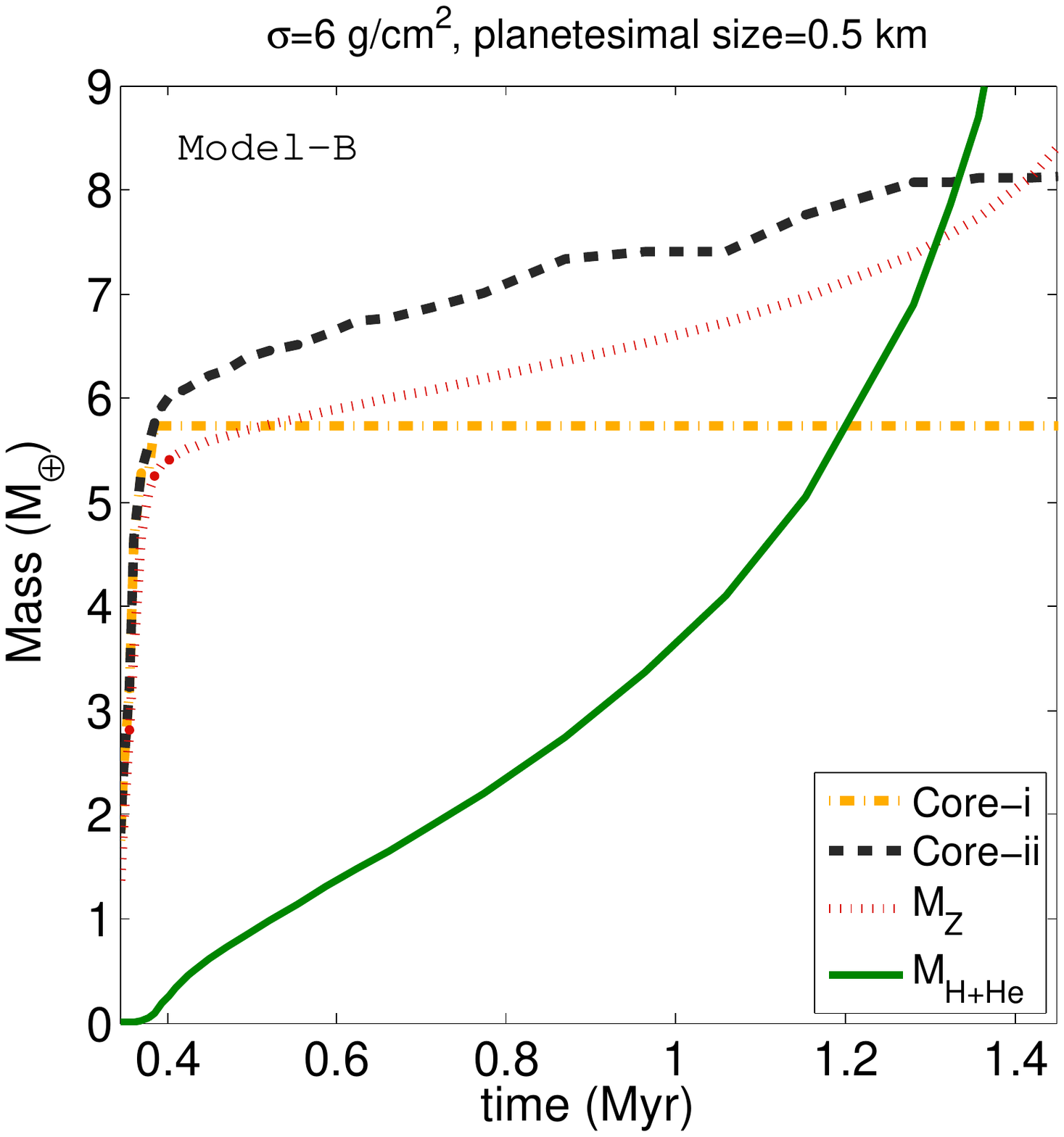}}\\
\vspace*{-1.2cm}
\subfloat{\includegraphics[trim=0 120 30 140,clip,width=0.55\columnwidth]{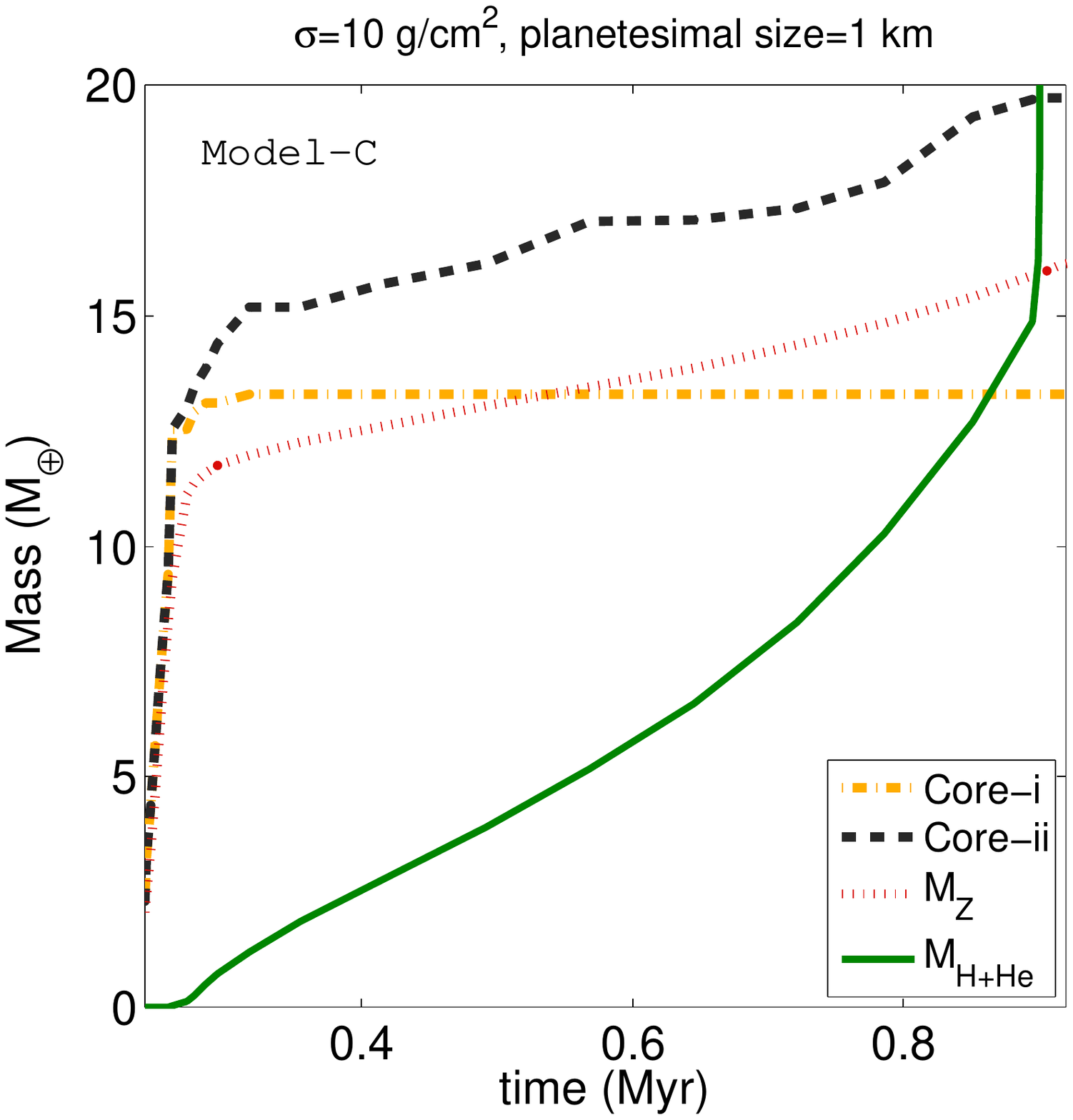}}
\subfloat{\includegraphics[trim=0 120 30 140,clip,width=0.55\columnwidth]{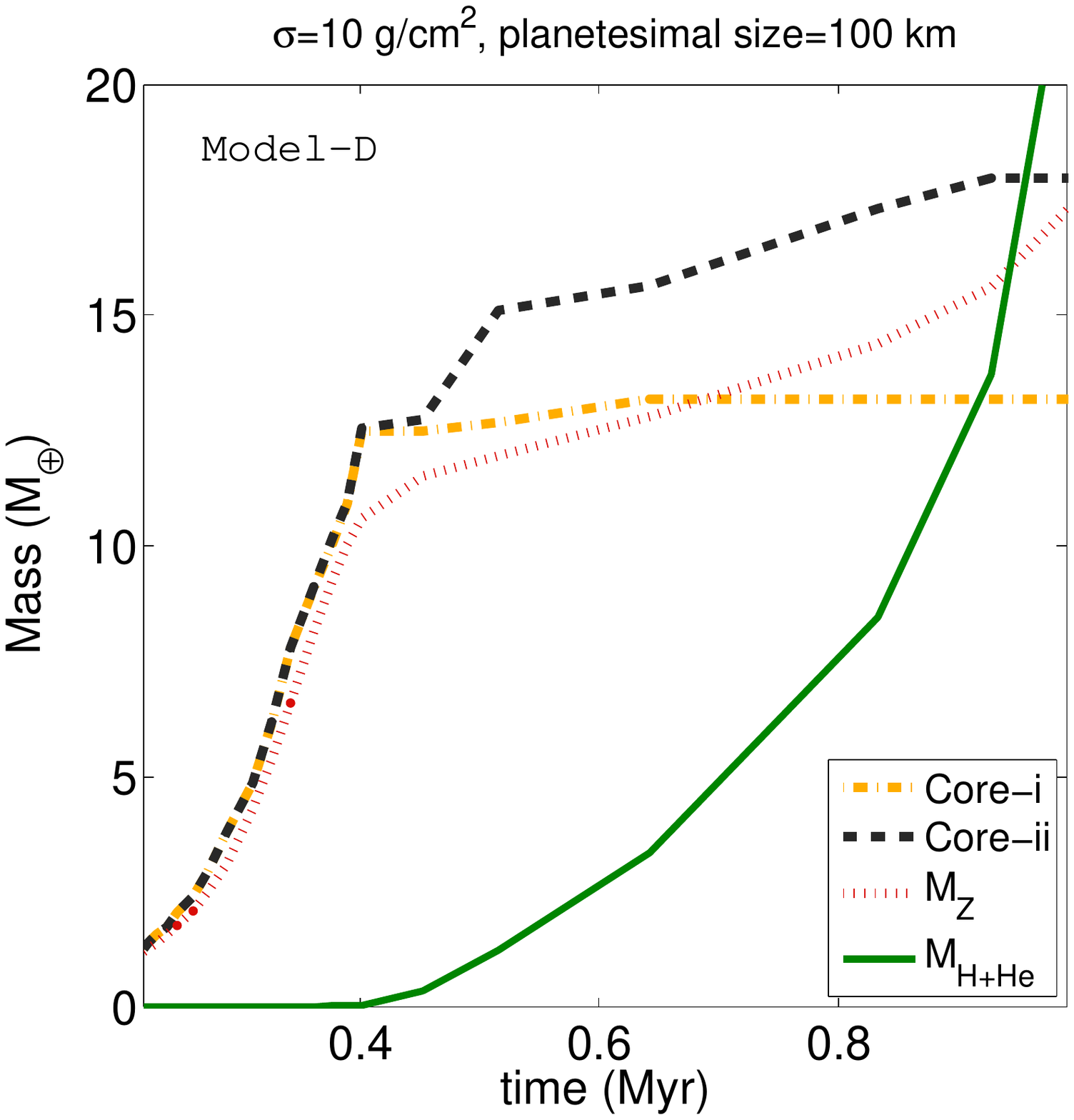}}
\caption{Evolution of the planetary mass for the four formation models. 
The masses of H+He and heavy elements are represented by the solid-brown and dotted-purple lines, respectively. Also here, the dashed-dotted orange and dashed-black curves correspond to Core-i, Core-ii, respectively. 
The gaseous (H+He) mass, total heavy-element mass, and core mass at crossover time are listed in Table \ref{tab:Crossover Z}.}\label{fig:Z05Z09, summary}
\end{figure}

\newpage
\begin{figure}[!h]\centering
\vspace*{-10mm}
\subfloat{\includegraphics[trim=10 750 0 70,clip,width=1\columnwidth]{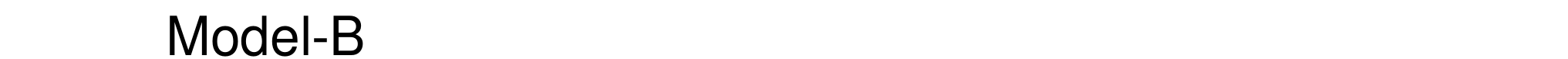}}\\
\vspace*{-10mm}
\subfloat{\includegraphics[trim=18 120 40 100,clip,width=0.32\columnwidth]{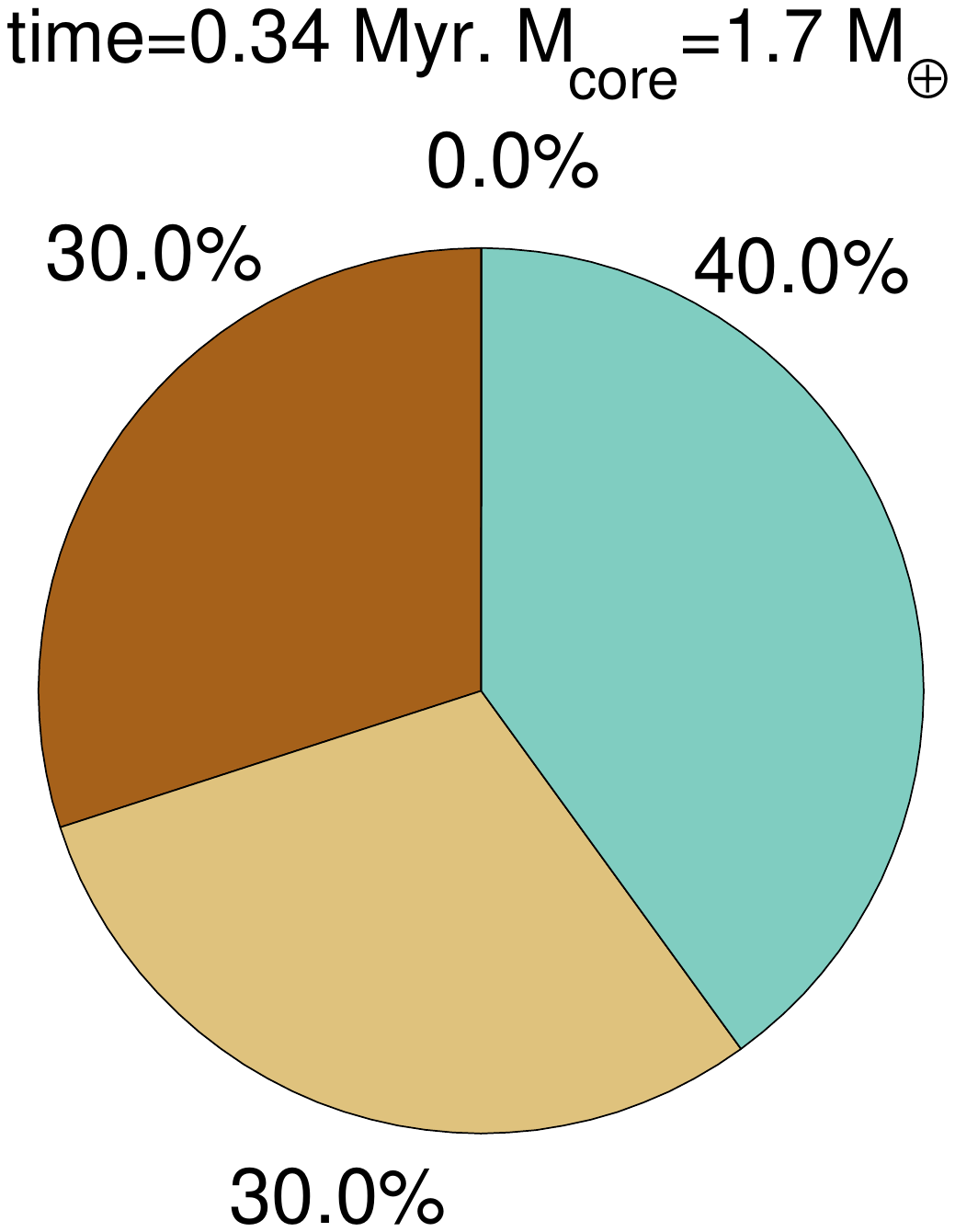}}
\subfloat{\includegraphics[trim=18 120 40 100,clip,width=0.32\columnwidth]{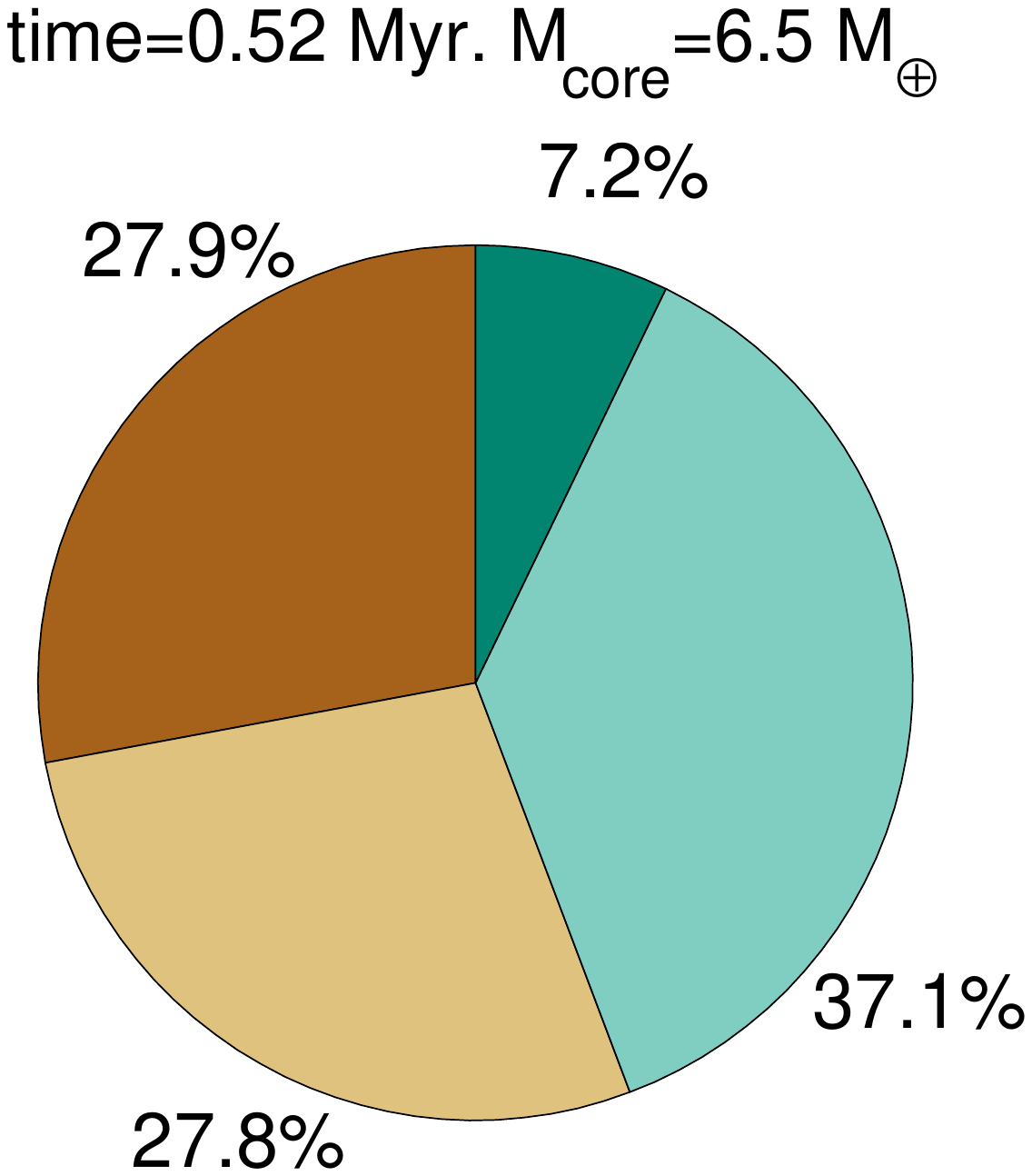}}
\subfloat{\includegraphics[trim=18 120 40 100,clip,width=0.32\columnwidth]{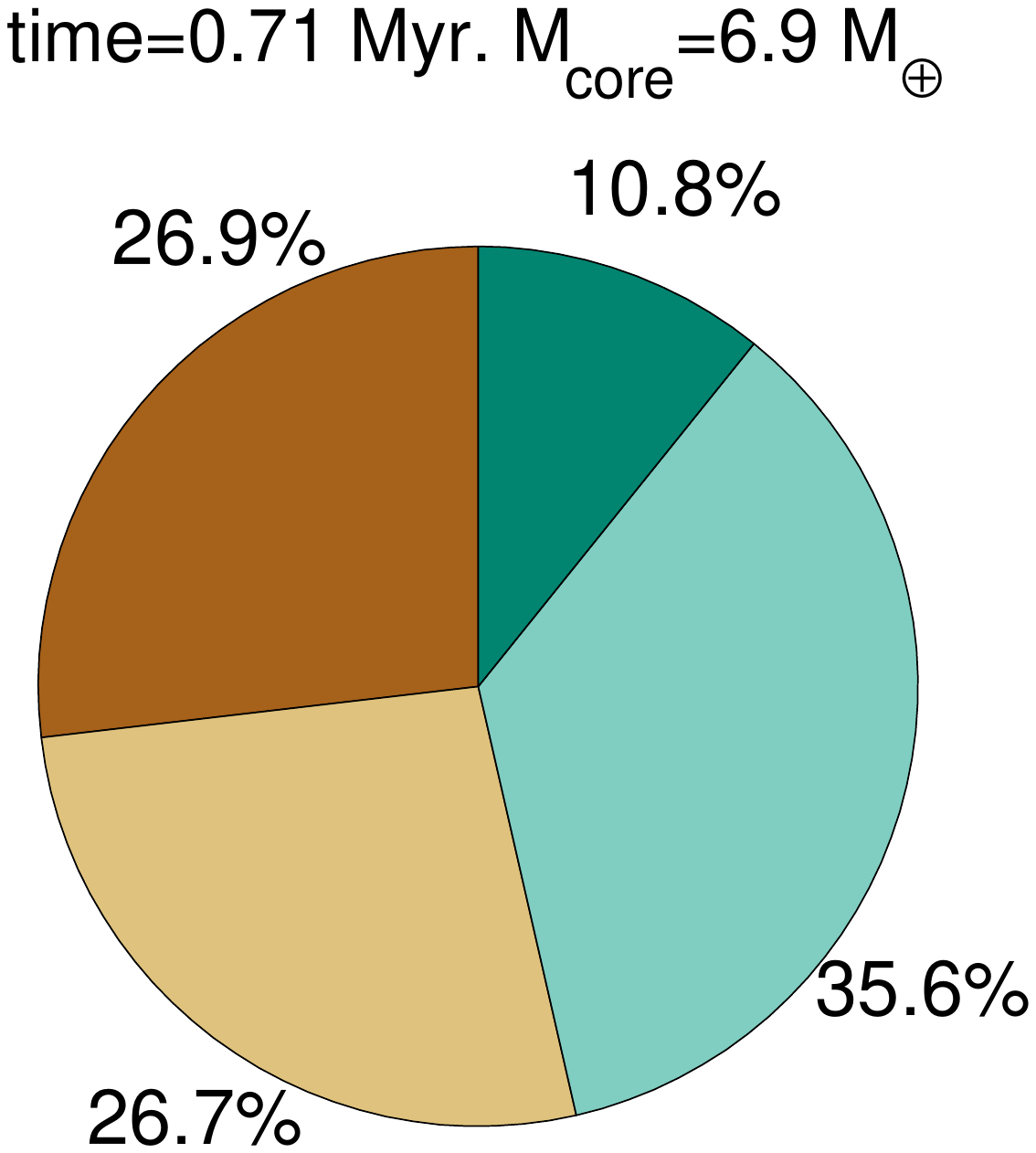}}\\
\vspace*{-1.in}
\subfloat{\includegraphics[trim=18 120 40 100,clip,width=0.32\columnwidth]{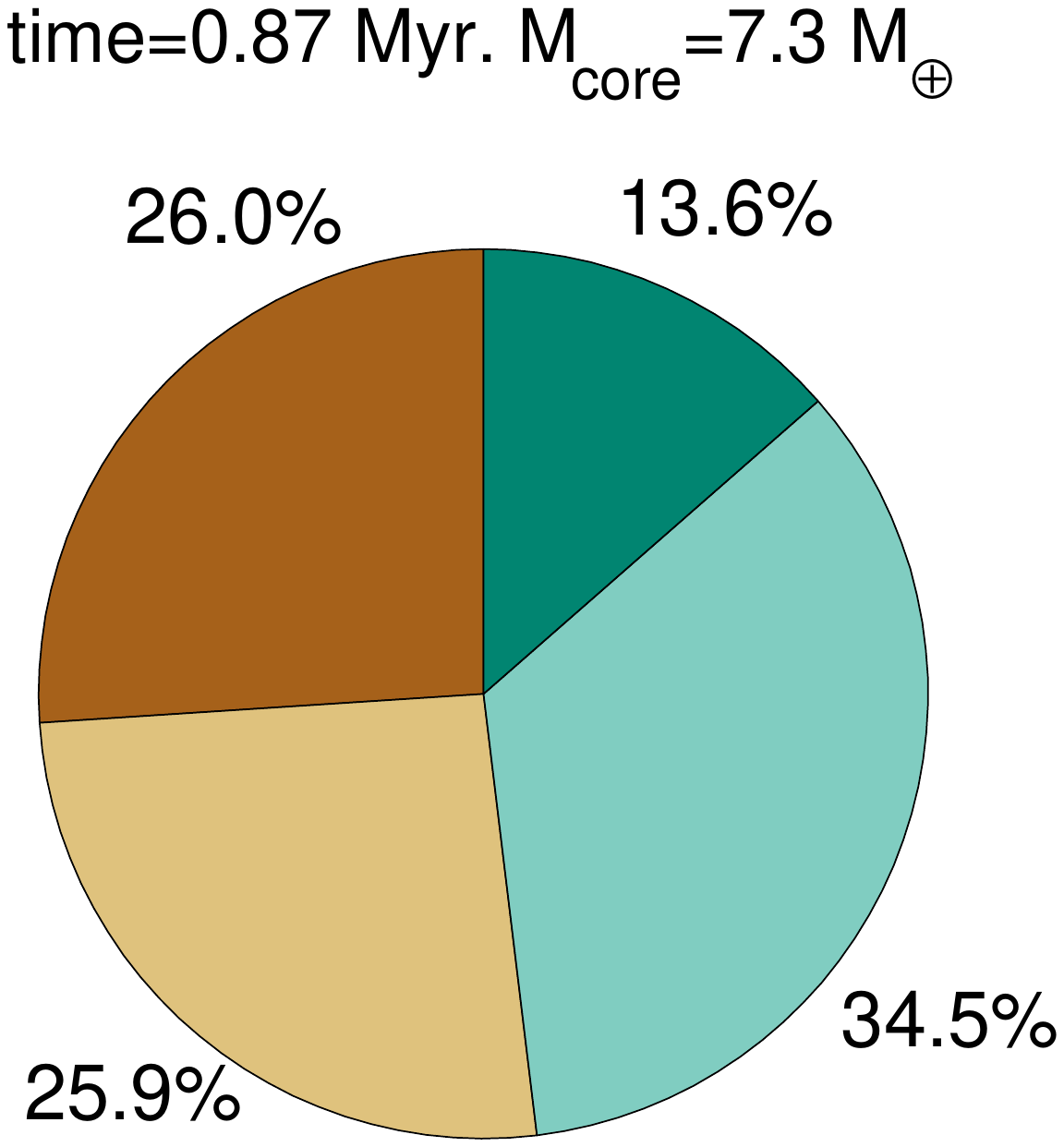}}
\subfloat{\includegraphics[trim=18 120 40 100,clip,width=0.32\columnwidth]{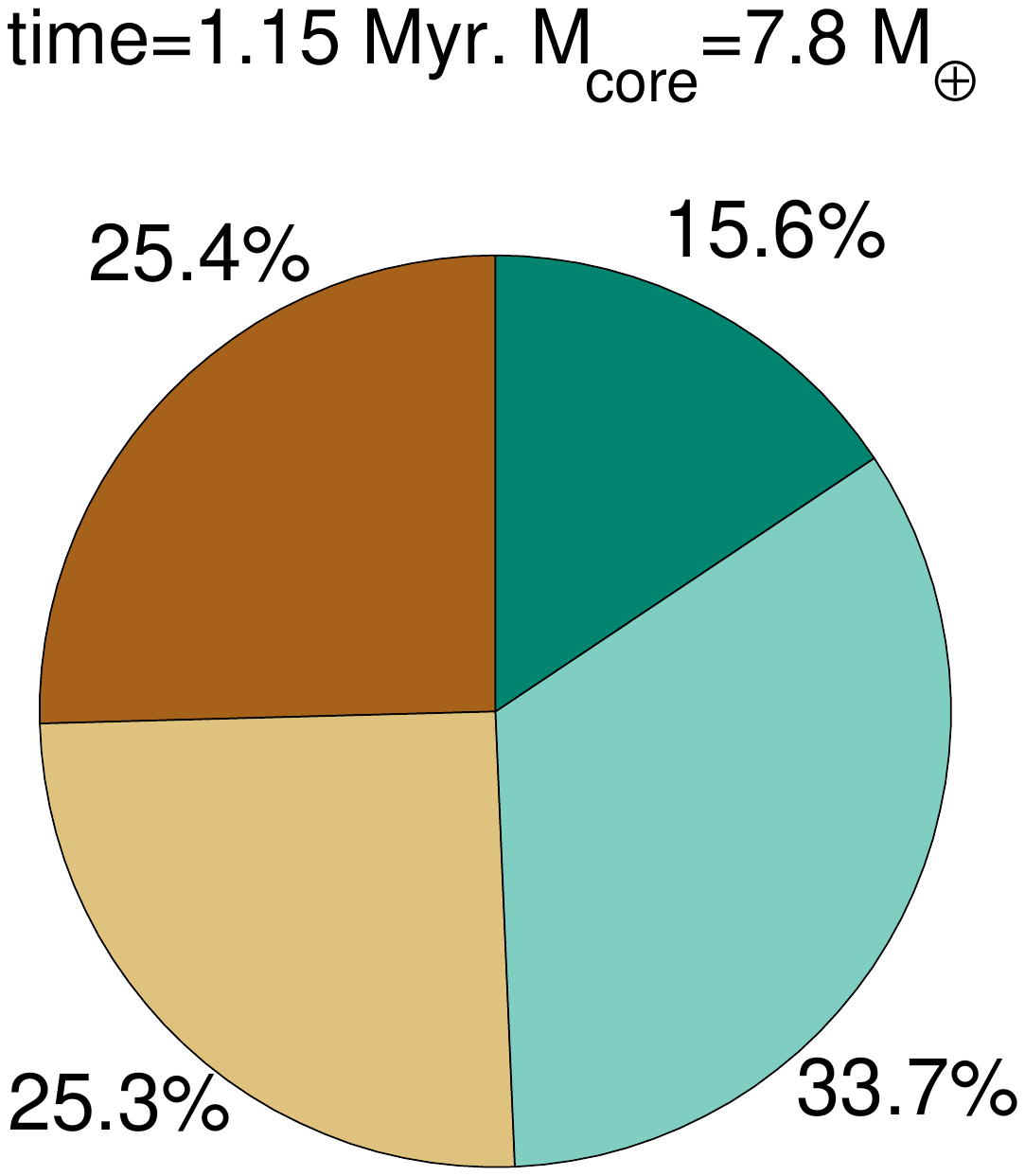}}
\subfloat{\includegraphics[trim=18 120 40 100,clip,width=0.32\columnwidth]{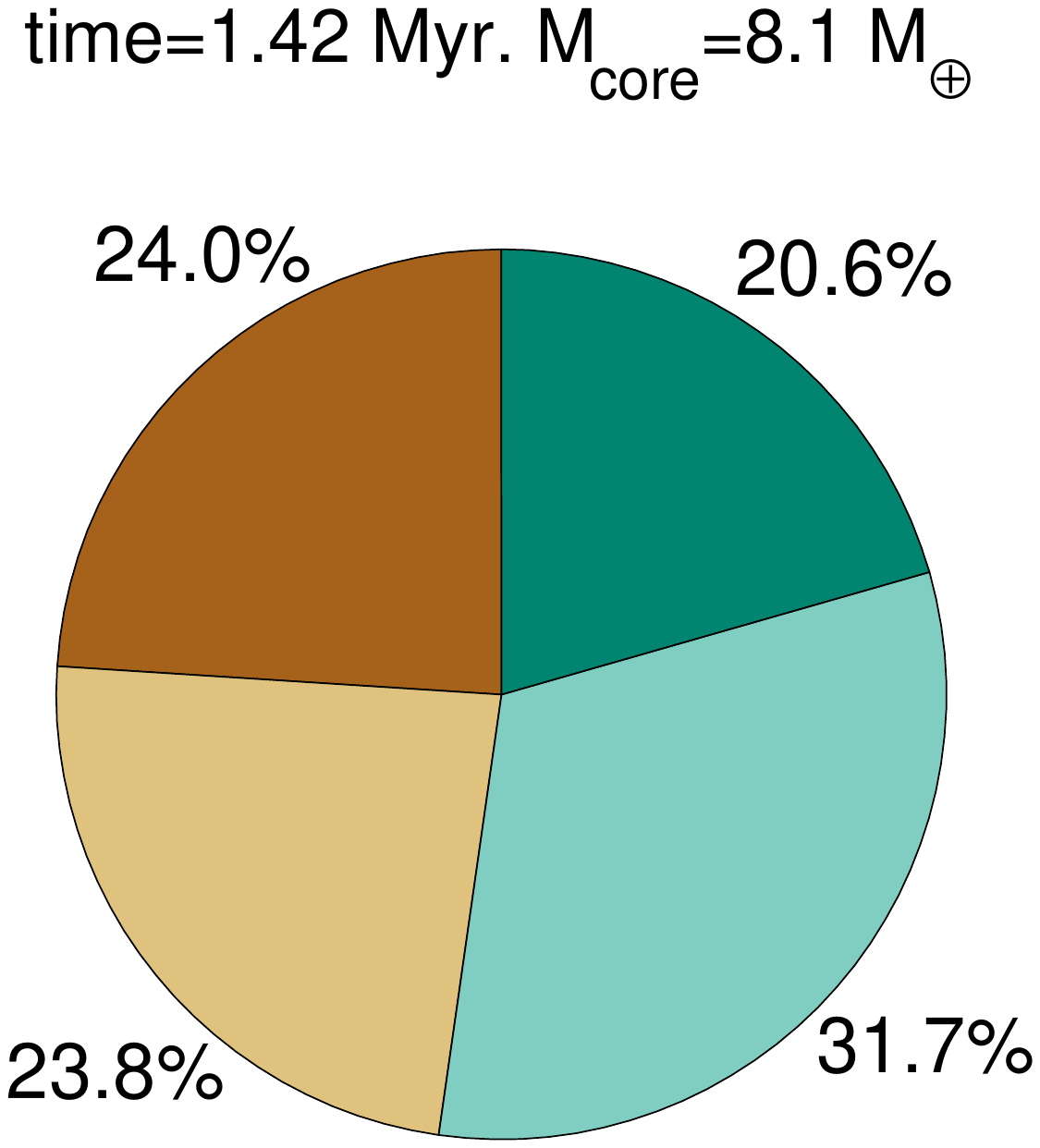}}\\
\vspace*{-10mm}
\subfloat{\includegraphics[trim=10 750 0 70,clip,width=1\columnwidth]{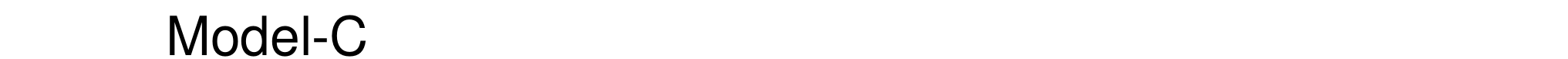}}\\
\vspace*{-10mm}
\subfloat{\includegraphics[trim=18 120 40 100,clip,width=0.32\columnwidth]{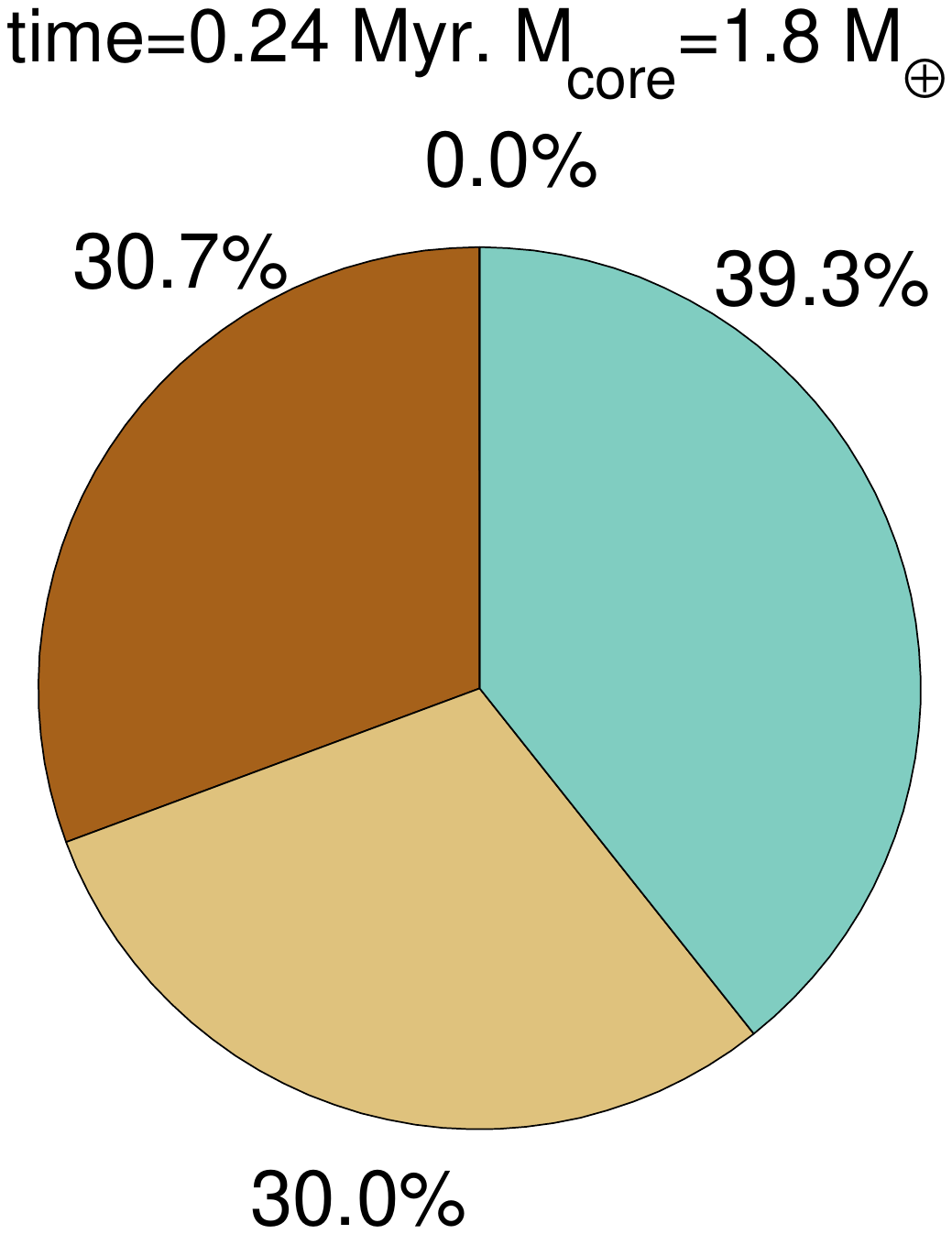}}
\subfloat{\includegraphics[trim=18 120 40 100,clip,width=0.32\columnwidth]{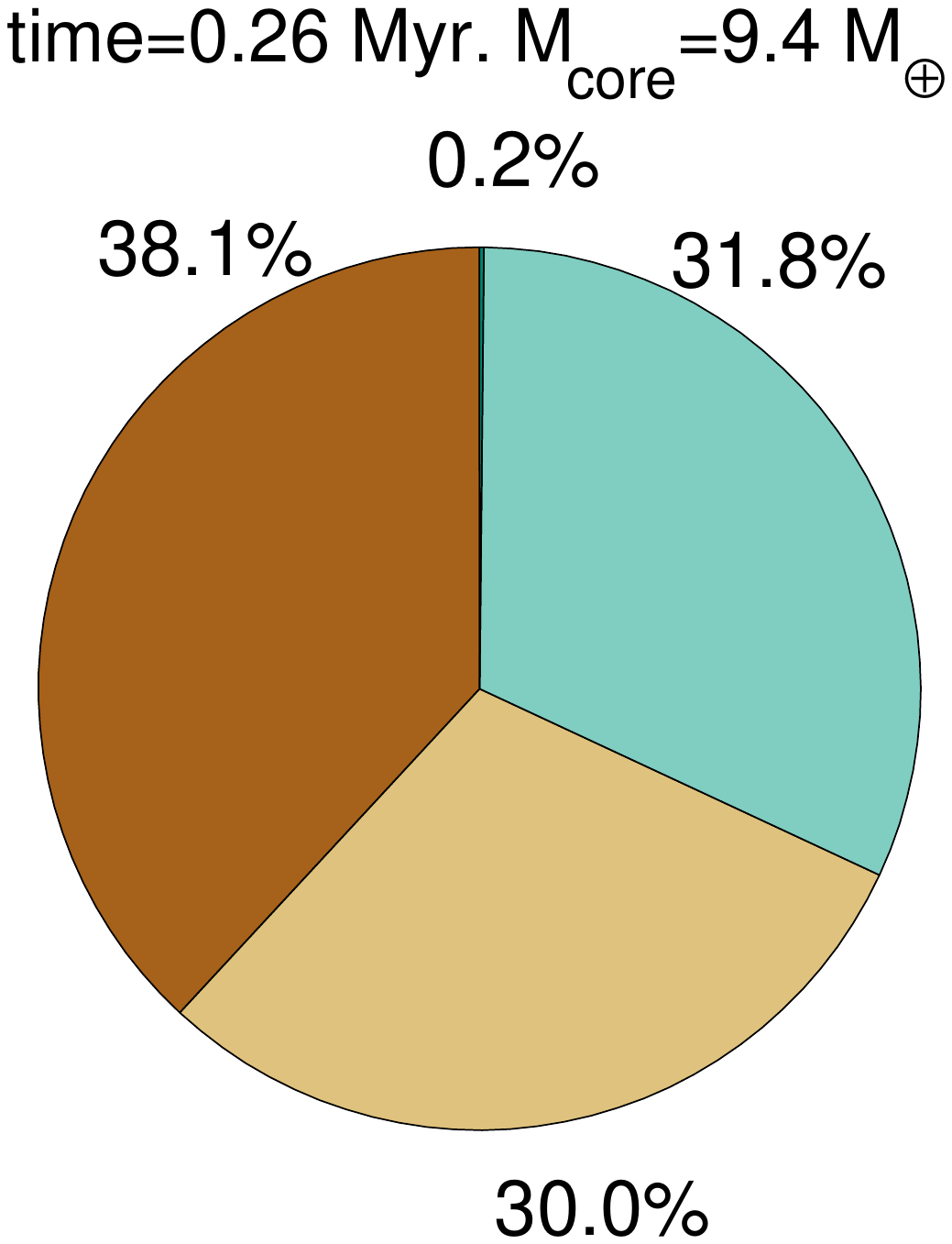}}
\subfloat{\includegraphics[trim=18 120 40 100,clip,width=0.32\columnwidth]{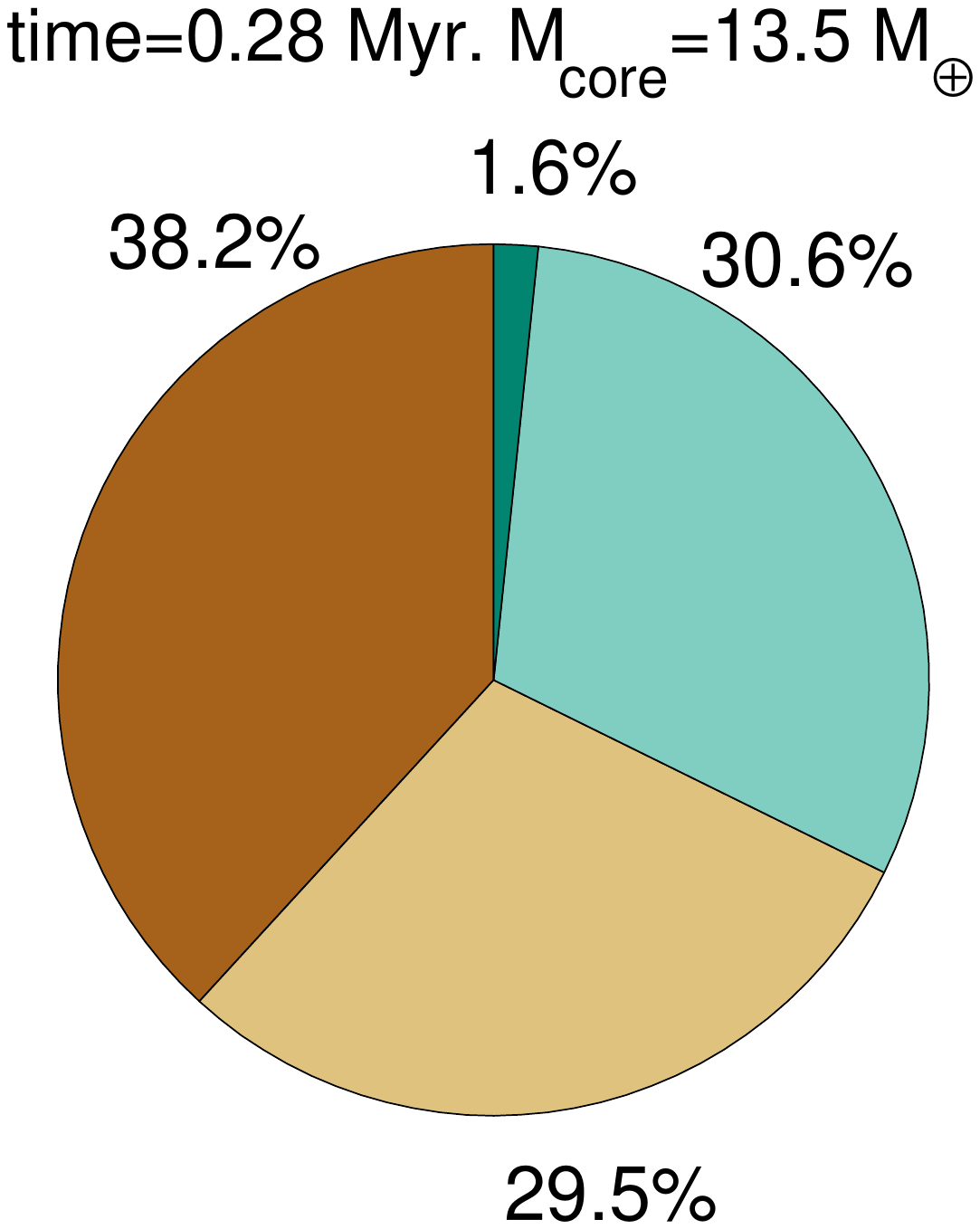}}\\
\vspace*{-1.in}
\subfloat{\includegraphics[trim=18 120 40 100,clip,width=0.32\columnwidth]{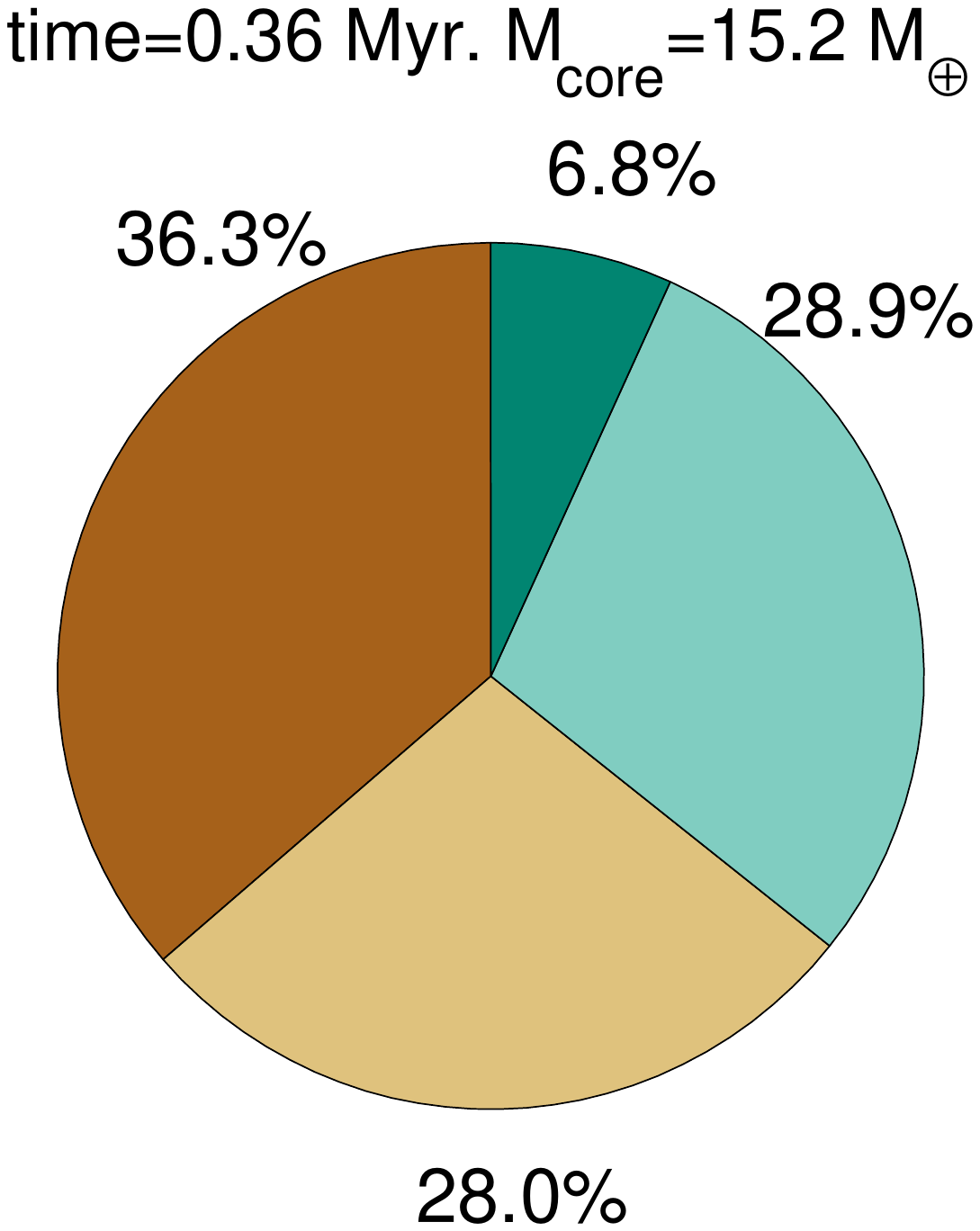}}
\subfloat{\includegraphics[trim=18 120 40 100,clip,width=0.32\columnwidth]{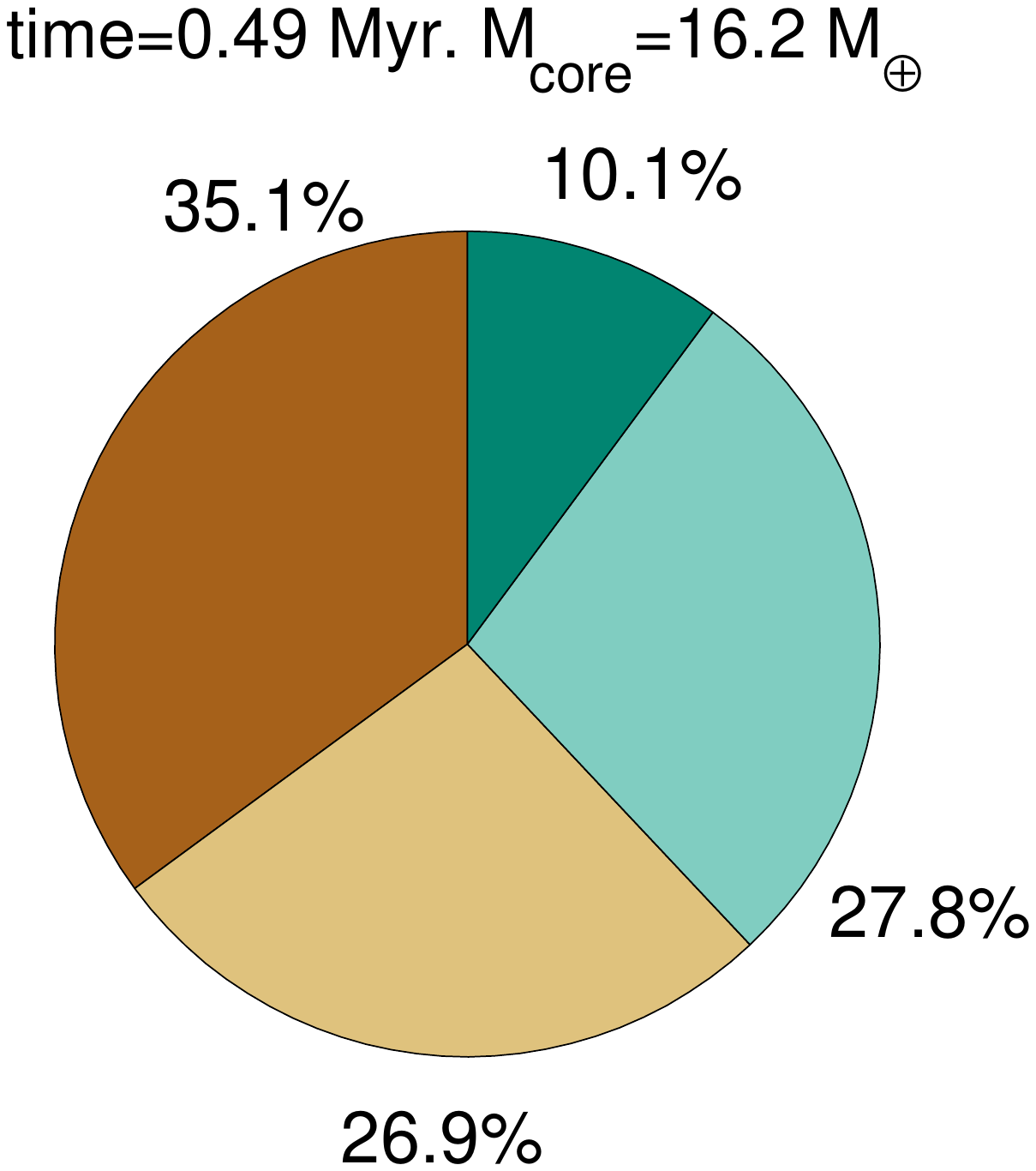}}
\subfloat{\includegraphics[trim=18 120 40 100,clip,width=0.32\columnwidth]{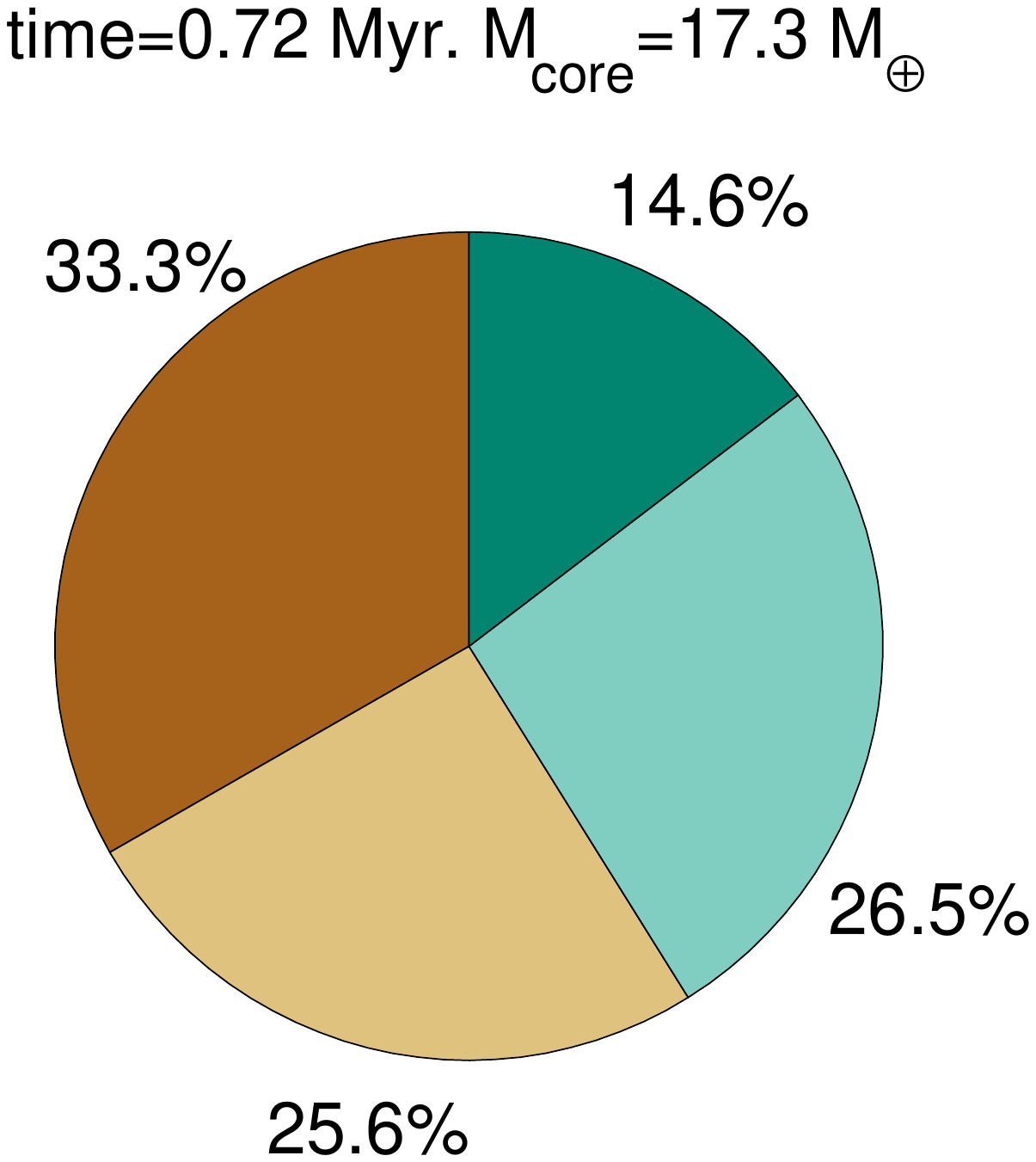}}\\
\vspace*{-8mm}
\subfloat{\includegraphics[trim=18 120 40 630,clip,width=0.6\columnwidth]{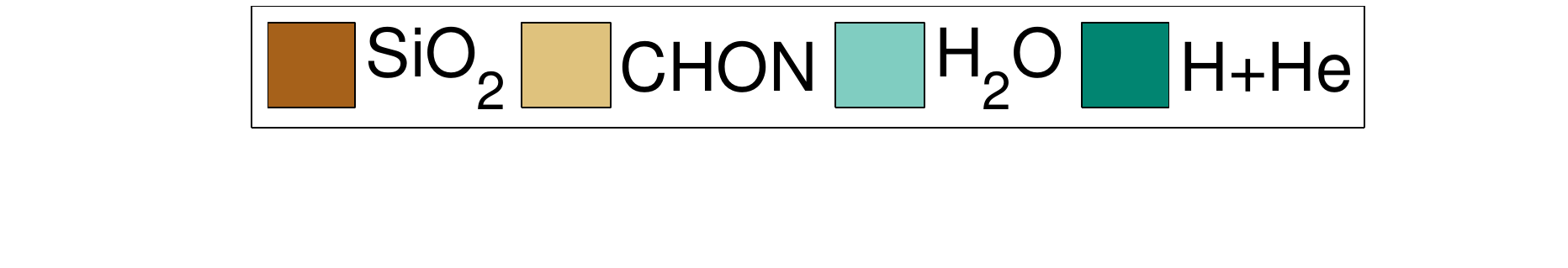}}\\
\vspace*{-8mm}
\caption{The derived core composition at various times for Model-B and Model-C. The core is defined as the innermost region with Z$>$0.5. H$_2$O, SiO$_2$, CHON and H+He are represented by the blue, brown, beige and turquoise colors, respectively.}\label{Pies, s=10, r=1}
\end{figure}

\newpage
\begin{figure}[!h]\centering
\subfloat{\includegraphics[trim=85 180 100 200,clip,width=0.38\columnwidth]{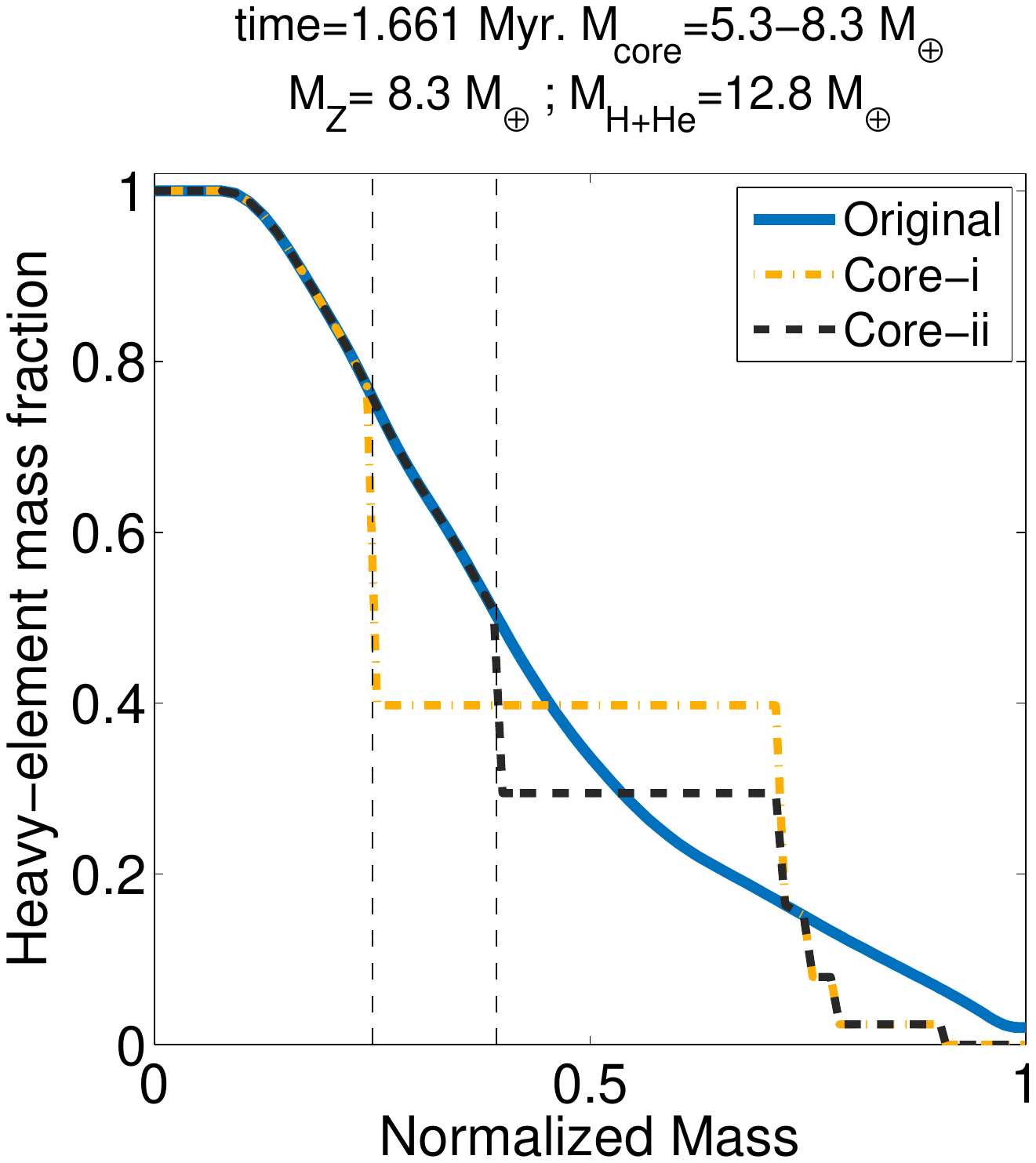}}
\subfloat{\includegraphics[trim=85 180 100 200,clip,width=0.38\columnwidth]{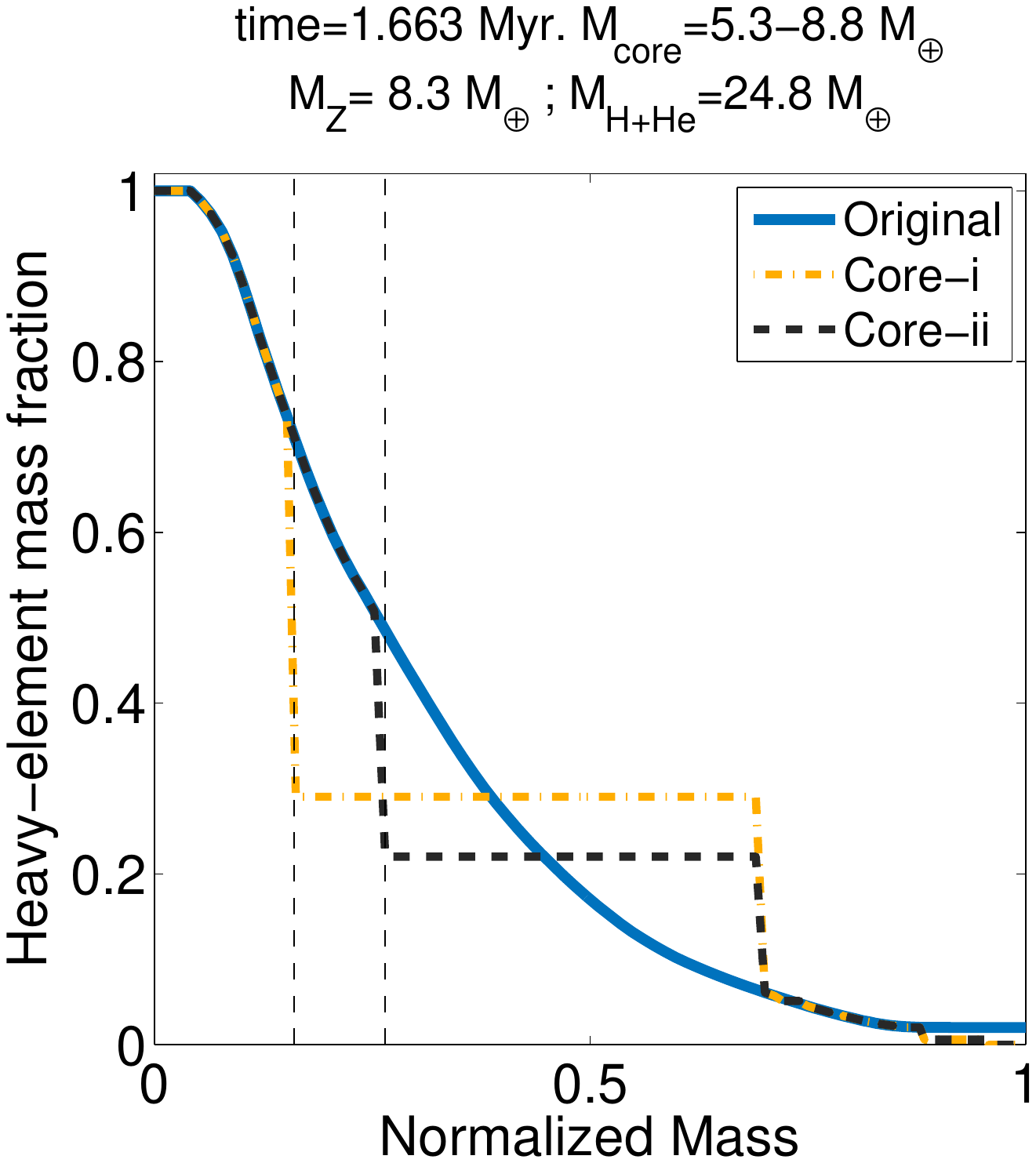}}\\
\vspace{-0.8cm}
\subfloat{\includegraphics[trim=85 180 100 200,clip,width=0.38\columnwidth]{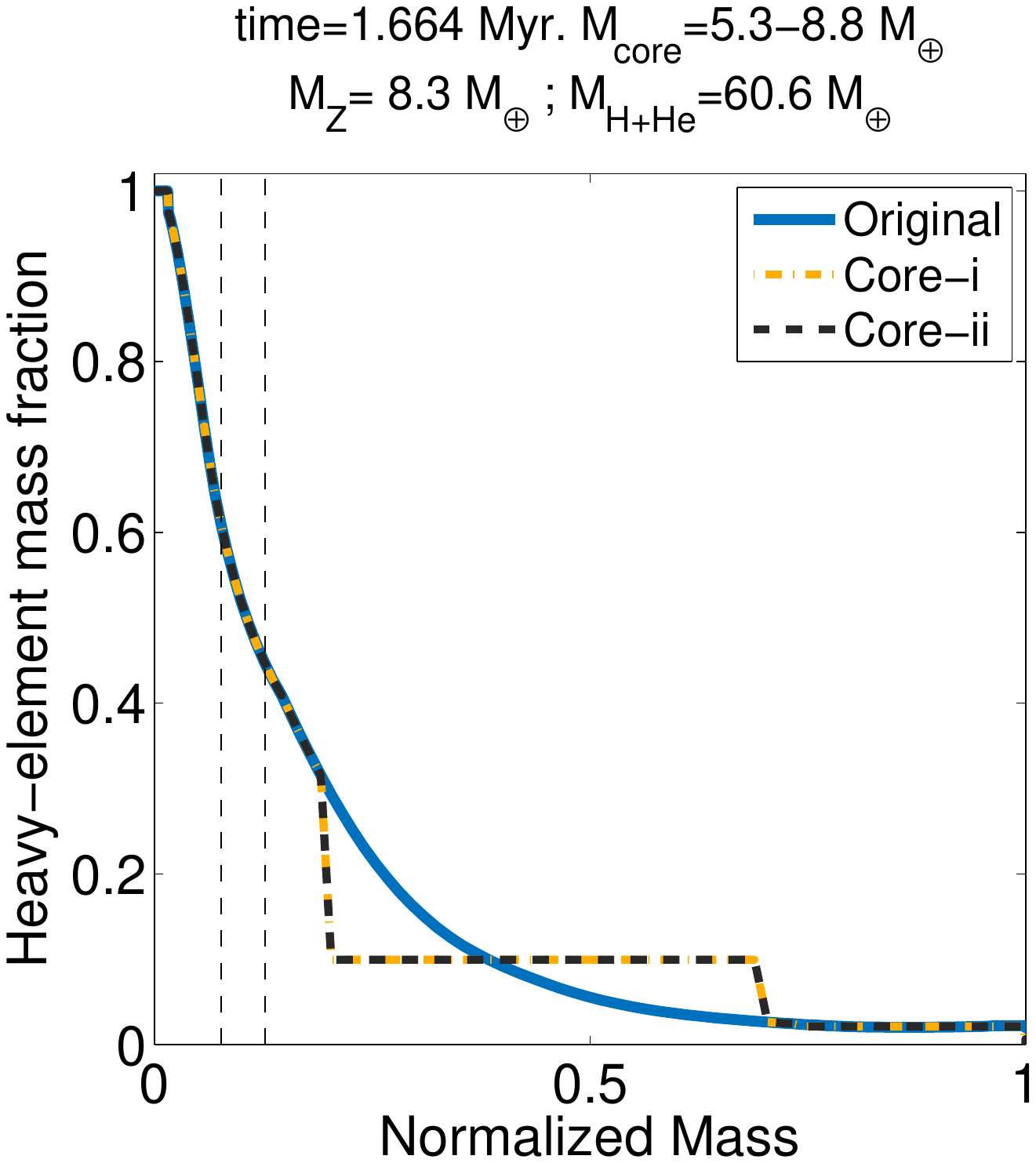}}
\subfloat{\includegraphics[trim=85 180 100 200,clip,width=0.38\columnwidth]{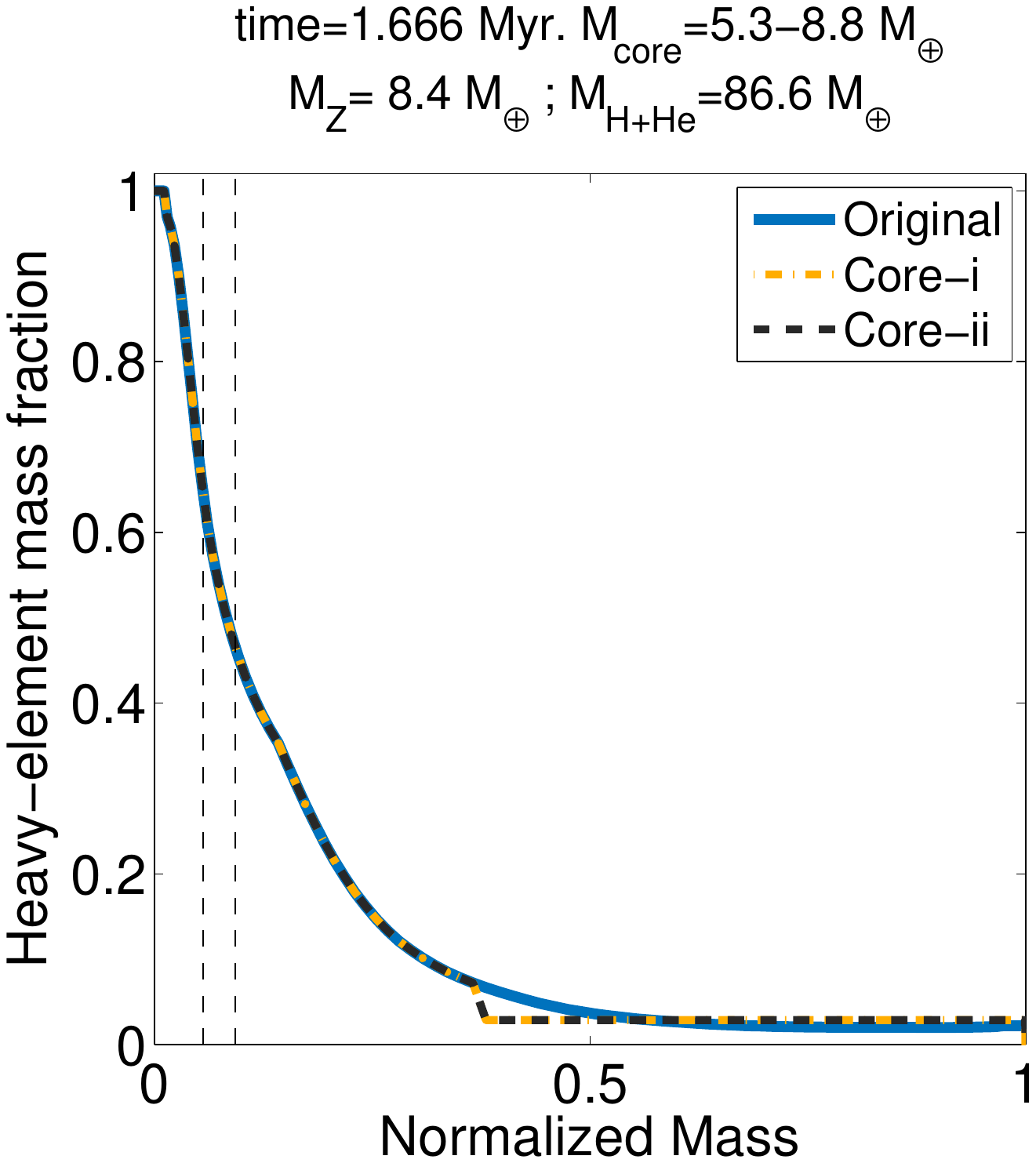}}\\
\vspace{-0.8cm}
\subfloat{\includegraphics[trim=85 180 100 200,clip,width=0.38\columnwidth]{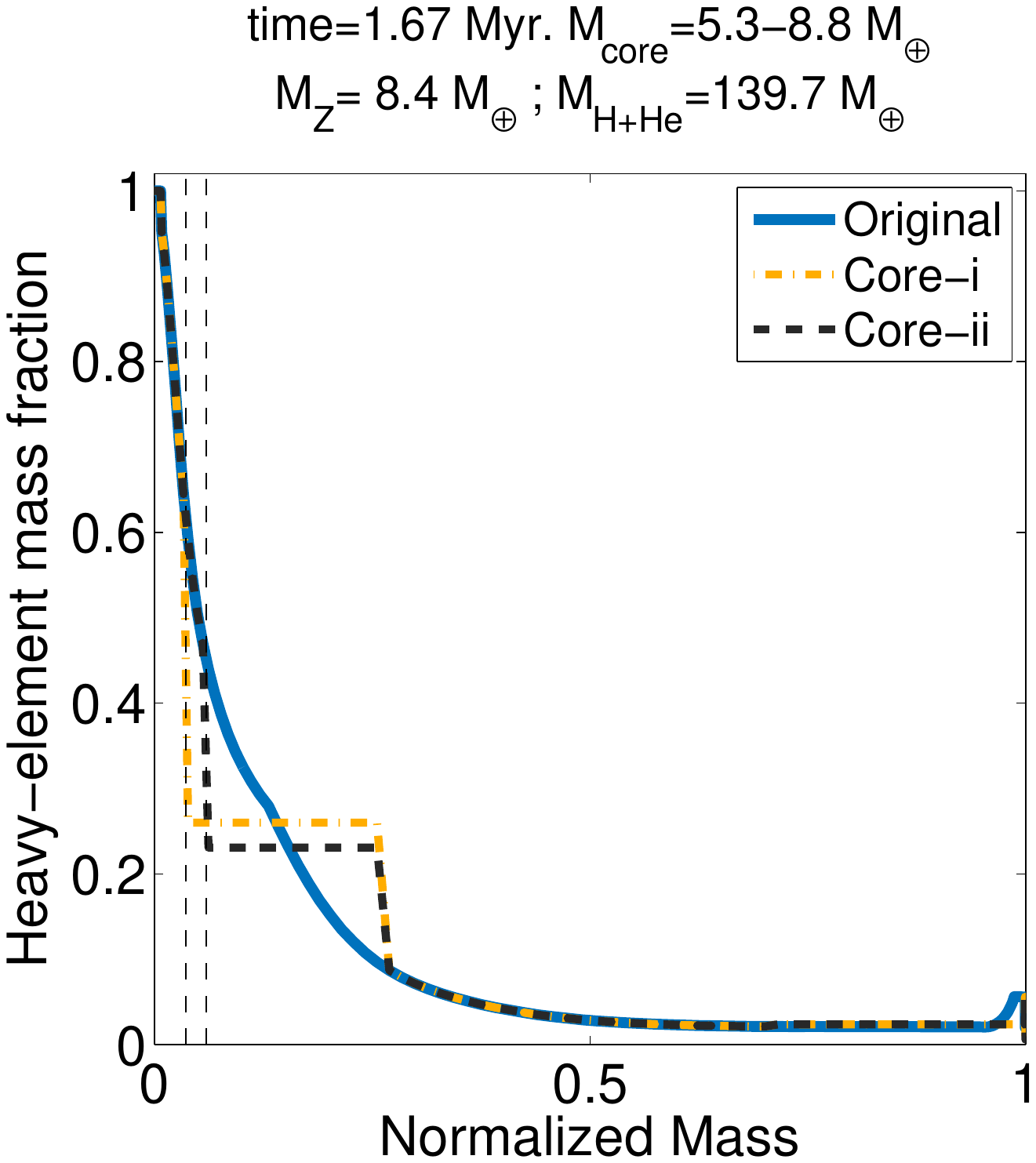}}
\subfloat{\includegraphics[trim=85 180 100 200,clip,width=0.38\columnwidth]{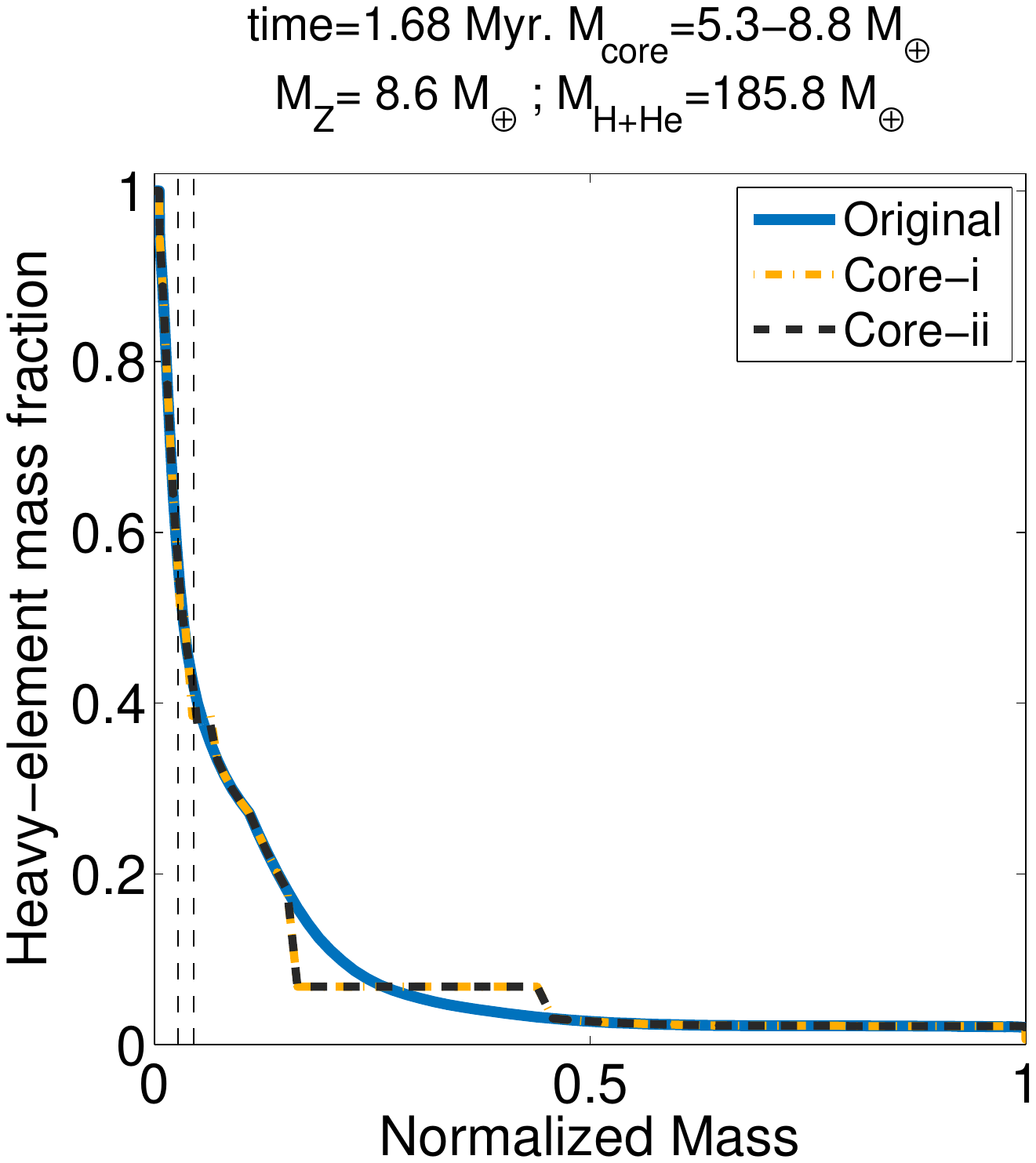}}
\caption{The distribution of the high-Z material vs.~normalized mass at different times for Model-A for Core-i (dashed-dotted orange) and Core-ii (dashed-black) during phase 3.}\label{fig:Z05Z09 - All the way, r6}
\end{figure}

\newpage
\begin{figure}[!h]\centering
\subfloat{\includegraphics[trim=85 180 100 200,clip,width=0.38\columnwidth]{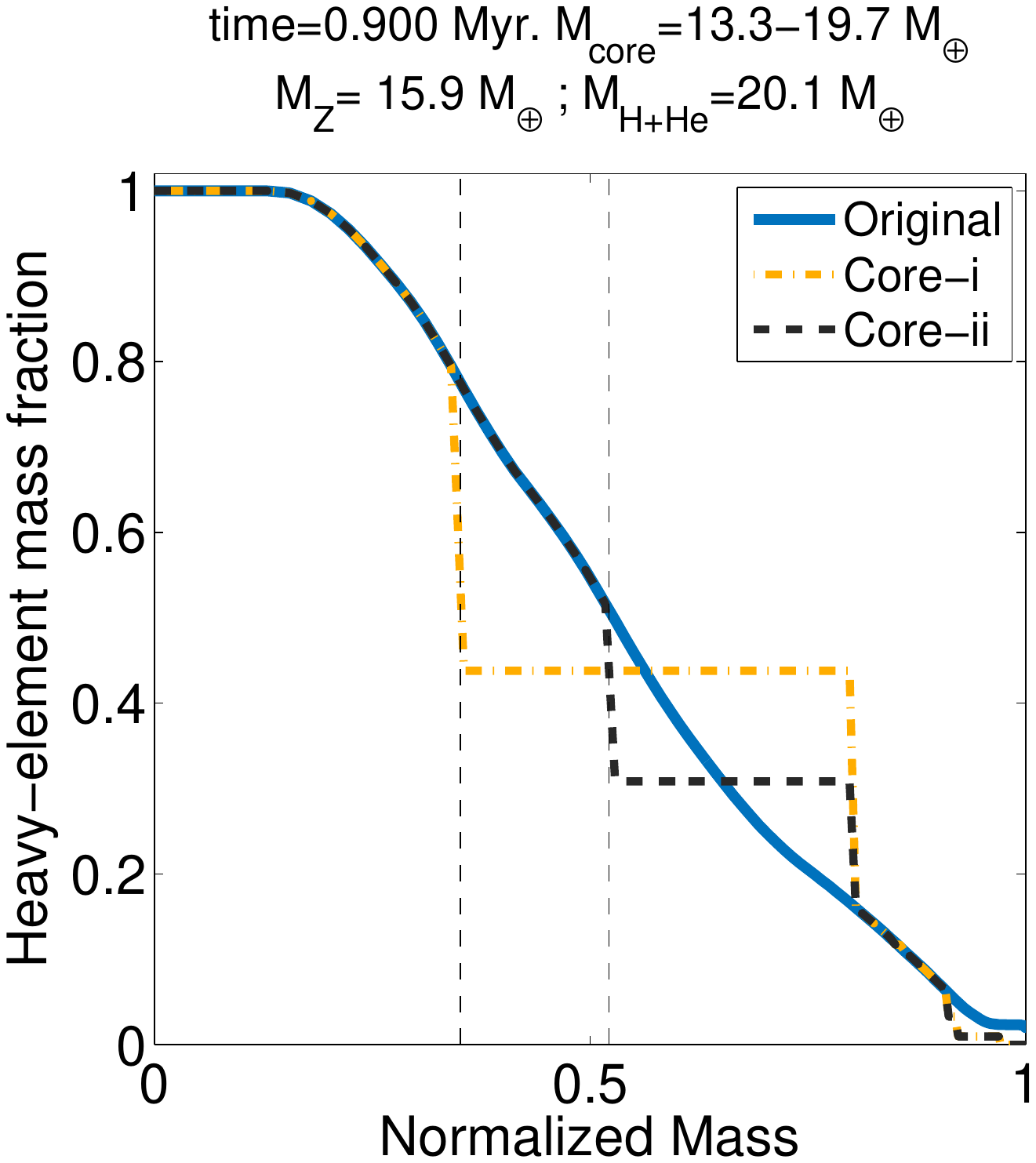}}
\subfloat{\includegraphics[trim=85 180 100 200,clip,width=0.38\columnwidth]{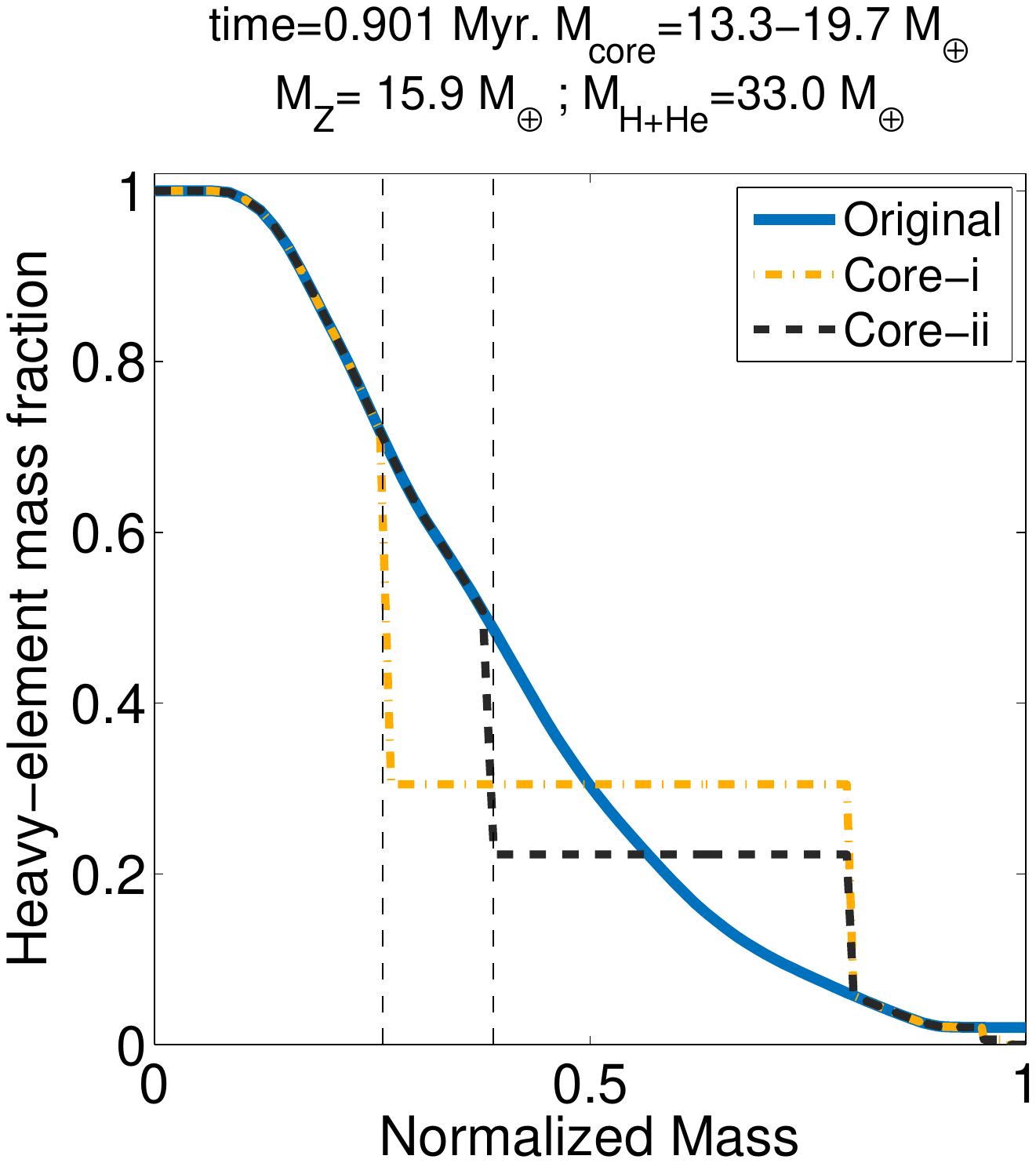}}\\
\vspace{-0.8cm}
\subfloat{\includegraphics[trim=85 180 100 200,clip,width=0.38\columnwidth]{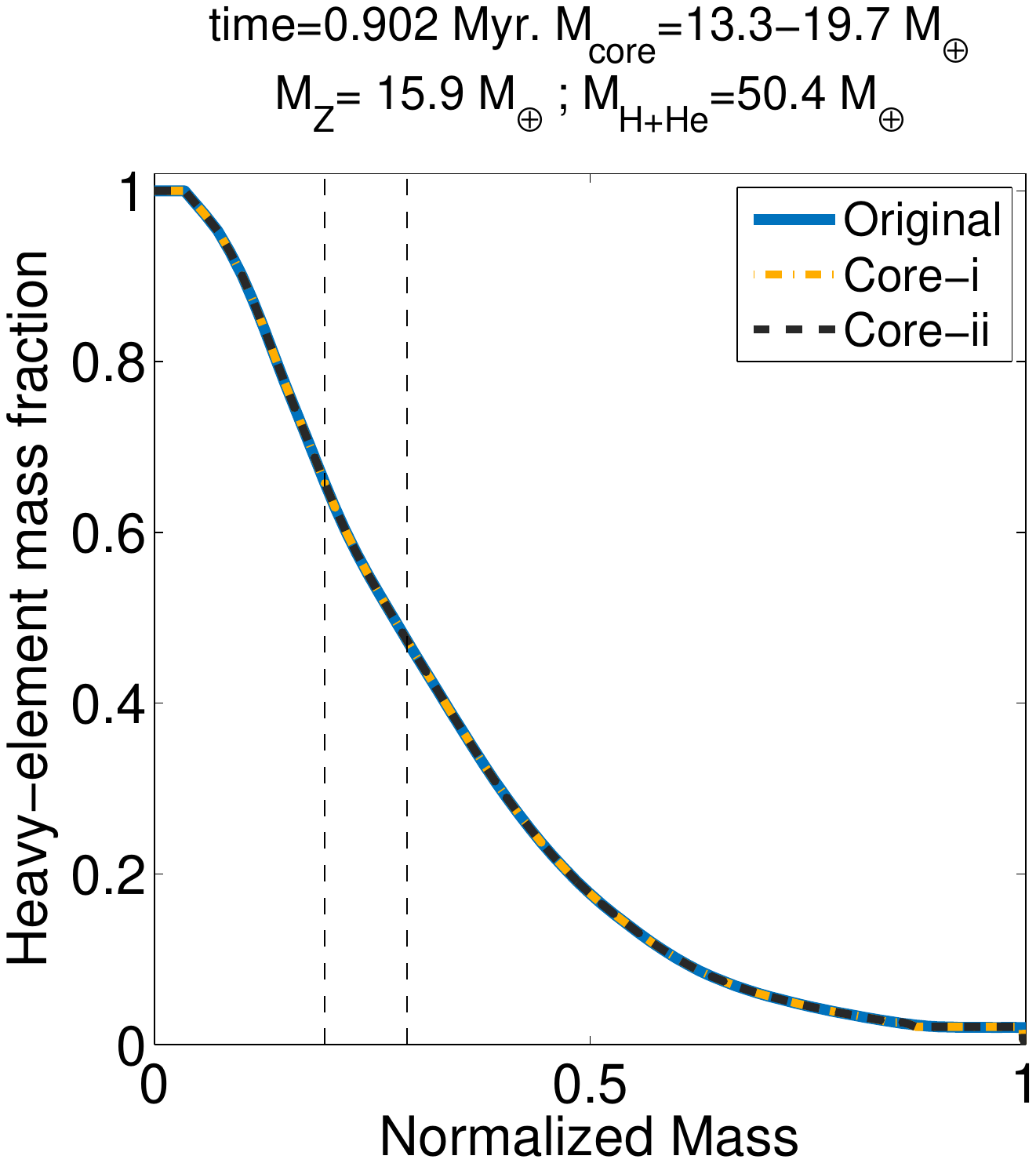}}
\subfloat{\includegraphics[trim=85 180 100 200,clip,width=0.38\columnwidth]{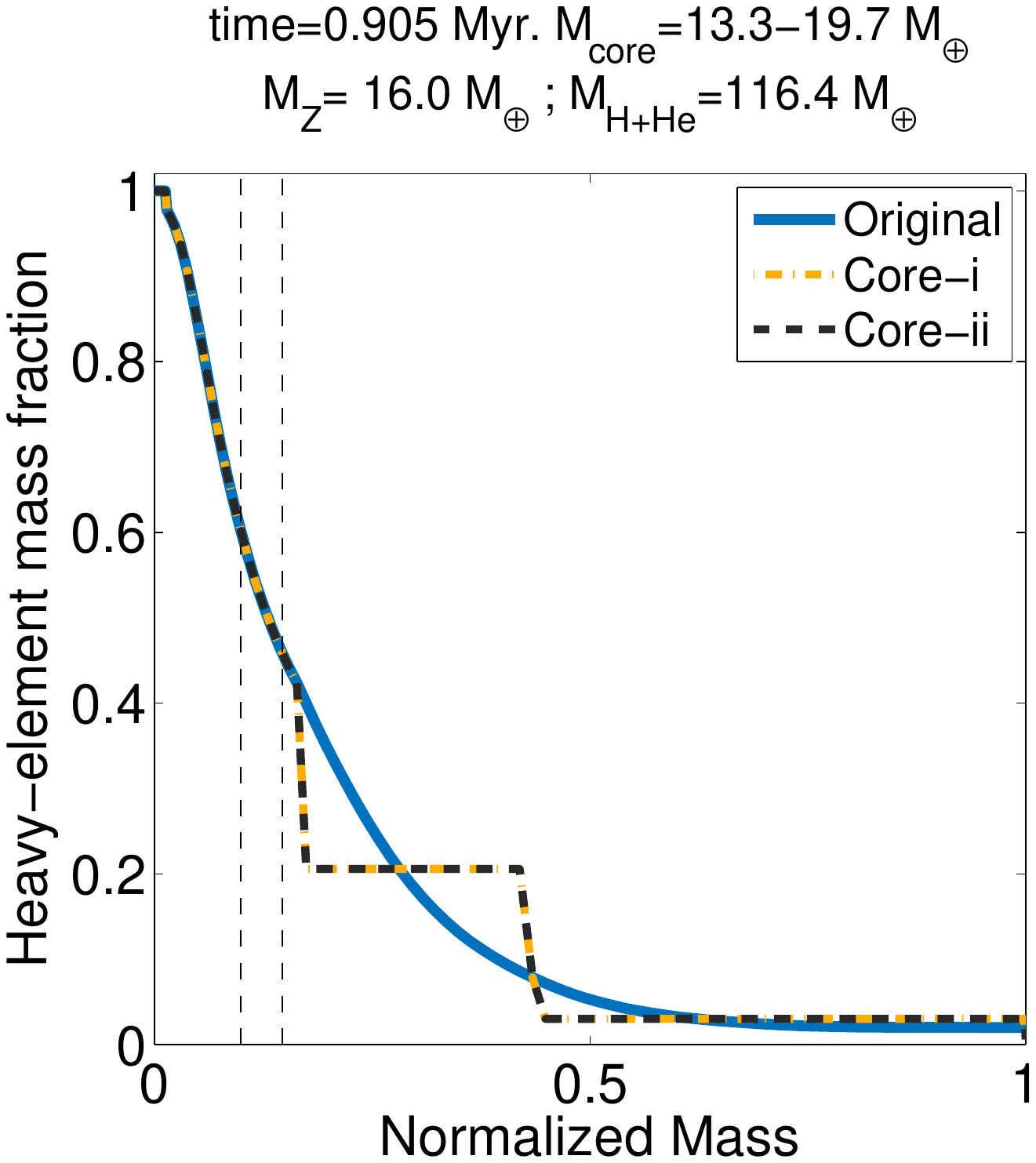}}\\
\vspace{-0.8cm}
\subfloat{\includegraphics[trim=85 180 100 200,clip,width=0.38\columnwidth]{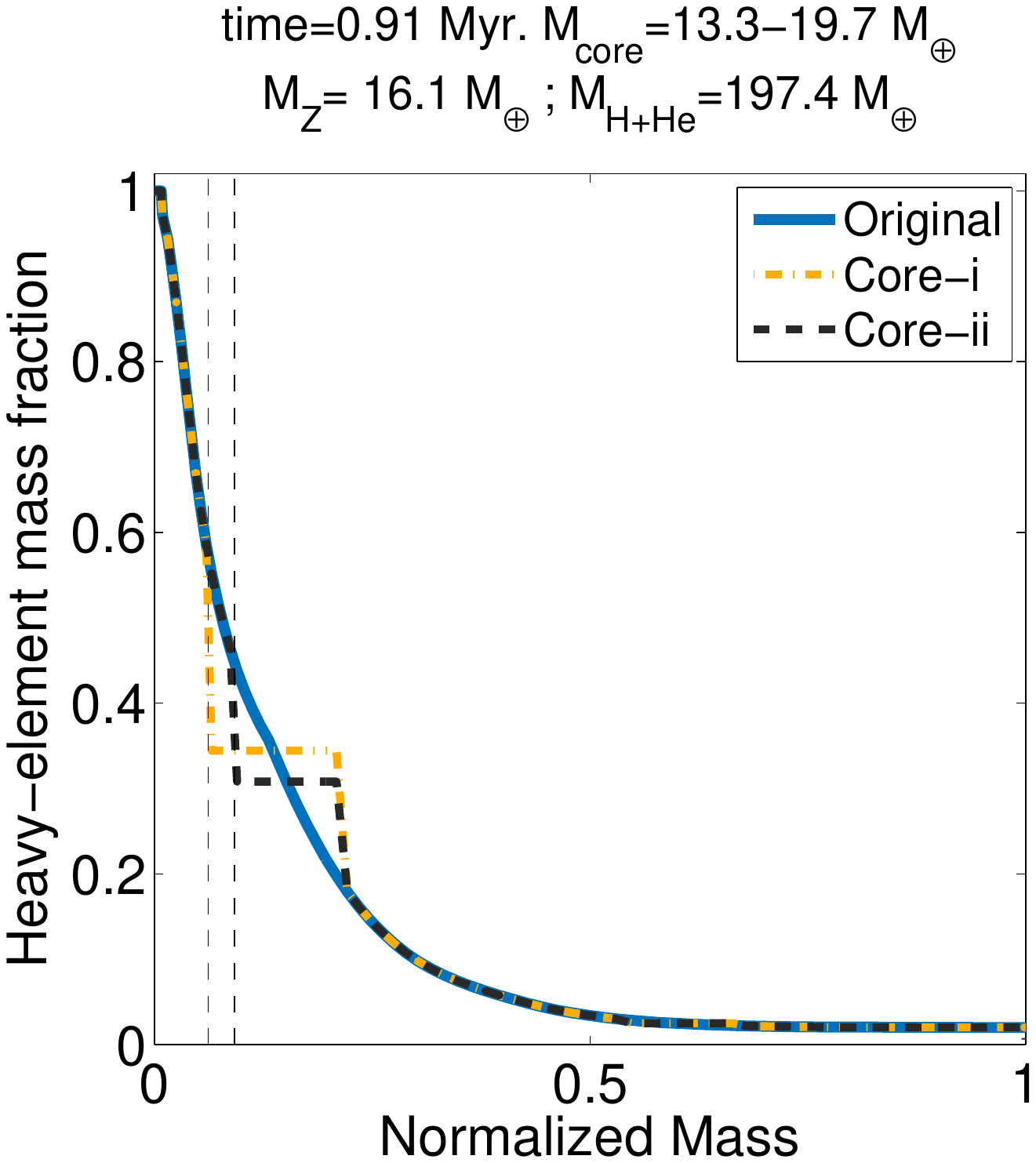}}
\subfloat{\includegraphics[trim=85 180 100 200,clip,width=0.38\columnwidth]{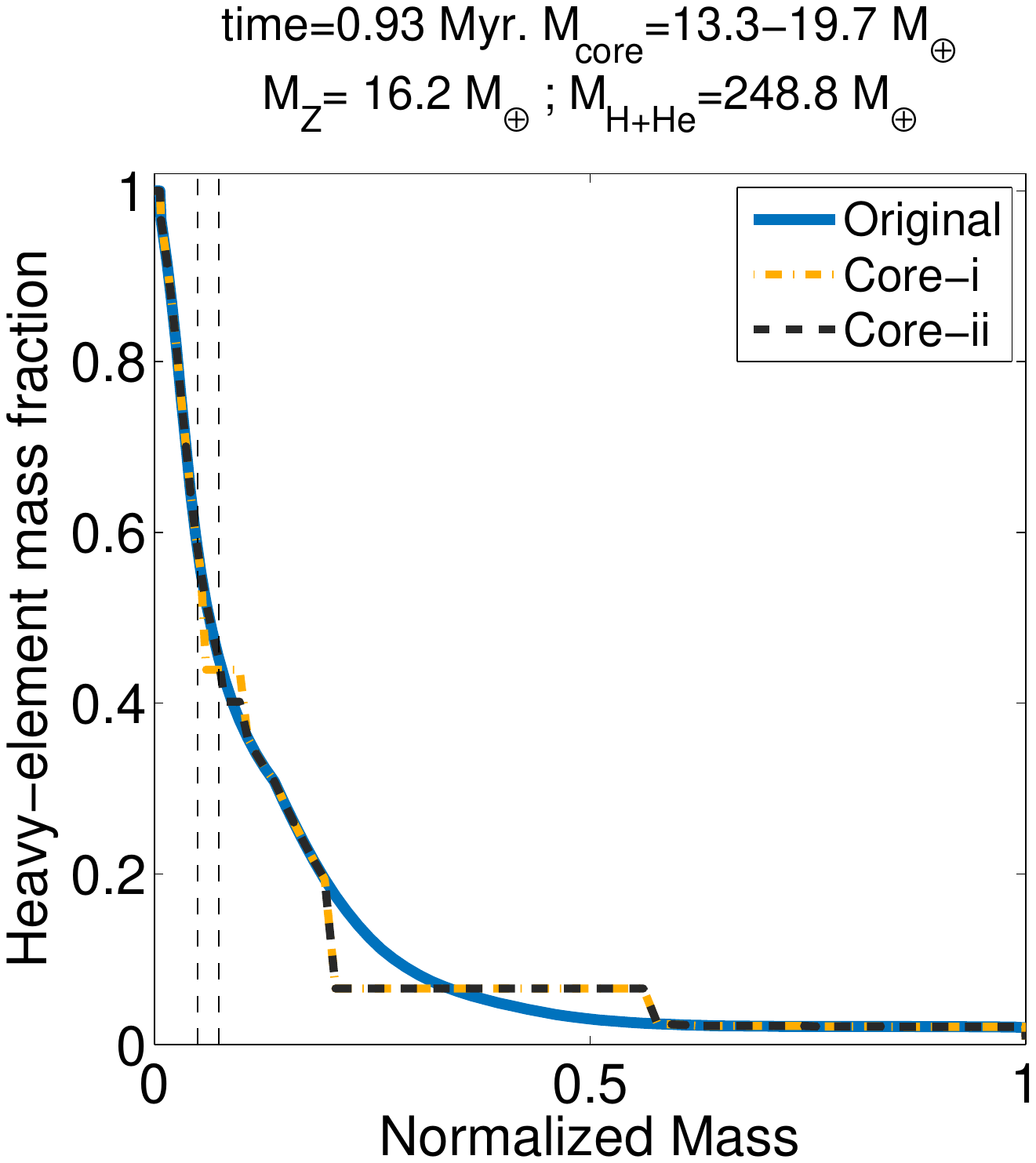}}
\caption{Same as \ref{fig:Z05Z09 - All the way, r6} for Model-C.}\label{fig:Z05Z09 - All the way, r10}
\end{figure}

\newpage

\end{document}